\documentclass[a4paper,12pt]{amsart}
\usepackage[left=1.2in, right=1.2in, bottom = 1.2in, top = 1.2in]{geometry}
\usepackage[utf8]{inputenc}
\usepackage{graphicx} % Required for inserting images
\usepackage[sort&compress]{natbib}
\usepackage{amsmath}
\usepackage{amsthm}
\usepackage{amsfonts}
\usepackage{amssymb}
\usepackage{enumitem}
\usepackage{algpseudocode}
\usepackage{algorithm}
\usepackage{booktabs}
\usepackage[table]{xcolor}
\usepackage{rotating}
\usepackage{pgfplots}
\usepackage{makecell}
\usepackage{centernot}

\usepackage{url}
\usepackage{setspace}
\newtheorem{definition}{Definition}

\newcommand{\independent}{\perp\mkern-9.5mu\perp}

\newtheorem{proposition}{Proposition}
\newtheorem{assumption}{Assumption}
\newtheorem*{assumption*}{Assumption}
\newtheorem{hypothesis}{Hypothesis}
\newtheorem{lemma}{Lemma}
\newtheorem{corollary}{Corollary}

\theoremstyle{definition}
\newtheorem{suppassumption}{Assumption}

\theoremstyle{remark}
\newtheorem{remark}{Remark}

\def\spacingset#1{\renewcommand{\baselinestretch}{#1}\small\normalsize}

\pgfplotsset{compat=1.18}

\usepackage{tikz}
\usetikzlibrary{arrows.meta,shapes}
\usetikzlibrary{arrows,shapes.arrows,shapes.geometric,shapes.multipart,
decorations.pathmorphing,positioning,shapes.swigs,}

\title{Causal Inference from Self-controlled Case Series}
\author{Gellért Perényi, Marco Piccininni, Mats J. Stensrud}
\date{November 2026}

\begin{document}

\begin{abstract}
    The self-controlled case series ($SCCS$) design is widely used in studies of rare events, for example of vaccine side effects. Conceptually, the design has two main features. First, it requires only data from individuals who have experienced the event of interest. Therefore, the design is especially useful when the outcome is rare and the available data come from surveillance or clinical records, say, of hospitalized patients. Second, because comparisons are made within single individuals, time-invariant confounding between treatment and outcome is eliminated. We formulate a set of sufficient identification assumptions for estimating the causal effect of vaccination on an adverse outcome under the $SCCS$ design, drawing on ideas from the causal inference literature. Unlike the existing SCCS literature, we do not consider a parametric model as the starting point. This allows us to establish when and why the $SCCS$ design allows identification of causal effects, using data from cases only. However, because the sufficient identifiability assumptions are implausible in some applications, we also develop statistical tests for the existence of a treatment effect that remain valid when only a subset of the identification assumptions holds.
\end{abstract}

\maketitle
\section{Introduction}

The causal effects of treatments on rare events are often of practical interest. Consider, for example, the effects of mRNA COVID-19 vaccination on myocarditis \citep{bozkurt_myocarditis_2021} and of the ChAdOx1 nCoV-19 vaccine on cerebral venous thrombosis \citep{greinacher_thrombotic_2021}. The presence of such effects would motivate preventive measures and inform future drug development. However, estimating treatment effects on rare events, like venous thrombosis shortly after vaccination, is difficult. Randomized trials are rarely powered for this purpose \citep{wittes_sample_2002, dai_empirical_2023}, as illustrated by the COVID-19 vaccine trials \citep{klein_surveillance_2021}. Even large observational studies often have insufficient precision or power for such questions, with the additional problem of possible confounding bias. Moreover, the costs and time associated with such studies are often prohibitive.

The self-controlled case series ($SCCS$) design \citep{farrington_relative_1995, whitaker_tutorial_2006, farrington_self-controlled_2018} has become a popular strategy for assessing treatment effects on rare outcomes, especially in drug safety studies \citep{grosso_use_2011, escolano_new_2013}. An $SCCS$ uses observational data from individuals who experienced the outcome only, i.e., a ``case series" \citep{petersen_self_2016}. The design requires repeated measurements from the same individual over time, allowing each individual to serve as their own ``self-control'' \citep{farrington_self-controlled_2018}. This is an attractive feature for a study design, because detailed clinical information is often available only for those patients who experienced the (rare) outcome, for example, from clinical records or health surveillance systems. The $SCCS$ is a case-only, self-controlled design, like the related but distinct case-crossover design \citep{maclure_case-crossover_1991}, whose causal interpretation has recently been formalized by \citet{shahn_formal_2023}.

The $SCCS$ design was described by \citeauthor{farrington_relative_1995} in 1995. Since then, the design has been used in several applications \citep{douglas_exposure_2008, minassian_invasive_2010, warren-gash_influenza_2012,  zenner_risk_2012, langan_risk_2014, shin_cardiovascular_2016, warren-gash_laboratory-confirmed_2018, lane_safety_2020, katsoularis_risk_2021}. Yet, a general causal justification that covers the structure and analysis of standard $SCCS$ studies remains lacking. 

The traditional literature on $SCCS$ motivates this design through a convenient property of the likelihood function under a specific parametric model \citep{farrington_relative_1995, whitaker_tutorial_2006, farrington_self-controlled_2018}. Using the parametric model as a starting point does not make explicit what the causal estimand of interest is, e.g., target trial specification, and what identification conditions are used. Without an explicit causal notation, it is difficult to develop a sound argument showing that the case-only selection permits a consistent estimator.

Recent attempts to give a causal justification for the $SCCS$ design have been limited to two time points and non-recurrent outcomes, and simple non-parametric estimators \citep{etievant_causal_2026}. Consequently, these results do not cover the data structure, implicit assumptions, or estimation techniques used in real-world $SCCS$ applications.

In this work, we give a causal justification for the $SCCS$ design in a realistic setting. We specify the target trial that the $SCCS$ design seeks to emulate, establish the identification argument without invoking parametric assumptions, and justify commonly used parametric estimators. Furthermore, we emphasize the differences between estimation and hypothesis testing and show that tests of certain causal null hypotheses can be valid under considerably weaker assumptions than those required for point estimation.

\section{A motivating target trial}\label{SEC: target}

Consider a population of independent and identically distributed  (i.i.d.) individuals. We are interested in whether a vaccine has an adverse effect in this population, specifically whether it changes the rate of thrombotic events. Suppose that the vaccine has a so-called reactogenicity window \citep{herve_hows_2019} defined here as the period during which vaccine-induced biological processes are believed to affect the rate of the adverse events. We denote the known length of the reactogenicity window by $m$. 

To fix ideas, suppose that the vaccine reactogenicity window for thrombosis is known to be 2 weeks ($m=14$ days) and, thus, consider a randomised controlled trial (RCT) with 14 days of follow-up after treatment initiation. Let $A_k$ indicate whether a participant receives the vaccine on day $k$ ($A_k=1$) or not ($A_k=0$), and let $Y_k \in \{0, 1, 2 \dots\}$ denote the number of thrombotic events experienced by the participant on day $k$. We choose days as the unit of time, so that within each day, individuals can receive at most one vaccination, ruling out multiple treatments on any given day $k$. To simplify the presentation, we ignore death and assume that all participants remain alive for the 14 days of follow-up.

In the RCT, individuals are randomised to either receive the vaccine on day $1$ only ($A_1=1, A_2= \dots=A_m=0$) or receive no vaccine ($A_1=0=A_2= \dots=A_m=0$). The rate ratio from such an RCT, comparing the expected rate of thrombotic events between the two arms, is defined as follows:

\begin{definition}[Cumulative rate ratio]
    $CRR_m:= \frac{\sum_{k=1}^{m}E[Y_k^{\overline{a}_m=(1, 0 \dots, 0)}]/m}{\sum_{k=1}^m E[Y_k^{\overline{a}_m=(0, \dots, 0)}]/m}$.
\end{definition}

Here, an overbar denotes the history of a (random) variable up to and including time $k$, with the convention $A_{k}=0$ for all $k\leq 0$, while superscripts indicate potential outcomes \citep{hernan_causal_2025}. Thus, $Y_k^{\overline{a}_m=(1, 0 \dots, 0)}$ denotes the number of thrombotic events a participant would have had on the $k$-th day had they, possibly contrary to fact, been vaccinated on day $1$ only and remained unvaccinated thereafter (intervention setting $\overline{A}_k=(1,0 \dots, 0)$).

We focus on the effect of a treatment (e.g., vaccination) vs. no treatment. However, our results can easily be adapted to extensions comparing different treatments, (e.g., comparison of two different vaccines), or assessing the effect of multiple doses (e.g., receiving two vaccine doses versus one). 
See Appendix \ref{APP: multi v} for further discussion.
.

\section{Data generation}\label{SEC: dat gen}

Although the $SCCS$ design uses data from cases only, we begin with a description of an underlying cohort from which the cases arise. This clarifies the assumptions required for an $SCCS$ analysis to identify the $CRR_m$. Consider a cohort of i.i.d. individuals followed up for $t \geq 2$ time points, with measurements taken at each $k \in \{1, \dots, t\}$.  To simplify the exposition, we assume that no individuals are lost to follow-up or die until time $t$, we relax this assumption in Appendix \ref{APP: death}.

The outcome is affected by time-fixed covariates, denoted by $U$, and time-varying covariates, denoted by $L_k$. These variables may also affect the probability of treatment. Consider, for example, a setting in which vaccination is more frequently recommended to individuals with certain time-invariant characteristics ($U$), e.g., sex, underlying health conditions, or residence in a specific geographic area, which also affect the rate of adverse events. In contrast, age and seasonality are typical examples of time-varying covariates $L_k$. We assume that all time-varying covariates are measured, but $U$ can be unmeasured.

Let $V_k$ indicate whether an individual is within the vaccine reactogenicity window at time $k$ \citep{herve_hows_2019}. Here, $V_k$ does not represent vaccine administration itself, but whether the individual has an altered ``susceptibility'' to the outcome because of vaccine-induced biochemical processes. Suppose that vaccination at time $k$ ($A_k=1$) initiates a period of altered susceptibility beginning at the time of the vaccination, and this vaccine-induced altered susceptibility period lasts for the next $m$ time points. More formally, $A_k = 1\implies V_{k}=\dots = V_{k+m-1}=1$ for some $m\geq1$.

Conversely, an individual has an unaltered susceptibility at time $k$ if they remained unvaccinated during the preceding $m$ relevant time points of the reactogenicity window, that is,
$ A_{k-m+1}= \dots = A_k=0 \implies V_{k}=0$, for some $m\geq1$. 

To summarise, $V_{k}$ is a deterministic function of the treatment history given by the $m$ preceding time points,
$$
V_{k}=\begin{cases}
    1 &\text{if } \sum_{j=k-m+1}^{k} A_j>0,\\
    0 &\text{if } \sum_{j=k-m+1}^{k} A_j=0,
\end{cases}
$$
for some fixed $m\geq 1$.

The $SCCS$ design uses only data for individuals who experience the event during the study period. We refer to these individuals as cases and denote case status by $S_t>0$, where 
$$
S_t=\sum_{k=1}^t Y_k,
$$
denotes the total number of outcome events observed up to and including time $t$. Therefore, an individual who experiences at least one thrombotic event during the study period has $S_t>0$.

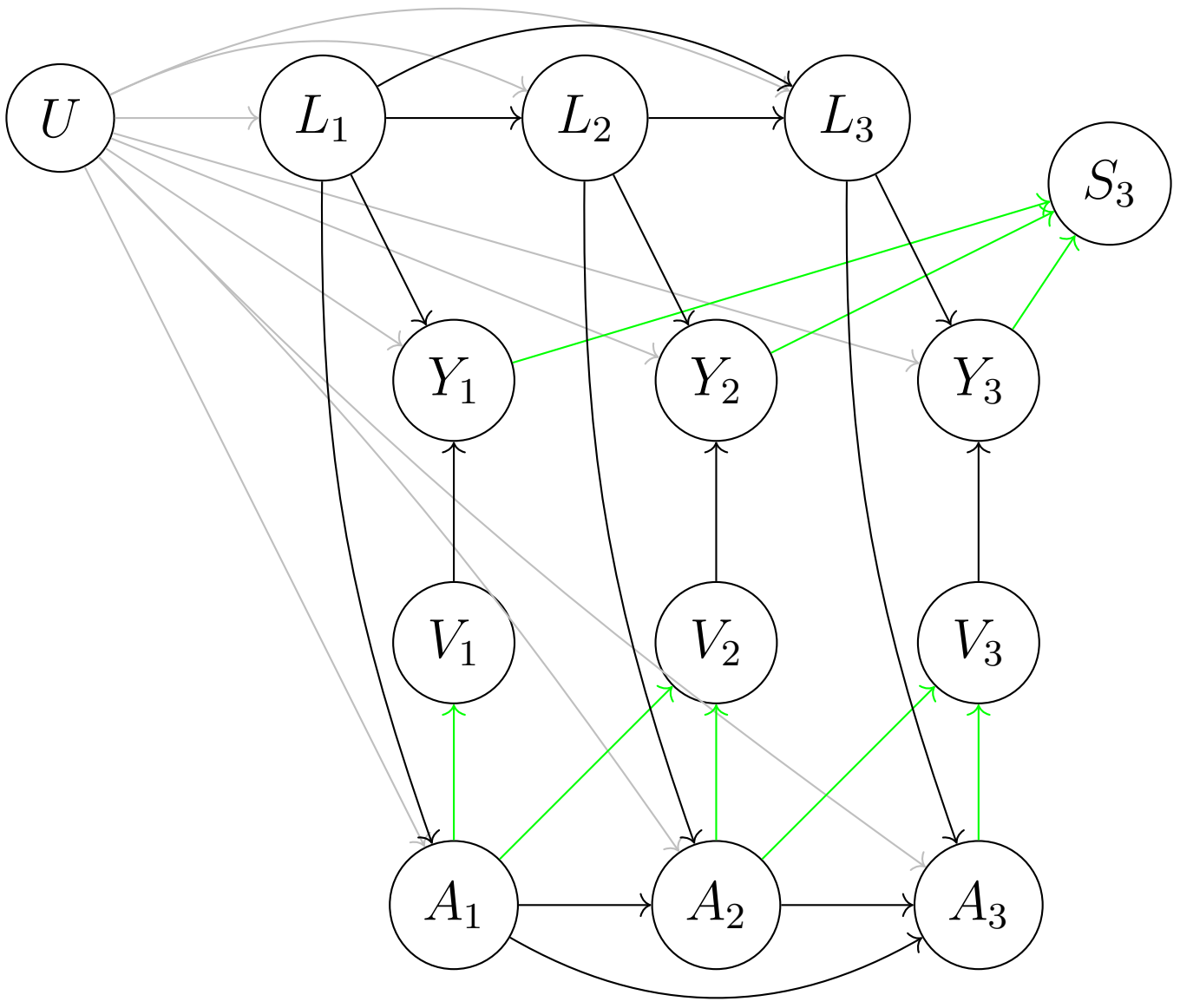
\begin{figure}[!h]
    \centering
    \begin{tikzpicture}[node distance=2cm, auto]

    % Nodes at time 1
    \node[draw, circle] at (1,1) (A1) {$A_{1}$};
    \node[draw, circle] at (1,3) (V1) {$V_{1}$};
    \node[draw, circle] at (1,5) (Y1) {$Y_{1}$};
    \node[draw, circle] at (0,7) (L1) {$L_{1}$};

    % Nodes at time 2
    \node[draw, circle] at (3,1) (A2) {$A_{2}$};
    \node[draw, circle] at (3,3) (V2) {$V_{2}$};
    \node[draw, circle] at (3,5) (Y2) {$Y_{2}$};
    \node[draw, circle] at (2,7) (L2) {$L_{2}$};

    % Nodes at time 3
    \node[draw, circle] at (5,1) (A3) {$A_{3}$};
    \node[draw, circle] at (5,3) (V3) {$V_{3}$};
    \node[draw, circle] at (5,5) (Y3) {$Y_{3}$};
    \node[draw, circle] at (4,7) (L3) {$L_{3}$};

    % U and S
    \node[draw, circle] at (-2,7) (U) {$U$};
    \node[draw, circle] at (6,6.5) (S) {$S_3$};

    % Arrows from U to all A_k, Y_k, L_k 
    \draw[->] (U) edge[lightgray] (A1);
    \draw[->] (U) edge[lightgray] (Y1);
    \draw[->] (U) edge[lightgray] (L1);
    \draw[->] (U) edge[lightgray, bend left=5] (A2);
    \draw[->] (U) edge[lightgray] (Y2);
    \draw[->] (U) edge[lightgray, bend left=25] (L2);
    \draw[->] (U) edge[lightgray, bend right=5] (A3);
    \draw[->] (U) edge[lightgray] (Y3);
    \draw[->] (U) edge[lightgray, bend left=25] (L3);

    % Arrows from outcomes to S
    \draw[->] (Y1) edge[green] (S);
    \draw[->] (Y2) edge[green] (S);
    \draw[->] (Y3) edge[green] (S);

    % Treatment to at-risk m=2
    \draw[->] (A1) edge[green] (V1);
    \draw[->] (A1) edge[green] (V2);
    \draw[->] (A2) edge[green] (V2);
    \draw[->] (A2) edge[green] (V3);
    \draw[->] (A3) edge[green] (V3);
    
    % Treatment to treatment
    \draw[->] (A1) edge (A2);
    \draw[->] (A1) edge[bend right=30] (A3);
    \draw[->] (A2) edge (A3);

    % Treatment to at-risk m=2
    \draw[->] (V1) edge (Y1);
    \draw[->] (V2) edge (Y2);
    \draw[->] (V3) edge (Y3);

    % L to L
    \draw[->] (L1) edge (L2);
    \draw[->] (L1) edge[bend left=30] (L3);
    \draw[->] (L2) edge (L3);

     % Arrows from L_k to A_k, Y_k
    \draw[->] (L1) edge[bend right=10] (A1);
    \draw[->] (L1) edge (Y1);
    % \draw[->] (L1) edge[bend left=15] (A2);
    % \draw[->] (L1) edge[bend left=10] (A3);
    \draw[->] (L2) edge[bend right=10] (A2);
    \draw[->] (L2) edge (Y2);
    % \draw[->] (L2) edge[bend left=15] (A3);
    \draw[->] (L3) edge[bend right=10] (A3);
    \draw[->] (L3) edge (Y3);

\end{tikzpicture}
    \caption{Directed acyclic graph (DAG) illustrating the variables in the underlying cohort study, with total follow-up $t=3$ and reactogenicity window $m=2$. Determinisms are denoted with green arrows. From the time-invariant covariate $U$, arrows originate to all nodes, except to $V_1, V_2, V_3$, and $S_3$, which are deterministic functions of their direct causes (arrows in green). To reduce clutter, arrows from $U$ are depicted in light gray, and arrows from $L_1$ to $A_2, A_3$ and from $L_2$ to $A_3$ are omitted, however, we do allow for direct effect of time-varying covariates on future treatment.}
    \label{FIG: DAG}    
\end{figure}

For any fixed time point $k$, we assume the temporal ordering $(L_k, A_k, V_k, Y_k)$, with $U$ preceding $L_1$, and $S_t$ deterministically realised after $Y_t$. We assume that the data are generated under the Finest Fully Randomized Causally Interpreted Structured Tree Graph (FFRCISTG) model with variables $(U, \overline{L}_t, \overline{A}_t, \overline{V}_t, \overline{Y}_t, S_t)$ \citep{robins_new_1986, richardson_single_2013-1}. We assume that the data-generating process is described by the causal directed acyclic graph (DAG) in Figure \ref{FIG: DAG}. Here, we interpret $V_k$ as a biological process deterministically induced by vaccination. Alternatively, it is possible to think of $V_k$ as a convenient way to impose restrictions on the functional relationship between $\overline{A}_k$ and $Y_k$. In this second interpretation, $V_k$ is included in the DAG to represent the functional restrictions. This use is similar to the inclusion of propensity scores in causal graphs to represent the balancing score condition, see Figure 15.2 of \citet{hernan_causal_2025}.

The proposed DAG encodes several assumptions that are central to the identification results for the $SCCS$ design. In what follows, we therefore assume that the data are generated according to the DAG in Figure \ref{FIG: DAG} for suitable values of $m$ and $t$. The assumptions implied by this graph are stated explicitly and discussed in Appendix \ref{APP: graph ass}, here we provide an informal summary:

    \begin{itemize}
        \item The susceptibility indicator at time $k$ affects only $Y_k$, and fully mediates the effect of treatment history (Assumption \ref{ASS: trans risk}).
       \item The time-fixed covariates $U$ and the time-varying covariates $\overline{L}_k$ are sufficient to adjust for confounding between treatment history and outcome trajectory at time $k$ (Assumption \ref{ASS: exch}). 
        \item Conditional on the full treatment and covariate histories and $U$, outcome counts at distinct time points are independent (Assumption \ref{ASS: ind out}(i)).
        \item Outcome variables do not directly affect any future time-varying covariates or treatments (Assumption \ref{ASS: ind out}(ii)-(iii)).
        \item Conditional on treatment history, the current measured covariates, and the time-fixed covariates, there is no association between earlier measured covariates and the current outcome (Assumption \ref{ASS: trans conf}).
    \end{itemize}

Even though some of these assumptions are required to justify the standard $SCCS$ model, they can be considerably relaxed when the goal is hypothesis testing of treatment effects. We elaborate on these distinctions in Appendix \ref{APP: graph ass}.

\section{Identification}\label{SEC: ident}

Consider first the conditional rate ratio,

\begin{definition}[Conditional rate ratio]
    $$c\text{-}RR(u,l)_k = \frac{E[Y_k^{\overline{a}^1_k }|U=u, L_k=l]}{E[Y_k^{\overline{a}^0_k }|U=u, L_k=l]},$$
where $\overline{a}^1_k \in  \{\overline{a}_k: \sum_{i=k-m+1}^{k} a_i >0 \}$ and $\overline{a}^0_k \in \{\overline{a}_k: \sum_{i=k-m+1}^{k} a_i =0 \}$, respectively. 
\end{definition}
The $c\text{-}RR(u,l)_k$ denotes the instantaneous rate ratio of the adverse outcome after receiving the vaccine in at least one of the last $m$ time points versus not receiving the vaccine in any of the last $m$ time points, conditional on fixed values of the covariates at time $k$. It follows from Assumption \ref{ASS: trans conf} and Corollary \ref{COR: int group}, that $c\text{-}RR(u,l)_k$ is well-defined, because any two interventions that give the same value of the treatment-induced susceptibility indicator at time $k$, give equal expected potential outcomes.

Consider a hypothetical experiment in which individuals have not received treatment before time $k$ and share the same values $U=u$ and $L_k=l$. Such individuals are then randomly assigned to receive the vaccine at time $k$ or not, and we assess the ratio of the average number of events at time $k$ in the two arms. Although this experiment cannot be run when $U$ is unmeasured, this conceptualization can still be useful. The $c\text{-}RR(u,l)_k$ can be interpreted as an ``individual-specific'' effect in the sense that it is conditioned on a potentially high-dimensional set of characteristics \citep{robins_estimation_1999, piccininni_causal_2025}. 

In addition to the assumptions encoded by the DAG and discussed in Appendix \ref{APP: graph ass}, we will invoke the following assumptions for identification of the $c\text{-}RR(u,l)_k$.

\begin{assumption}[Stationarity of potential outcomes]\label{ASS: stat all}
    $$P[Y_1^{\overline{a}^1_1}=y|U=u, L_1=l]=\dots = P[Y_t^{\overline{a}^1_t}=y|U=u, L_t=l]$$
    and  
    $$P[Y_1^{\overline{a}^0_1}=y'|U=u, L_1=l]=\dots = P[Y_t^{\overline{a}^0_t}=y'|U=u, L_t=l]$$
    for any $y, y' \in \{0, 1, 2,\dots\}$ for all $u\in \mathcal{U}$, $l\in \mathcal{L}$, where $\overline{a}^1_k \in  \{\overline{a}_k: \sum_{i=k-m+1}^{k} a_i >0 \}$ and $\overline{a}^0_k \in \{\overline{a}_k: \sum_{i=k-m+1}^{k} a_i =0 \}$, respectively.
\end{assumption}

Assumption \ref{ASS: stat all} states that, among individuals with the same values of the measured and unmeasured covariates, the distributions of the potential outcomes are the same at all time points under interventions that set $V_k = 1$. Similarly, the distributions of potential outcomes are the same at all time points under interventions that set $V_k = 0$.

Assumption \ref{ASS: stat all} will be violated if, for example, the distribution of the outcome at time $k$ , given $U=u$, $L_k=l$ and never having received treatment, differs from this conditional distribution at $k'$. In this sense, the assumption is, in principle, falsifiable.

Stationarity assumptions are often invoked when making causal inference from studies including only one participant \citep{piccininni_causal_2025}. The stationarity assumption is strong, but it becomes more reasonable when $L_k$ contains some measurement of time, such as age or calendar time. Consider, for example, a study where time points correspond to weeks and $L_k$ denotes the calendar month at time $k$, then Assumption \ref{ASS: stat all} implies that the distributions of potential outcomes under a treatment inducing $V_k=0$ in March are the same for every week within the month, conditional on $U=u$. However, Assumption \ref{ASS: stat all} still allows the potential outcome distributions to differ between, say, March and February.

We will also invoke a positivity assumption.

\begin{assumption}[Time point positivity]\label{ASS: pos}
\leavevmode
\begin{enumerate}[label=(\roman*)]
    \item For a given $u \in \mathcal{U}$ and $l \in \mathcal{L}$
    \begin{align*}
        &\sum_{k=1}^t \mathbb{I}(V_k=1, U=u, L_k=l)>0& &\text{and}& &\sum_{k=1}^t \mathbb{I}(V_k=0, U=u, L_k=l)>0.&
    \end{align*}
    \item Given $U=u$ and $\overline{L}_k=\overline{l}_k$, $P(\overline{A}_k=\overline{a}_k|\overline{L}_k=\overline{l}_k, U=u)>0$ for all treatment regimes $\overline{a}_k$ of interest.
\end{enumerate}

\end{assumption}

Assumption \ref{ASS: pos} requires that there exist two time points at which the covariates had the same value and the the susceptibility indicator differs, that is, at which the individual is susceptible ($V_k=1$) at one time point and not susceptible ($V_k=0$) at the other.
If no such two time points exist for a given $u$ and $l$, the $c\text{-}RR(u, l)$ cannot be identified, for example, if $L_k$ denotes age, and after the age of $60$ the vaccine must be administered once every 6 months, we cannot identify the conditional rate ratio for individuals over the age of $60$, if the reactogenicity window is one year ($m=365$).

Under these additional assumptions, the $c\text{-}RR(u,l)$ is identified from the cases ($S_t>0$), based on the variables $(\overline{Y}_t, \overline{L}_t, \overline{A}_t, U)$ (see Proposition \ref{PROP: c-RR(u,l)}). However, such an identification is not particularly useful in practice unless $U$ is measured, because it requires expectations conditional on $U=u$. 

Despite $U$ being unobserved, we can still estimate expectations conditional on $U=u$ when we take the mean of several observations from the same individual.
But in doing so, we do not only condition on $U$, but also on the exact number of events experienced by the individual during the study period ($S_t=s$). Such conditioning, which differs from conditioning on $S_t>0$, can introduce bias for the $c\text{-}RR(u,l)$. This bias is examined in Appendix \ref{APP: bias}, where we discuss its behaviour heuristically. To ensure that the introduced bias vanishes, we rely on asymptotic assumptions on the frequency of the outcome as the number of time points goes to infinity, approximating a continuous time formulation. To simplify notation, the fixed treatment and covariate history is denoted by 
    $$
    (\overline{A}_t, U, \overline{L}_t)=(\overline{a}_t, u, \overline{l}_t):=\overline{H}_t=\overline{h}_t.
    $$

\begin{assumption}\label{ASS: rare event}
   For any history $\overline{h}_t \in \mathcal{A}\times \mathcal{U}\times \mathcal{L}$
    assume that as the number of time points $t\rightarrow \infty$, the following assumptions hold
    \begin{enumerate}[label=(\roman*)]
        \item $\max_{k \in \{1, \dots, t\}} E[Y_k|\overline{H}_t=\overline{h}_t] \rightarrow 0$.
        \item $\sum_{k \in  \{1, \dots , t\}} E[Y_k|\overline{H}_t=\overline{h}_t] \rightarrow \lambda(\overline{h_t})$ for some $\lambda \in [0, \infty)$.
        \item Define $\mathcal{K}_{>0}:=\{k: P(Y_k>0|\overline{H}_t=\overline{h}_t)>0\}$, then $\sup_{k \in \mathcal{K}_{>0}} E[Y_k|Y_k>0, \overline{H}_t=\overline{h}_t]\rightarrow 1$.
    \end{enumerate}
\end{assumption}
Assumption \ref{ASS: rare event} states that as the number of time points goes to infinity over a fixed follow-up period, the expected number of events per interval vanishes, and the probability of multiple events within an interval becomes negligible, thus the outcomes $Y_k$-s converge to ``essentially Bernoulli" random variables. That is, even conditional on $Y_k$ taking a value greater than $0$, the probability of it taking a value greater than $1$ goes to $0$. Assumption \ref{ASS: rare event} is plausible, if the increase in the number of time points is not achieved by extending the follow-up period, but rather the frequency of the data collection is increased. As the data-generating process approximates a continuous time process, it is plausible that at any (tiny) time interval, at most one event occurs, and the set of event times has measure zero.
By invoking Assumption \ref{ASS: rare event}, we derive an asymptotically unbiased identification formula, which relies on individual-level data, without the need to measure the covariate $U$, as we implicitly condition on it, by conditioning on the individual. Thus it is possible to identify the rate ratio for a fixed and known $l$ and a fixed and unobserved $u$ (see Proposition \ref{PROP: c-RR(u,l) St=st}). Yet, this does not by itself identify the marginal rate ratio ($CRR_m$).

To identify the marginal rate ratio, we require the following additional assumption:

\begin{assumption}[Semi-invariance across measured and unmeasured covariates]\label{ASS: Stat U}

Suppose that the support of the measured and unmeasured covariates can be divided into two disjoint subsets, respectively, $\mathcal{L}' \cup \mathcal{I}_L=\mathcal{L}$, $\mathcal{U}' \cup \mathcal{I}_U=\mathcal{U}$, such that
    $$\frac{E[Y_k^{\overline{a}^1_k}|U=u, L_k=l]}{E[Y_k^{\overline{a}^0_k}|U=u,  L_k=l]}= \frac{E[Y_k^{\overline{a}^1_k}|U=u', L_k=l']}{E[Y_k^{\overline{a}^0_k}|U=u',L_k=l']}$$
    for all $u, u' \in \mathcal{U'}$ and $l,l' \in \mathcal{L}'$, where $\overline{a}^1_k \in  \{\overline{a}_k: \sum_{i=k-m+1}^{k} a_i >0 \}$ and $\overline{a}^0_k \in \{\overline{a}_k: \sum_{i=k-m+1}^{k} a_i =0 \}$, respectively and
    \begin{align*}
        &E[Y_k^{\overline{a}_k}|U=u, L_k=l']=E[Y_k^{\overline{a}_k}|U=u', L_k=l]=0
    \end{align*}
    for all $u \in \mathcal{I}_U$, $l' \in \mathcal{L}$ and $u' \in \mathcal{U}$, $l \in \mathcal{I}_L$, for any $\overline{a}_k\in \{0,1\}^{k}$
\end{assumption}

Assumption \ref{ASS: Stat U} enforces no effect heterogeneity on the relative scale. In the $SCCS$ literature, it is often assumed that the rate ratio for the treatment is the same for any individual, at any time. For example, \citet{farrington_self-controlled_2018} define the proportional incidence model, in which the incidence rate is a product of baseline, age, treatment and covariate effects, such that the age and treatment effects are fixed across individuals and time. Assumption \ref{ASS: Stat U} relaxes this by allowing an immune subpopulation and by requiring homogeneity only among the non-immune. Indeed, Assumption \ref{ASS: Stat U} allows for some individuals in the underlying population to be immune, leveraging the immune-selection stability property of this effect measure \citep{piccininni_immune-selection_2025}. Specifically, Assumption \ref{ASS: Stat U} means that at any time $k$, we have two subpopulations: 1) a subpopulation of immune individuals, $u \in \mathcal{I}_U$ or $l \in\mathcal{I}_L$, who cannot develop the outcome under any intervention on treatment; and 2) a subpopulation in which the rate ratio between interventions that set the vaccine-induced susceptibility indicator at time $k$ to $1$ and $0$ is the same regardless of the values taken by the measured and unmeasured covariates.

Assumption \ref{ASS: Stat U} imposes a strong homogeneity restriction on the treatment effect, and may therefore fail in practice. For example, it has been suggested that the effect of the ChAdOx1 nCoV-19 vaccine on cerebral venous thrombosis among women is larger in younger age groups \citep{whiteley_association_2022}. Nevertheless, we will show in Section \ref{SEC: test} that Assumption \ref{ASS: Stat U} is not needed if we are just interested in testing for an effect.

Lack of effect heterogeneity is rarely explicitly justified in the $SCCS$ literature, although this assumption  is often implied by the chosen parametric model. In Section \ref{SEC: Est}, and in more detail in Appendix \ref{APP: pies}, we show that the lack of effect heterogeneity can be justified under a specific class of sufficient-component cause models \citep{hernan_causal_2025, rothman1976causes}. 

Under Assumptions \ref{ASS: stat all}-\ref{ASS: Stat U} and \ref{ASS: exch}-\ref{ASS: trans conf}, the same within-person functional identifies the $CRR_m$ as the $c\text{-}RR(u,l)$. 

Moreover, by Assumption \ref{ASS: Stat U}, the conditional rate ratio is constant across all non-immune covariate strata. Consequently, any normalized weighted average of the $c\text{-}RR(u,l)$ values over $(u,l)\in\mathcal U'\times\mathcal L'$ equals this same common rate ratio, and hence the $CRR_m$. Thus, the particular distribution used to weight the non-immune strata does not affect the resulting causal contrast. This invariance also allows information to be pooled across individuals and covariate strata for estimation. 

\begin{proposition}\label{PROP: CRR}
    Under Assumptions \ref{ASS: stat all}-\ref{ASS: Stat U} and \ref{ASS: exch}-\ref{ASS: trans conf},
    $$
    \frac{\sum_{k \in \mathcal{K}_{1,l}} E[Y_{k}|U=u, \overline{A}_t=\overline{a}_t, \overline{L}_t=\overline{l}_t, S_t=s_t]/|\mathcal{K}_{1,l}|}{\sum_{k \in \mathcal{K}_{0,l}} E[Y_{k}|U=u, \overline{A}_t=\overline{a}_t, \overline{L}_t=\overline{l}_t, S_t=s_t]/|\mathcal{K}_{0,l}|} \rightarrow CRR_m
    $$
        where $\mathcal{K}_{1,l}:=\{k: \sum_{i=k-m+1}^{k} a_i >0, l_k=l\}$, $\mathcal{K}_{0,l}:=\{k: \sum_{i=k-m+1}^{k} a_i =0, l_k=l\}$, for $u \in \mathcal{U}'$, $l \in \mathcal{L}'$.

    Moreover, the $CRR_m$ can be identified in the limit by
    \begin{align*}
            &\sum_{u, l \in \mathcal{U}', \mathcal{L}'} \phi(u,l)
            \cdot \frac{\sum_{k \in \mathcal{K}_{1,l}} E[Y_{k}|U=u, \overline{A}_t=\overline{a}_t, \overline{L}_t=\overline{l}_t, S_t=s_t]/|\mathcal{K}_{1,l}|}{\sum_{k \in \mathcal{K}_{0,l}} E[Y_{k}|U=u, \overline{A}_t=\overline{a}_t, \overline{L}_t=\overline{l}_t, S_t=s_t]/|\mathcal{K}_{0,l}|},
    \end{align*}
    where $\phi: \mathcal{U}'\times \mathcal{L}' \rightarrow [0,1]$ is a generic probability mass function.
\end{proposition}

\section{Estimation}\label{SEC: Est}
In Section \ref{SEC: target} we defined the causal estimand of interest, in Section \ref{SEC: ident} we made explicit under which assumptions this causal estimand can be identified by a functional potentially estimable in a $SCCS$ study.
Our approach differs from the existing $SCCS$ literature, where the effect of interest is often defined as a regression coefficient and the $SCCS$ assumptions are formulated primarily as parametric modelling assumptions \citep{whitaker_tutorial_2006, farrington_self-controlled_2018}. Specifically, standard $SCCS$ analyses assume that the data are generated by a specific Poisson regression model. While this parametric formulation provides a convenient basis for estimation, it does not make explicit either the causal estimand being targeted or the identification assumptions required for a causal interpretation. We now show that, under additional structural assumptions, the outcome process admits a limiting log-linear Poisson representation in which the treatment coefficient identifies the $CRR_m$, thereby linking our results to common estimation techniques in applied $SCCS$ studies.

\subsection{Poisson approximation}\label{SUBSEC: Pois}

In Section \ref{SEC: ident} we showed that under suitable assumptions, both the $CRR_m$ and the $c\text{-}RR(u, l)$ are identified through contrasts between outcomes at vaccine-induced susceptibility and non-susceptibility time points, conditional on fixed covariate values, treatment history and the number of events experienced during the follow-up. We showed that even though the within-person functional is biased for the $c\text{-}RR(u, l)$ in finite $t$, the bias vanishes as the number of time points increases, and the instantaneous expectation of the outcome decreases. A non-parametric estimator of $c\text{-}RR(u,l)$ can be obtained by replacing the corresponding numerator and denominator with empirical means computed over time for the same individual. This will be a consistent estimator as the number of observations over time goes to infinity. However, in a setting with finite data this is not an efficient strategy.
One might then try to estimate the $CRR_m$ by combining these empirical estimates, according to some weight function. However, the optimal weights depend on the joint distribution of $U, \overline{L}_t, \overline{A}_t,$ and $\overline{Y}_t$, which is unobserved. Thus, naive estimators of the $CRR_m$ may suffer from both finite-sample bias and high variance, and their finite-sample performance is generally difficult to characterize without further assumptions on the data-generating process.

We therefore turn to a parametric approximation and give a causal justification for the log-linear Poisson models commonly used in $SCCS$ analyses.

The multiplicative outcome model requires two homogeneity assumptions. First, the susceptibility effect must be homogeneous across levels of both measured and unmeasured covariates, as guaranteed by Assumption \ref{ASS: Stat U}. Second, the effects of measured covariates must be homogeneous across levels of time-fixed unmeasured covariates, as formalized by the following assumption.

\begin{assumption}[$U$-$L_k$ homogeneity]\label{ASS: no int}
    $$E[Y_k^{\overline{a}_k^0}|U=u, L_k=l] \cdot E[Y_k^{\overline{a}_k^0}|U=u^*, L_k=l^*] = E[Y_k^{\overline{a}_k^0}|U=u, L_k=l^*]  \cdot E[Y_k^{\overline{a}_k^0}|U=u^*, L_k=l] ,$$
    for all $(l, u) \in \mathcal{L}' \times \mathcal{U}'$, where $u^* \in \mathcal{U}'$ and $l^*\in \mathcal{L}'$ are some reference levels, where $\overline{a}_k^0 \in \{\overline{a}_k: \sum_{k'=k-m+1}^k a_{k'}=0\}$.
\end{assumption}

By Assumption \ref{ASS: Stat U}, Assumption \ref{ASS: no int} implies that the $U$-$L_k$ homogeneity holds for treatment interventions $\overline{a}_k^1$ as well, which makes individuals susceptible at time $k$. In Appendix \ref{APP: pies}, we show using a sufficient-component cause model \citep{hernan_causal_2025, rothman1976causes} that Assumption \ref{ASS: Stat U} and Assumption \ref{ASS: no int} can be justified when the outcome is a treatment's side effect, under appropriate mechanistic assumptions. In fact, these outcomes can be often assumed to arise as an ``interaction'' of treatment-induced processes, individual time-invariant characteristics, and time-varying factors. 

\begin{proposition}[Poisson model of homogeneous intervals]\label{PROP: pois}
    Under Assumptions \ref{ASS: stat all}-\ref{ASS: no int} and \ref{ASS: trans risk}-\ref{ASS: trans conf} and the resolution convention below, let $t\rightarrow\infty$ through increasingly fine partitions of a fixed follow-up period, and let $I$ be a fixed subinterval on which $V_k$, $U$ and $L_k$ take the constant values $v$, $u$ and $l$. Then, conditional on $\overline{H}_t=\overline{h}_t$, the number of events $S_I$ in $I$ converges in distribution to
    $$
    \mathrm{Poisson}\left(\exp(\lambda_0+\alpha_v+\beta_l+\beta_u+\sigma_I)\right),
    $$
    and counts over non-overlapping intervals are conditionally independent.
\end{proposition}

The resolution convention requires that, for any fixed physical interval, the conditional expected event count does not depend on the resolution. Here, $\lambda_0$ is the log baseline event rate per reference time unit, $\beta_u$ and $\beta_l$ are the log rate ratios relative to reference levels $u^*$ and $l^*$, and $\sigma_I=\log|I|$ is a known offset, with $|I|$ the length of $I$ in the reference time unit. The at-risk parameters satisfy $\alpha_0=0$ and $\exp(\alpha_1)=CRR_m$. This parameterization applies to non-immune strata with positive rates, otherwise, the limit is $\mathrm{Poisson}(0)$.

See the extended discussion on the parameters of the model in Appendix \ref{APP: Pois reg}, and the proof of the proposition in Appendix \ref{APP_Sub: pois proof}. Proposition \ref{PROP: pois} justifies the use of Poisson regressions to estimate the $CRR_m$.

\begin{remark}
    Assumption \ref{ASS: pos} (i) requires that each individual is observed both susceptible ($V_k=1$) and not susceptible ($V_k=0$) at the same level of $L_k$, which is needed for the nonparametric identification of $c\text{-}RR(u,l)$. Under the parametric model of Proposition \ref{PROP: pois}, this can be relaxed. As the effects of $V_k$ and $L_k$ are then multiplicative and common across individuals (Assumptions \ref{ASS: Stat U} and \ref{ASS: no int}), the effect of $L_k$ can also be estimated from individuals whose susceptibility does not change during follow-up, such as unexposed cases \citep{farrington_self-controlled_2018}. The effect $\alpha_1$ is then identified as long as, pooled across individuals, the within-individual changes in $V_k$ are not collinear with those in $L_k$. This identification, however, rests on the modelling assumptions rather than on direct comparisons within strata.
\end{remark}

\section{Testing}\label{SEC: test}
As shown in Section \ref{SEC: ident}, a within-individual contrast implicitly conditions on the number of events. As a result, the quantity identified by the weighted average of the within-individual contrasts generally differs from the $CRR_m$, thus inducing bias on the relative scale.
Under the sharp null introduced below, however, valid tests can be constructed without relying on the asymptotic conditions in Assumption \ref{ASS: rare event}.
In addition, under the null, Assumptions \ref{ASS: Stat U} and \ref{ASS: trans risk} (iii) are satisfied and therefore need not be imposed separately. In this sense, testing provides a more assumption-lean approach than point estimation, while also avoiding reliance on a continuous-time approximation.

However, when Assumption \ref{ASS: rare event} is not imposed, we no longer ensure that the outcomes at each time point are ``essentially" binary. Therefore, to derive valid tests of no effect of susceptibility on the outcome, without relying on the approximation as $t\rightarrow \infty$, we explicitly impose the assumptions that the outcomes are binary.

\begin{assumption}[Binary outcome]\label{ASS: bin}
    $Y_k \in \{0,1\}$,
    for all $k=1, \dots, t$.
\end{assumption}

We consider the following sharp null hypothesis of no causal effect of treatment history on the outcome process. Under the definition of $V_k$, this is equivalently a sharp null of no effect of treatment-induced susceptibility.

\begin{hypothesis}\label{HYP: null}
    $$\overline{Y}_k^{\overline{a}_k}=\overline{Y}_k^{\overline{a}'_k}$$ for any two $\overline{a}_k\neq \overline{a}_k'$. In particular,
    $$Y_k^{\overline{a}^1_k}=Y_k^{\overline{a}^0_k}$$ for any $\overline{a}^1 \in \{\overline{a}_k: \sum_{i=k-m+1}^{k} a_i >0 \}$, $\overline{a}^0 \in \{\overline{a}_k: \sum_{i=k-m+1}^{k} a_i =0 \}$.
\end{hypothesis}

The standard $SCCS$ method relies on modelling the outcomes coming from an inhomogeneous Poisson process, thus independence between the outcomes must hold, as stated in Assumption \ref{ASS: ind out} (i). To derive a valid test of the treatment effect, such assumption is not necessary, instead we can assume the following:

\begin{assumption}[Partial outcome interchangeability]\label{ASS: null strong inter}
    Under the null hypothesis, Hypothesis \ref{HYP: null}, conditional on $U$ and $\overline{L}_t$, suppose that the outcome vector is invariant under all permutations that preserve measured-covariate stratum. That is for every permutation $\pi =(\pi_1, \dots ,\pi_t)$ that preserves 
    $$
        L_{\pi(k)}=L_k \quad \text{for every $k$},
    $$

    $$\overline{Y}_t|U, \overline{L}_t \overset{d}{=} \overline{Y}_{\pi(t)}|U, \overline{L}_{\pi(t)} \overset{d}{=} \overline{Y}_{\pi(t)}|U, \overline{L}_t, $$
    where $\overline{Y}_{\pi(t)}=(Y_{\pi(1)}, \dots, Y_{\pi(t)})$.
\end{assumption}

Assumption \ref{ASS: null strong inter} is stronger than assuming that in each subvector $(Y_k)$ for all $\{k: L_k=l\}$ the outcomes are marginally exchangeable. Exchangeability must hold for the whole outcome vector, but only for a restricted set of permutations. Under the null hypothesis, Hypothesis \ref{HYP: null} and stationarity, Assumption \ref{ASS: stat all}, Assumption \ref{ASS: ind out} (i) implies Assumption \ref{ASS: null strong inter}.
We propose an exact testing algorithm that is valid under Assumption \ref{ASS: null strong inter}, however, in Appendix \ref{APP: boot test}, we introduce an asymptotic test, Algorithm \ref{ALG: Dependent} that is valid even when Assumption \ref{ASS: null strong inter} fails. Its validity instead relies on equality of the relevant conditional means under the null and on independent sampling of individuals.

\begin{proposition}\label{PROP: test hyper}
    Suppose that Assumptions \ref{ASS: bin}, \ref{ASS: null strong inter}, and \ref{ASS: exch} hold, the testing procedure proposed in Algorithm \ref{alg: hyper} is a valid level-$\alpha$ test of Hypothesis \ref{HYP: null}.
\end{proposition}

Under the null hypothesis and the assumptions stated in Proposition \ref{PROP: test hyper}, for a fixed level of $U=u$ and $L_k=l$, the time points are exchangeable, regardless of the treatment history. Thus, conditional on the total number of events for a fixed level of the covariates, the number of events under treatment-induced susceptibility is hypergeometrically distributed, with parameters governed by the time spent being susceptible, the total number of time points with these covariates, and the total number of events. By Assumption \ref{ASS: null strong inter}, these hypergeometrically distributed random variables are independent; thus, we can simulate from the exact distribution. See the proof of the validity of Algorithm \ref{alg: hyper} in Appendix \ref{APP: proof hyper}.

\begin{algorithm}[!ht]
\caption{Conditional Monte Carlo permutation test across $U\times L$ strata}
\label{alg: hyper}
\begin{algorithmic}[1]
\Require Observed cases $i=1, \dots, n$; informative strata $l\in \mathcal{L}_i^+=\{l:n_{1,il}>0,\;n_{0,il}>0\}$ with:
\Statex \hspace{\algorithmicindent} $\mathcal{K}_{1, il}$ = number of at-risk intervals  for given $l$ for and individual $i$
\Statex \hspace{\algorithmicindent} $\mathcal{K}_{0, il}$ = number of non-risk intervals for given $l$ for and individual $i$
\Statex \hspace{\algorithmicindent} $s_{\mathcal{K}_{il}}$ = total number of events for given $l$ for and individual $i$
\Statex \hspace{\algorithmicindent} $s_{\mathcal{K}_{1,il}}$ = observed number of events in at-risk intervals for given $l$ for and individual $i$
\Require Number of Monte Carlo replicates $B$
\Ensure Observed statistic $D$, Monte Carlo $p$-value, permutation distribution $\{D^{(b)}\}_{b=1}^B$

\For{$i\in \{1, \dots, n\}$ and $l\in\mathcal L_i^+$}
    \State Compute the observed centred score:
    \[
    D_{\mathcal{K}_{il}} := s_{\mathcal{K}_{1,il}} - s_{\mathcal{K}_{il}}\frac{|\mathcal{K}_{1,il}|}{|\mathcal{K}_{1,il}|+|\mathcal{K}_{0,il}|}
    \]
\EndFor
\State Compute the observed test statistic:
\[
D :=  \sum_{i=1}^n \sum_{l\in\mathcal L_i^+} D_{\mathcal{K}_{il}}.
\]

\For{$b = 1,\dots,B$}
    \For{$i\in \{1, \dots, n\}$ and $l\in\mathcal L_i^+$}
        \State Draw
        \[
        s_{\mathcal{K}_{1,il}}^{(b)} \sim \mathrm{Hypergeometric}(|\mathcal{K}_{1, il}|+|\mathcal{K}_{0, il}|, |\mathcal{K}_{1, il}|, s_{\mathcal{K}_{il}})
        \]
        \State Compute the centred permuted score:
        \[
        D_{\mathcal{K}_{il}}^{(b)} := s_{\mathcal{K}_{1,il}}^{(b)} - s_{\mathcal{K}_{il}}\frac{|\mathcal{K}_{1,il}|}{|\mathcal{K}_{1,il}|+|\mathcal{K}_{0,il}|}
        \]
    \EndFor
    \State Compute the permuted test statistic:
    \[
    D^{(b)} := \sum_{i=1}^n \sum_{l\in\mathcal L_i^+}D_{\mathcal{K}_{il}}^{(b)}
    \]
\EndFor

\State Compute the Monte Carlo $p$-value:
    \[
    p_{\text{two-sided}}^{B}:=\frac{\sum_{b=1}^B \mathbb{I}(|D^{(b)}|\geq |D|)+1}{B+1}
    \]

\State \Return $D, p_{\text{two-sided}}^{B}, \{D^{(b)}\}_{b=1}^B$
\end{algorithmic}
\end{algorithm}

\subsection{Simulation results}\label{SUBSEC: sim}

We conducted a simulation study, with $5,000$ replications per scenario to compare three procedures for testing the null hypothesis in Hypothesis \ref{HYP: null}: Algorithm \ref{alg: hyper}, Algorithm \ref{ALG: Dependent}, and the standard $SCCS$ procedure. Across several simulation scenarios in which one or more of the sufficient assumptions underlying the standard $SCCS$ procedure were violated, the standard $SCCS$ test did not maintain the nominal type I error rate of $0.05$.
Algorithm \ref{alg: hyper} maintained the nominal level ($0.04$ to $0.05$), whenever partial interchangeability (Assumption \ref{ASS: null strong inter}) held.
Algorithm \ref{ALG: Dependent} remained close to the nominal level ($0.05$ to $0.06$) in all cases, except under the violation of stationarity (Assumption \ref{ASS: stat all}) and in a scenario with severe exposure imbalance and sparse informative contrasts ($0.12$ in both cases).

When all sufficient assumptions were satisfied, however, the standard $SCCS$ procedure generally had greater power, reflecting its ability to exploit the additional structural assumptions. The simulation design and results are described in detail in Appendix \ref{APP: sim}.

\section{Data example}

We reanalyzed the \textit{itpdat} and \textit{dtpdat} datasets from the \texttt{SCCS} package \citep{weldeselassie_sccs_2024}. Detailed descriptions of these datasets are given in \citet{farrington_self-controlled_2018}. Because our goal was to test the sharp null hypothesis defined in Section \ref{SEC: test}, we used a single risk period rather than distinguishing multiple post-vaccination risk periods. To make our analysis comparable with that of \citet{farrington_self-controlled_2018}, we therefore collapsed the at-risk periods into a single interval and excluded washout periods. Specifically, we defined the following at-risk periods, and time-varying confounders:
\begin{itemize}
    \item \textit{itpdat}: 42 days following measles, mumps, and rubella (MMR) vaccination. Each individual was exposed exactly once. 6 age groups of equal length (61 days). 
    \item \textit{dtpdat}: 14 days following any diphtheria, tetanus, and pertussis (DTP) vaccination. Most cases received all three doses, although some received two, one, or no dose. 12 age groups of equal length (28 days).
\end{itemize}

For the \textit{itpdat} dataset, it appears reasonable that the assumptions required for the standard $SCCS$ analysis are approximately satisfied. As noted by \citet{farrington_self-controlled_2018}, there is strong evidence of an effect of MMR vaccination on the outcome, and little evidence against the reduced model with a single risk period. In our reanalysis, both Algorithm \ref{ALG: Dependent} and the standard $SCCS$ method provided strong evidence of an effect, whereas Algorithm \ref{alg: hyper} did not reach the conventional threshold for statistical significance ($p = 0.052$). Given the small number of cases, this borderline result may reflect the lower efficiency of the exact conditional procedure relative to the more model-based alternative.

\begin{figure}
    \centering
    \includegraphics[width=0.9\linewidth]{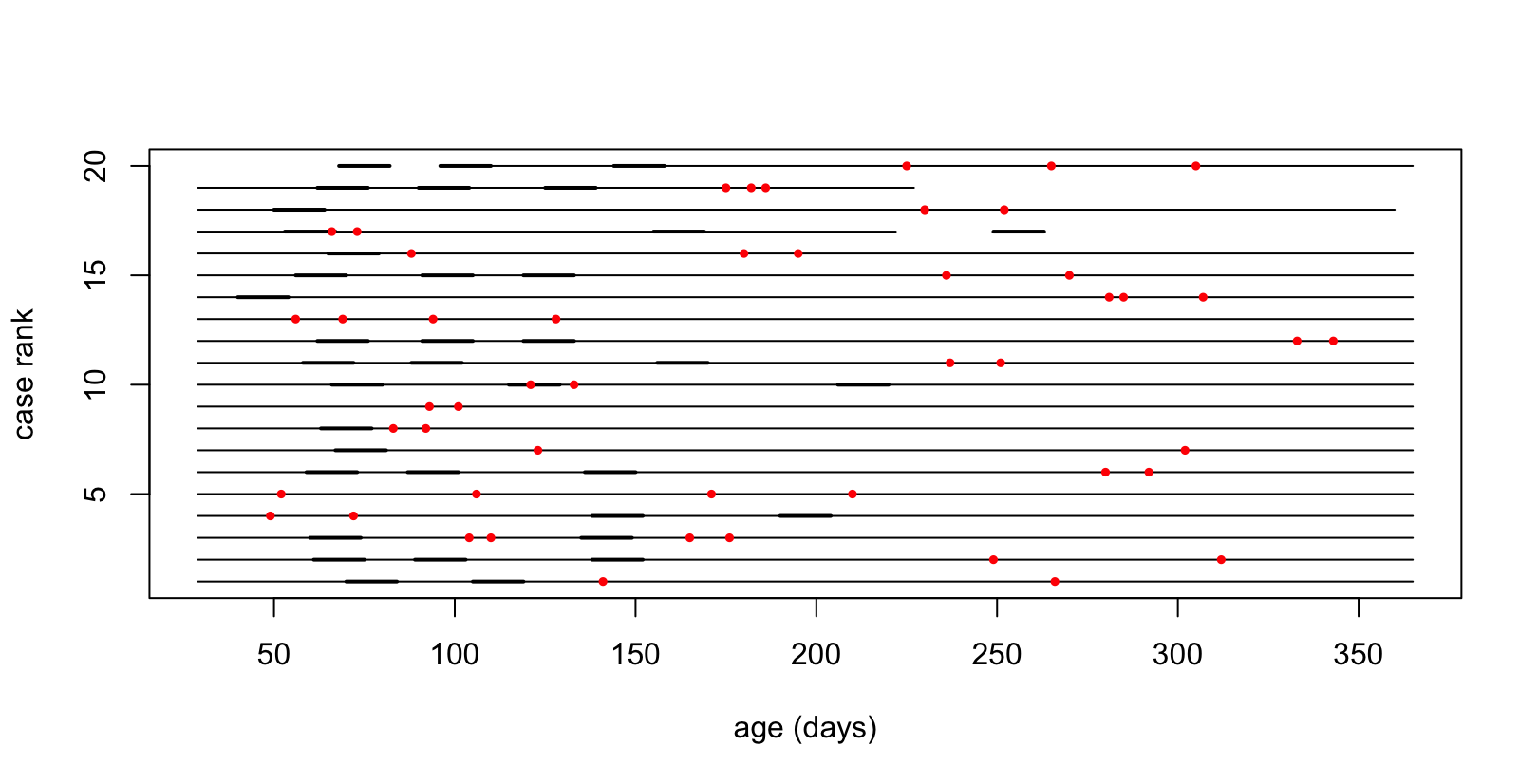}
    \caption{Random sample of cases with multiple outcome events. Horizontal lines represent the follow-up periods, thicker segments standing for the at-risk periods, that is, 14 days after each treatment, while red dots denote the time of the outcomes. For the majority of the sampled individuals, the events cluster around each other, even outside of the treatment periods, which suggests a violation of Assumption \ref{ASS: ind out}.}
    \label{FIG: clustered dtp}
\end{figure}

For the \textit{dtpdat} dataset, however, Assumptions \ref{ASS: ind out} (i) and \ref{ASS: null strong inter} may be violated. Among the 1214 individuals, 130 experienced repeated convulsions, and these recurrent events appear to cluster closely in time, see Figure \ref{FIG: clustered dtp}. Under the violation of Assumptions \ref{ASS: ind out} (i)  and \ref{ASS: null strong inter} the validity of the standard $SCCS$ method and of Algorithm \ref{alg: hyper} is not guaranteed. In our analysis, the standard $SCCS$ method found insufficient evidence for a treatment effect in the 14-day risk window, whereas both Algorithm \ref{alg: hyper} and, more notably, Algorithm \ref{ALG: Dependent} rejected the null hypothesis of no effect. This example illustrates that when exploratory data analysis suggests violation of the independence assumption, and thus casts doubt on the Poisson model, methods that do not require conditional independence of outcomes over time might be preferable.

\begin{table}
\centering
\begin{tabular}{llrrr}
\arrayrulecolor{black}
\toprule
\rowcolor{white}Dataset & \makecell{Number of \\ cases} & \makecell{$SCCS$\\ p-value} & \makecell{Algorithm \ref{alg: hyper} \\ p-value} & \makecell{Algorithm \ref{ALG: Dependent}\\ p-value}\\
\specialrule{1.1pt}{0pt}{0pt}
\rowcolor{black!5}\textit{itpdat} & $35$ & $0.002$ & $0.052$ & $0.017$ \\
\arrayrulecolor{black}
\specialrule{0.45pt}{0pt}{0pt}
\rowcolor{white}\textit{dtpdat} &  $1214$ & $0.453$ & $<0.001$ & $0.001$ \\
\arrayrulecolor{black}
\bottomrule
\end{tabular}
\caption{P-values for the three tests calculated for 2 $SCCS$ package datasets.}
\label{TAB: real test}
\end{table}

\section{Discussion}
The $SCCS$ design is convenient for estimating treatment effects from observational data, particularly when outcomes are rare and standard estimators are underpowered in conventional studies. This gain in power comes at the cost of imposing strong restrictions to the data-generating process. Instead of invoking a conventional, but restrictive, parametric model as the starting point, we have motivated the $SCCS$ estimator from first principles and made its underlying assumptions explicit. This motivation clarifies the causal estimand targeted by $SCCS$ analysis and helps researchers to assess whether the $SCCS$ design is appropriate for their analyses.

In particular, we have shown that, under a collection of causal assumptions, the data-generating process converges to the conventional parametric model proposed by \citet{farrington_relative_1995} in the continuous-time limit.

Making the assumptions transparent also reveals that some of them are likely to be violated in practice, thereby introducing bias in the estimation of the causal quantity of interest. However, even when some of these assumptions are in doubt, we have derived statistical tests that can be used to test the null hypothesis of no effect.

\bibliographystyle{biometrika_mod}
\bibliography{references}

% =====================================================================
% SUPPLEMENTARY MATERIALS
% (\title, \date, \begin{document} and \maketitle of the separate
%  supplement cannot be repeated in one file; the title is set by hand)
% =====================================================================
\clearpage
\def\spacingset#1{\renewcommand{\baselinestretch}%
{#1}\small\normalsize} \spacingset{1}

\appendix
\spacingset{2} 

\begin{center}
{\LARGE Supplementary Materials for Causal Inference from Self-controlled Case Series\par}
\end{center}

\section*{}
\addcontentsline{toc}{section}{Appendices}
\renewcommand{\thesection}{\Alph{section}}

% Numbering and styles of the separate supplement
\let\assumption\suppassumption
\let\endassumption\endsuppassumption
\setcounter{proposition}{0}\renewcommand{\theproposition}{S\arabic{proposition}}
\setcounter{corollary}{0}\renewcommand{\thecorollary}{S\arabic{corollary}}
\setcounter{remark}{0}\renewcommand{\theremark}{S\arabic{remark}}
\setcounter{figure}{0}\renewcommand{\thefigure}{S\arabic{figure}}
\setcounter{table}{0}\renewcommand{\thetable}{S\arabic{table}}
\setcounter{algorithm}{0}\renewcommand{\thealgorithm}{S\arabic{algorithm}}

% Fraction macro of the separate supplement (applies to the supplement only)
\DeclareRobustCommand{\frac}[3][0pt]{%
  {\begingroup\hspace{#1}#2\hspace{#1}\endgroup\over\hspace{#1}#3\hspace{#1}}}

\section{Effect of the vaccine-induced susceptibility}\label{APP: V_k eff}
The primary reason we introduced the vaccine-induced susceptibility $V_k$, to use a convenient summary for the class of treatment interventions, that we believe to have the same causal effect on the outcome. However, the interpretation of $V_k$, and in particular its causal relationship to $Y_k$ is subtle. Since $V_k$ is an indicator of a state (increased susceptibility), which is not intervenable, defining an effect of $V_k$ on $Y_k$ is questionable, when one commits to the interventionist notion of causes and effects. However, building on the deterministic relationship between $\overline{A}_k$ and $V_k$, we can informally define what we understand under the expression ``the effect of $V_k$ on $Y_k$".

\begin{corollary}\label{COR: int group}
Under Assumptions \ref{ASS: trans risk}, \ref{ASS: exch}, positivity and  consistency
    $$
    E\left[ Y_k^{\overline{a}_k^{1'}}\middle| U=u, \overline{L}_k=\overline{l}_k\right]=E\left [Y_k^{\overline{a}_k^{1''}} \middle| U=u, \overline{L}_k=\overline{l}_k\right]
    $$
    and thus integrating over the distribution of $(U,L_k)$

    $$
    E\left[ Y_k^{\overline{a}_k^{1'}}\right]=E\left [Y_k^{\overline{a}_k^{1''}} \right]
    $$
    for any $\overline{a}_k^{1'}, \overline{a}_k^{1''} \in \{\overline{a}_k: \sum_{i=k-m+1}^{k} a_i >0 \}$.
    Similarly,
    $$
    E\left[ Y_k^{\overline{a}_k^{0'}}\middle| U=u, \overline{L}_k=\overline{l}_k\right]=E\left [Y_k^{\overline{a}_k^{0''}}\middle| U=u, \overline{L}_k=\overline{l}_k\right]
    $$
    and thus
    $$
    E\left[ Y_k^{\overline{a}_k^{0'}}\right]=E\left [Y_k^{\overline{a}_k^{0''}}\right]
    $$
    for any $\overline{a}_k^{0'}, \overline{a}_k^{0''} \in \{\overline{a}_k: \sum_{i=k-m+1}^{k} a_i =0 \}$.
\end{corollary}
See the proof of Corollary \ref{COR: int group} in Appendix \ref{APP_sub: proof of cor}.

    Under the proposed data-generating mechanism encoded by the DAG in Figure \ref{FIG: DAG} we can give an informal interpretation of the ``effect of being at-risk on the outcome", which would correspond to the contrast $E[Y_k^{v_k=1}]$ vs $E[Y_k^{v_k=0}]$, had the intervention on $V_k$ been well-defined.
     There exists intervenable random variables, $\overline{A}_k$-s, that have a deterministic effect on $V_k$ and the effect of them on $Y_k$ is fully mediated through $V_k$ \citep{hernan_causal_2025}. Thus, the ill-defined effect $E[Y_k^{v_k=1}]-E[Y_k^{v_k=0}]$ can be interpreted as $E[Y_k^{\overline{a}_k^{1}}]-E[Y_k^{\overline{a}^{0}_k}]$, where $\overline{a}_k^{1}\in \{\overline{a}_k: \sum_{i=k-m+1}^{k} a_i>0 \}$ and $\overline{a}_k^{0} \in \{\overline{a}_k: \sum_{i=k-m+1}^{k} a_i=0\}$, under the results of Corollary \ref{COR: int group}.  Thus, the effect of $V_k$ on $Y_k$ can be interpreted informally, as the contrast between the expected potential outcomes under interventions on $\overline{A}_k$ that set the value of $V_k$ to $1$ and $0$, respectively, in the setting where $\overline{A}_k$ has no direct effect on $Y_k$, except through $V_k$.

\begin{figure}[!h]
    \centering
    \begin{tikzpicture}[node distance=2cm, auto]

    % Nodes at time 1
    \node[draw, circle] at (1,1) (A1) {$A_{1}$};
    \node[draw, circle] at (1,3) (V1) {$V_{1}$};
    \node[draw, circle] at (1,5) (Y1) {$Y_{1}$};
    \node[draw, circle] at (0,7) (L1) {$L_{1}$};

    % Nodes at time 2
    \node[draw, circle] at (3,1) (A2) {$A_{2}$};
    \node[draw, circle] at (3,3) (V2) {$V_{2}$};
    \node[draw, circle] at (3,5) (Y2) {$Y_{2}$};
    \node[draw, circle] at (2,7) (L2) {$L_{2}$};

    % Nodes at time 3
    \node[draw, circle] at (5,1) (A3) {$A_{3}$};
    \node[draw, circle] at (5,3) (V3) {$V_{3}$};
    \node[draw, circle] at (5,5) (Y3) {$Y_{3}$};
    \node[draw, circle] at (4,7) (L3) {$L_{3}$};

    % U and S
    \node[draw, circle] at (-2,7) (U) {$U$};
    \node[draw, circle] at (6,6.5) (S) {$S_3$};

    % Arrows from U to all A_k, Y_k, L_k 
    \draw[->] (U) edge[lightgray] (A1);
    \draw[->] (U) edge[lightgray] (Y1);
    \draw[->] (U) edge[lightgray] (L1);
    \draw[->] (U) edge[lightgray, bend left=5] (A2);
    \draw[->] (U) edge[lightgray] (Y2);
    \draw[->] (U) edge[lightgray, bend left=25] (L2);
    \draw[->] (U) edge[lightgray, bend right=5] (A3);
    \draw[->] (U) edge[lightgray] (Y3);
    \draw[->] (U) edge[lightgray, bend left=25] (L3);

    % Arrows from outcomes to S
    \draw[->] (Y1) edge[green] (S);
    \draw[->] (Y2) edge[green] (S);
    \draw[->] (Y3) edge[green] (S);

    % Treatment to at-risk m=2
    \draw[->] (A1) edge[green] (V1);
    \draw[->] (A1) edge[green] (V2);
    \draw[->] (A2) edge[green] (V2);
    \draw[->] (A2) edge[green] (V3);
    \draw[->] (A3) edge[green] (V3);
    
    % Treatment to treatment
    \draw[->] (A1) edge (A2);
    \draw[->] (A1) edge[bend right=30] (A3);
    \draw[->] (A2) edge (A3);

    % Treatment to outcome m=2
    \draw[->] (A1) edge[bend right=25] (Y1);
    \draw[->] (A2) edge[bend right=25] (Y2);
    \draw[->] (A3) edge[bend right=25] (Y3);

    \draw[->] (A1) edge (Y2);
    \draw[->] (A2) edge (Y3);

    % L to L
    \draw[->] (L1) edge (L2);
    \draw[->] (L1) edge[bend left=30] (L3);
    \draw[->] (L2) edge (L3);

     % Arrows from L_k to A_k, Y_k
    \draw[->] (L1) edge[bend right=10] (A1);
    \draw[->] (L1) edge (Y1);
    \draw[->] (L2) edge[bend right=10] (A2);
    \draw[->] (L2) edge (Y2);
    \draw[->] (L3) edge[bend right=10] (A3);
    \draw[->] (L3) edge (Y3);

\end{tikzpicture}
    \caption{Modified version of Figure \ref{FIG: DAG}. The determinisms between $V_k$-s and $A_k$-s still hold, however, $V_k$ has no effect on $Y_k$, while the last $m=2$, $A_k$-s have a direct effect on the outcomes.}
    \label{FIG: mod DAG}    
\end{figure}
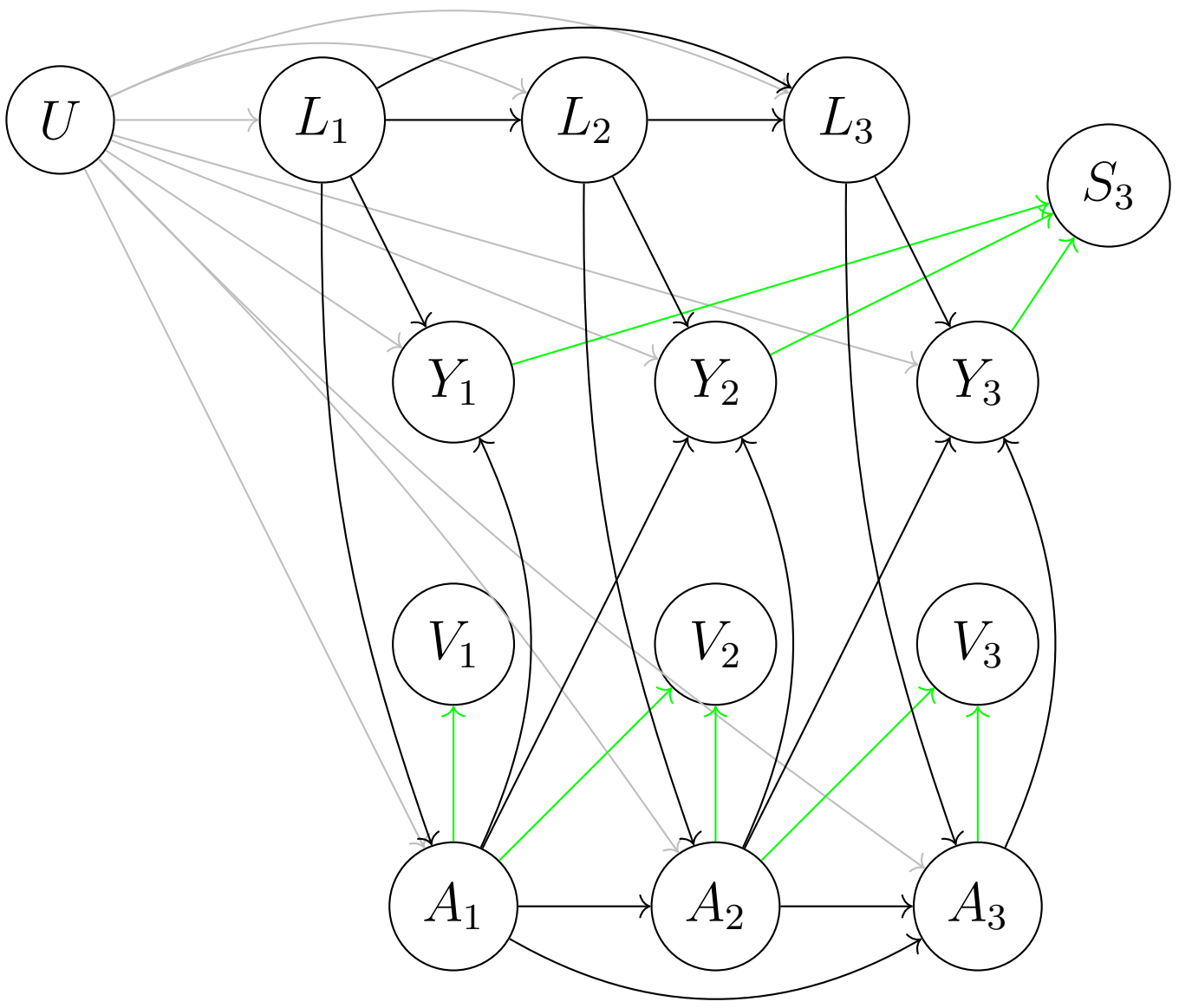

    Assumption \ref{ASS: trans risk} holds under the DAG depicted in Figure \ref{FIG: DAG}, and the relevant conditional independences can be read off the DAG using d-separation \citep{pearl_chapter_1988}. However, a reasonable critique of the DAG in Figure \ref{FIG: DAG} could be, that since $V_k$ is not intervenable, the arrow from $V_k$ to $Y_k$ is ill-defined. Instead it is the treatment history $\overline{A}_k$ that affects $Y_k$. To address this critique, we may instead consider the modified DAG in Figure \ref{FIG: mod DAG}. 

Under the DAG in Figure \ref{FIG: mod DAG}, Assumption \ref{ASS: trans risk} (iii) is no longer a graphical assumption and must be stated explicitly. Alternatively, Corollary \ref{COR: int group} can be formulated as an assumption, stating that, under all interventions $\overline{a}_k$ that result in the same susceptibility indicator $V_k^{\overline{a}_k}$, the expected potential outcome is the same. Under this formulation, $V_k$ represents a functional restriction on the potential outcomes rather than a cause of $Y_k$.

\section{Graphical Assumptions}\label{APP: graph ass}

We state and explicitly discuss the assumptions on which our justification of the SCCS design relies. All of these assumptions hold under the DAG in Figure \ref{FIG: DAG}, however, some can be relaxed when the goal is testing rather than estimation.

\begin{assumption}[Transient risk]\label{ASS: trans risk}
\leavevmode
\begin{enumerate}[label=(\roman*)]
        \item $V_{k'}\independent Y_k|V_k, \overline{L}_{k}, U$
        \item $V_{k'}\independent L_k|\overline{L}_{k'}, U$
        \item $\overline{A}_k\independent Y_k|V_k, \overline{L}_k, U$
    \end{enumerate}
    for all $k'<k \in \{1, \dots t\}$. 
\end{assumption}

Informally, Assumption \ref{ASS: trans risk} (i) and (ii) states that $V_{k'}$ has no effect on future outcomes or future covariates, while Assumption \ref{ASS: trans risk} (iii) posits that $V_k$ fully mediates the effect of the whole treatment history ($\overline{A}_k$) on $Y_k$.

Since the vaccine-induced susceptibility indicator $V_k$ is defined as a deterministic function of $A_{k-m+1}, \dots, A_k$, there are no other incoming arrows into $V_k$. This rules out confounding between $V_k$ and any other variables in the DAG when conditioning on treatment history.

The more general case can be defined with the reactogenicity window starting $m_s\geq 0$ time points after vaccination, resulting in $A_k = 1\implies V_{k+m_s}=\dots = V_{k+m_s+m-1}=1$ and $A_{k-m+1}= \dots = A_k=0 \implies V_{k+m_s}=0$ for some $m_s\geq 0, m\geq1$. To simplify notation, we assume hereafter $m_s=0$, however, the same arguments apply analogously to the more general case $m_s\geq 0$.

We also assume no treatment-outcome confounding when conditioning on the measured time-varying covariates and the unmeasured time-invariant covariates.

\begin{assumption}[Treatment exchangeability]\label{ASS: exch}
    $$
    \overline{A}_k\independent \overline{Y}_k^{\overline{a}_k}|\overline{L}_k, U,
    $$ for all $k \in \{1, \dots, t\}$.
\end{assumption}
Assumption \ref{ASS: exch} can be read from the Single World Intervention Graph \citep{richardson_single_2013-1, richardson_single_2013}. Assumption \ref{ASS: exch} essentially requires that there are no unmeasured common causes of outcome and treatment measurements: every confounder is either time-invariant (possibly unmeasured, $U$) or time-varying and measured ($L_k$). Under Assumption \ref{ASS: exch} and the null hypothesis of no causal effect of susceptibility on the outcome, see Section \ref{SEC: test}, Assumption \ref{ASS: trans risk} (iii) holds automatically. Although the null hypothesis does not directly imply Assumption \ref{ASS: trans risk} (i), it is natural to require a corresponding condition for past susceptibility measurements: if $V_k$ has no effect on $Y_k$, then previous values $V_{k'}$ should not affect $Y_k$ either. The absence of uncontrolled confounding for these associations is ensured by Assumption \ref{ASS: exch}.

\begin{remark}\label{COR: point exch}
    By decomposition, under Assumption \ref{ASS: exch}, exchangeability holds at the index level, that is
    $$
    \overline{A}_k\independent Y_k^{\overline{a}_k}|\overline{L}_k, U
    $$ for all $k \in \{1,...t\}$.
\end{remark}

Under Assumptions \ref{ASS: trans risk}, \ref{ASS: exch} and consistency, we showed in Appendix \ref{APP: V_k eff} that if two treatment interventions $\overline{a}_k$, $\overline{a}_k'$ give the same vaccine-induced susceptibility status at time $k$, then they lead to the same expected outcome at time $k$. In this sense, we informally refer to the contrast between expected potential outcomes under interventions that induce $V_k=1$ versus $V_k=0$ as an effect of $V_k$ on $Y_k$. This interpretation relies on the deterministic definition of $V_k$ and on the fact that the effect of $A_{k-m+1}, \dots, A_k$ on $Y_k$ is fully mediated by $V_k$.

\begin{assumption}[Outcome independence]\label{ASS: ind out}
\leavevmode
\begin{enumerate}[label=(\roman*)]
        \item $Y_k \independent (Y_{k+1}, \dots Y_t) |\overline{A}_{t}, \overline{L}_{t}, U$
        \item $Y_k \independent (A_{k+1}, \dots A_t)|\overline{A}_k, \overline{L}_{k}, U$
        \item $Y_k\independent (L_{k+1}, \dots L_t)|\overline{A_t}, \overline{L_k}, U$
    \end{enumerate}
    for all $k\in \{1, \dots, t-1\}$. 
\end{assumption}

Assumption \ref{ASS: ind out} formalises two standard SCCS conditions: (i) within-individual event occurrences are independent given treatment and covariates history (Assumption A1 in \citet{farrington_self-controlled_2018}), and (ii) event occurrence does not affect subsequent treatment (Assumption A3 in \citet{farrington_self-controlled_2018}).
Finally, Assumption \ref{ASS: ind out} (iii) ensures that the outcome has no effect on the future covariates. As in most applications $L_k$ stands for age or seasonality, it is reasonable to expect Assumption (iii) to hold.
Assumption \ref{ASS: ind out} (ii) and (iii) jointly imply that 
$$
Y_k \independent (A_{k+1}, \dots, A_{t}, L_{k+1}, \dots , L_t)|\overline{A}_k, \overline{L}_{k}, U,
$$

for all $k \in \{1, \dots, t-1 \}$, that is, conditional on treatment and covariate history, the outcome is independent from the joint distribution of future treatments and covariates.

The assumptions required for testing depend on the procedure. The standard $SCCS$ based analysis relies on Assumption \ref{ASS: ind out} (i) to obtain the Poisson model. In contrast, Algorithm \ref{alg: hyper} requires under the sharp null partial outcome interchangeability, Assumption \ref{ASS: null strong inter}. Finally, Algorithm \ref{ALG: Dependent} further relaxes the requirement of outcome interchangeability and allows arbitrary within-person dependence.

\begin{assumption}[Transient time-varying confounding]\label{ASS: trans conf}
    \begin{align*}
        Y_k \independent \overline{L}_{k-1}|\overline{A}_k, L_k, U
        % &\text{and}& &A_k \independent \overline{L}_{k-1}|\overline{Y}_k, L_k, U&
    \end{align*}
    for all $k \in \{1, \dots, t\}$.
\end{assumption}

Assumption \ref{ASS: trans conf} states that $L_k$ acts as a confounder between treatment and outcome only at time $k$, and not across multiple time points. In many applications of the SCCS, this holds by design for the user chosen $L_k$, because $L_k$ represents variables such as age or seasonality, once $L_k$ is specified, the history $\overline{L}_{k-1}$ is also determined. Assumption \ref{ASS: trans conf} can be relaxed by allowing a confounding window $m'$, such that $L_k$ affects 
$Y_k, \dots Y_{k+m-1}$ and $A_{k}, \dots, A_{k+m-1}$. However, this would complicate the discussion, and is therefore only discussed in Appendix \ref{APP: L_k ext}.

\begin{remark}
    Under Assumption \ref{ASS: trans risk} (ii) treatment $A_k$ cannot have a direct effect on any future covariate $L_{k'}$ for $k'>k$.
    
    Suppose that such an effect exists. As $V_k$ is a deterministic function of the treatment history, including $A_k$, we  the open path 
    $$
    V_k \leftarrow A_k \rightarrow L_{k'}
    $$
    exists as well. As none of $\overline{L}_k$ and $U$ are on this path, conditioning on them would not block the path above, contradicting Assumption \ref{ASS: trans risk} (ii).
\end{remark}

\section{Intermediate Propositions}\label{APP: inter props}
We showed in Section \ref{SEC: ident} that under a well-defined set of assumptions, the standard $SCCS$ method identifies the causal estimand of interest, the $CRR_m$. While we have informally stated the intermediate results, that is, what quantities can be identified, at least in principle, whenever only a subset of the sufficient assumptions hold, here we provide the formal statements and their interpretations and limitations. The proofs of these propositions are provided in Appendix \ref{APP: PROOFS}.

\begin{proposition}\label{PROP: c-RR(u,l)}
    Under Assumptions \ref{ASS: stat all}-\ref{ASS: pos} and \ref{ASS: exch}-\ref{ASS: trans conf}, the $c\text{-}RR(u,l)_k$ is identified as
    $$
    \frac{E[Y_{k^1}|U=u, \overline{A}_t=\overline{a}_t, \overline{L}_t=\overline{l}_t, S_t>0]}{E[Y_{k^0}|U=u, \overline{A}_t=\overline{a}_t, \overline{L}_t=\overline{l}_t, S_t>0]},
    $$
    for any time points $k^1, k^0 \in \{1, 2, \dots ,t\}$, where
    $\overline{a}_t, \overline{l}_t$ are defined such that 
    $\sum_{k=k^1-m+1}^{k^1} a_k >0$, $\sum_{k=k^0-m+1}^{k^0} a_k =0 $, and $l_{k^1}=l_{k^0}=l$.
    Moreover, the $c\text{-}RR(u,l)$ is equivalently identified by
    $$
    \frac{\sum_{k \in \mathcal{K}_{1,l}} E[Y_{k}|U=u, \overline{A}_t=\overline{a}_t, \overline{L}_t=\overline{l}_t, S_t>0]/|\mathcal{K}_{1,l}|}{\sum_{k \in \mathcal{K}_{0,l}} E[Y_{k}|U=u, \overline{A}_t=\overline{a}_t, \overline{L}_t=\overline{l}_t, S_t>0]/|\mathcal{K}_{0,l}|},
    $$
    where $\mathcal{K}_{1,l}:=\{k: \sum_{i=k-m+1}^{k} a_i >0, l_k=l\}$, $\mathcal{K}_{0,l}:=\{k: \sum_{i=k-m+1}^{k} a_i =0, l_k=l\}$, for any $u$ and $l$ for which the ratio is well-defined.
\end{proposition}

The time points $k^1$ and $k^0$ are chosen so that, for fixed treatment and covariates histories $\overline{A}_t$, $\overline{L}_t$, the value of the covariate is the same at both times, whereas the treatment-induced susceptibility indicator $V_k$ equals $1$ at one time point and $0$ at the other.
See the proof of Proposition \ref{PROP: c-RR(u,l)} in Appendix \ref{APP_sub: proof c-rr(u,l)}.

The identification formula of Proposition \ref{PROP: c-RR(u,l)} requires observing $U$ for all cases ($S_t>0$), however, $U$ is typically unobserved. Alternatively, instead of conditioning on the unobserved variable $U$, we can condition on individuals, as it is done implicitly in the standard $SCCS$ design. However, by conditioning on the individual, we do not only condition on being a case, $S_t>0$, but we fix its value, $S_t=s_t>0$, which in general introduces bias. In the following proposition, we show, that under Assumption \ref{ASS: rare event}, the introduced bias vanishes as the time-scale approaches continuous time.

\begin{proposition}\label{PROP: c-RR(u,l) St=st}
    Under Assumptions \ref{ASS: stat all}-\ref{ASS: rare event} and \ref{ASS: exch}-\ref{ASS: trans conf},
    $$
    \frac{E[Y_{k^1}|U=u, \overline{A}_t=\overline{a}_t, \overline{L}_t=\overline{l}_t, S_t=s]}{E[Y_{k^0}|U=u, \overline{A}_t=\overline{a}_t, \overline{L}_t=\overline{l}_t, S_t=s]} \rightarrow c\text{-}RR(u,l)
    $$
    as $t \rightarrow \infty$ for any time points $k^1, k^0 \in \{1, 2, \dots ,t\}$, where
    $\overline{a}_t, \overline{l}_t$ are defined such that 
    $\sum_{k=k^1-m+1}^{k^1} a_k >0$, $\sum_{k=k^0-m+1}^{k^0} a_k =0 $, and $l_{k^1}=l_{k^0}=l$.

    Moreover, the $c\text{-}RR(u,l)$ is equivalently identified, in the limit without bias, by
    $$
    \frac{\sum_{k \in \mathcal{K}_{1,l}} E[Y_{k}|U=u, \overline{A}_t=\overline{a}_t, \overline{L}_t=\overline{l}_t, S_t=s]/|\mathcal{K}_{1,l}|}{\sum_{k \in \mathcal{K}_{0,l}} E[Y_{k}|U=u, \overline{A}_t=\overline{a}_t, \overline{L}_t=\overline{l}_t, S_t=s]/|\mathcal{K}_{0,l}|},
    $$
    where $\mathcal{K}_{1,l}:=\{k: \sum_{i=k-m+1}^{k} a_i >0, l_k=l\}$, $\mathcal{K}_{0,l}:=\{k: \sum_{i=k-m+1}^{k} a_i =0, l_k=l\}$, for any $u$ and $l$ for which the ratio is well-defined.
\end{proposition}
See the proof of Proposition \ref{PROP: c-RR(u,l) St=st} in Appendix \ref{APP_sub: St=st proof}.

\section{Simulation}\label{APP: sim}
The three testing procedures were compared were a hypergeometric test, Algorithm \ref{alg: hyper}, a bootstrap-based rate-difference test, Algorithm \ref{ALG: Dependent}, and the standard $SCCS$ procedure, by fitting a conditional logistic regression model on the cloned data, as described by \citet{farrington_self-controlled_2018}.
For each simulated dataset, longitudinal person-time was generated for a fixed cohort of  $N=5000$ independent individuals observed over $t=200$ discrete time intervals, without censoring. All these procedures use the cases, that is, individuals with $S_t>0$, thus the number of cases is random. Each individual had a baseline covariate $U$, and follow-up time was partitioned into four age or calendar-time strata $L_{k}$. At each interval, treatment $A_{k}$ was generated, depending on the covariates $U$ and $L_{k}$. The susceptibility indicator $V_{k}$ was then defined based on the treatment history in the reactogenicity window of the last $m=30$ timepoints. Outcome events $Y_{k}$ were generated under several data-generating mechanisms, described in detail as follows. All tests were two-sided, at level $0.05$, and Algorithms \ref{alg: hyper} and \ref{ALG: Dependent} used $B=9,999$ resamples. Each scenario was replicated $5,000$ times.

\subsection{Description of the simulated scenarios}\label{APP: scen desc}
 We evaluated Algorithms \ref{alg: hyper} and \ref{ALG: Dependent}, and the standard SCCS model in some simulated datasets in Subsection \ref{APP: sim}. The various scenarios included data-generating mechanisms, when all or just a subset of the sufficient assumptions for valid testing by the procedures hold. 

 For each simulated scenario, we generated data for $N=5000$ individuals observed over $t=200$ discrete timepoints. Even though we simulated from a discrete process, we can assume that Assumption \ref{ASS: rare event} approximately holds.
 In the \textbf{Independent} scenarios, for example, the largest per-time-point probability was approximately $0.005$, and the sum of the squared per-time-point event probabilities of an individual, which by Equation \eqref{EQ: le cam} bounds the total variation distance between the individuals' event count and their Poisson approximation (see Appendix \ref{APP_Sub: pois proof}), was about $0.0014$ on average and at most $0.0028$.
 Similarly to the main text, we forgo indexing each individual, and we index time with $k$. Each individual had a baseline covariate $U \sim \mathcal{N}(0,1)$ truncated to the interval $[-3,3]$, unless otherwise specified, and the follow-up was divided into four equal deterministic time strata $L_k \in \{0,1,2,3\}$.

 At each time point $k$, the treatment indicator $A_k$ was generated from a Bernoulli model:
 
$$
A_k \sim \mathrm{Bernoulli}\left(\mathrm{logit^{-1}}\left(\gamma_0+\gamma_U U+\gamma_L( L_k)
\right)\right).
$$

In the base scenario
the treatment model parameters are $\gamma_0=-2.8$, $\gamma_U=0.25$ and $\gamma_L=(0, 0.15, 0.3, 0.45)$, corresponding to the 4 levels of the time-varying covariate $L_k$.
 
The binary at risk indicator $V_k$ was defined as a deterministic function of $A$, with $m=30$, specifically, $V_k= \mathbb I \left(\sum_{j=\max(1, k-29)}^k A_j>0\right)$. 

The outcome model was changing with the specified scenarios, to investigate the consequences of the violated sufficient assumptions. In the simplest scenario, with independent outcomes, the outcomes were generated as

$$
Y_k \sim \mathrm{Bernoulli}\left(\mathrm{exp}\left(\beta_0+\alpha_VV_k+ \beta_U U+\beta_L( L_k)
\right)\right).
$$

Here, the covariates act multiplicatively on the outcome probability, with $\beta_U=0.10$ and $\beta_L=(0, 0.25, 0.5, 0.75)$, while $r=\exp(\alpha_V)$ varied according to whether we considered the null scenario, or if not, the strength of the signal.
The intercept $\beta_0$ depended on $r$, and it was fixed before the simulation to obtain approximately $2,500$ events per dataset: $\beta_0=-6.41$, $-6.6$, $-6.77$ and $-7.03$ for $r=1$, $1.25$, $1.5$ and $2$, respectively. The same intercepts were used in the \textbf{CorrelatedNoise} and \textbf{PolyaUrn} scenarios. Consequently, the effect scenarios differ from the null scenarios both in $\alpha_V$ and in $\beta_0$.

More complex scenarios modified this baseline mechanism, as described below.

\subsubsection{Scenario definitions}
\begin{itemize}
    \item \textbf{Independent\_Null}
    \begin{itemize}
        \item Base scenario for both the treatment and the outcome models.
        \item  $\alpha_V=0$.
        \item All sufficient assumptions hold for the three (Standard SCCS model, Algorithms \ref{alg: hyper} and \ref{ALG: Dependent}) proposed methods.
    \end{itemize}
        \item \textbf{Independent\_SmallEffect}:
    \begin{itemize}
        \item Identical to \textbf{Independent\_Null}, except $\alpha_V=\mathrm{log}(1.25)$.
        \item No assumption is violated; however, the null hypothesis is false.
    \end{itemize}
    \item \textbf{Independent\_MildEffect}:
    \begin{itemize}
        \item Identical to \textbf{Independent\_Null}, except $\alpha_V=\mathrm{log}(1.5)$.
        \item No assumption is violated; however, the null hypothesis is false.
    \end{itemize}
    \item \textbf{Independent\_StrongEffect}:
    \begin{itemize}
        \item Identical to \textbf{Independent\_Null}, except $\alpha_V=\mathrm{log}(2)$.
        \item No assumption is violated; however, the null hypothesis is false.
    \end{itemize} 
    \item \textbf{ULInteraction\_Null}:
    \begin{itemize}
        \item The baseline covariate \(U\) was made binary rather than Gaussian:
        $$
        U \in \{-1.5,\,1.5\}, \qquad P(U=-1.5)=P(U=1.5)=0.5.
        $$
        
        \item A $U \times L_k$ interaction term was added to the outcome model:
        $$
        Y_k \sim \mathrm{Bernoulli}\left(\mathrm{exp}\left(\beta_0+\alpha_VV_k+ \beta_U U+\beta_L( L_k)+\beta_{UL}(L_k)U
        \right)\right).
        $$
        with
        $\alpha_V=0$,
        $\beta_U=0.05$,
        $\beta_{UL}=(-1.2,\,-0.4,\,0.4,\,1.2)$, where the value of $\beta_{UL}$ is determined by the category of $L$, and $\beta_0=-7.21$.

        \item Assumption \ref{ASS: no int} is violated, which we relied on in justifying the standard SCCS model.
    \end{itemize} 

    \item \textbf{ULInteraction\_WeakNull}:
    \begin{itemize}
        \item Same data-generating mechanism as for \textbf{ULInteraction\_Null}, except the interaction term is weaker, $\beta_{UL}=(-0.6,\,-0.2,\,0.2,\,0.6)$, and $\beta_0=-6.64$.
        \item Assumption \ref{ASS: no int} is violated, to a lesser extent.
    \end{itemize}

    \item \textbf{ULInteraction\_AlignedNull}:
    \begin{itemize}
        \item Same data-generating mechanism as for \textbf{ULInteraction\_Null}, except the same interaction pattern is added to the treatment model
        $$
        A_k \sim \mathrm{Bernoulli}\left(\mathrm{logit}^{-1}\left(\gamma_0+\gamma_U U+\gamma_L( L_k)+\gamma_{UL}(L_k)U\right)\right),
        $$
        with $\gamma_{UL}=(-1.2, -0.4, 0.4, 1.2)$.
        \item Assumption \ref{ASS: no int} remains violated for the outcome model, but now the violation is more pronounced in the bias, as the same interaction raises both the treatment and the outcome model, acting as an unmeasured confounder.
    \end{itemize}

    \item \textbf{ULInteraction\_AlignedStrongNull}:
    \begin{itemize}
        \item Same data-generating mechanism as for \textbf{ULInteraction\_AlignedNull}, except the interaction term is made more extreme: 
        $$
            U \in \{-2,2\}, \quad P(U=-2)=P(U=2)=0.5
        $$
        and
        $$\gamma_{UL}=\beta_{UL}=(-1.8, -0.6, 0.6, 1.8),$$
        ad $\beta_0=-8.78$.
        \item Assumption \ref{ASS: no int} remains violated, stronger omitted alignment between the treatment and the outcome models.
    \end{itemize}

    \item \textbf{CorrelatedNoise\_Null}:
    \begin{itemize}
        \item Marginal event probabilities are the same as in \textbf{Independent\_Null}.
        \item Serial dependence is introduced by a latent Gaussian $\mathrm{AR}(1)$ process:
        $$
        Z_{1}\sim N(0,1),\qquad Z_{k}=\rho Z_{k-1}+\sqrt{1-\rho^2}\,\varepsilon_{k}, \qquad \varepsilon_{k}\stackrel{iid}{\sim}N(0,1),
        $$
        with $\rho =0.7$. $\varepsilon_k$ is an exogenous random variable, thus independent of $A_k$, $L_k$, and $U$. $\rho$ is the correlation of the latent process, not the outcomes.

        Outcomes were generated as 
        $$
        Y_{k}=\mathbb{I}(\Phi(Z_{k})<\mathrm{exp}\left(\beta_0+\alpha_VV_k+ \beta_U U+\beta_L( L_k)
\right)).
        $$

        \item Assumption \ref{ASS: ind out} (i) is violated, as the outcomes are dependent due to the latent error process. 
        Marginal stationarity within the covariate strata still holds.
        \item Assumption \ref{ASS: ind out} (i) was amongst the sufficient assumptions for the standard SCCS method.Algorithm \ref{alg: hyper} instead requires Assumption \ref{ASS: null strong inter}, which fails as well, because the dependence decays based on the temporal distance of the outcomes. Algorithm \ref{ALG: Dependent} requires neither.
    \end{itemize}
    
    \item \textbf{CorrelatedNoise\_MildEffect}:
    \begin{itemize}
        \item Same data-generating mechanism as for \textbf{CorrelatedNoise\_Null}, except $\alpha_V=\mathrm{log}(1.5)$.

        \item Assumption \ref{ASS: ind out} (i) remains violated.
    \end{itemize}

    \item \textbf{CorrelatedNoise\_StrongEffect}:
    \begin{itemize}
        \item Same data-generating mechanism as for \textbf{CorrelatedNoise\_Null}, except $\alpha_V=\mathrm{log}(2.0)$.

        \item Assumption \ref{ASS: ind out} (i) remains violated.
    \end{itemize}

     \item \textbf{CorrelatedNoise\_Rho30\_Null} and \textbf{CorrelatedNoise\_Rho90\_Null}:
    \begin{itemize}
        \item Same data-generating mechanism as for \textbf{CorrelatedNoise\_Null}, except $\rho=0.3$ and $\rho=0.9$, respectively.
        \item Assumptions \ref{ASS: ind out} (i) and \ref{ASS: null strong inter} remain violated, to a lesser and a greater extent, respectively.
    \end{itemize}
 
    \item \textbf{CorrelatedNoise\_Rho30\_MildEffect} and \textbf{CorrelatedNoise\_Rho90\_MildEffect}:
    \begin{itemize}
        \item Same data-generating mechanism as for \textbf{CorrelatedNoise\_MildEffect}, except $\rho=0.3$ and $\rho=0.9$, respectively.
    \end{itemize}
 
    \item \textbf{PolyaUrn\_Null}:
    \begin{itemize}
        \item Treatment model and marginal event probabilities as in \textbf{Independent\_Null}.
        \item For each person-by-$L_k$ stratum, a separate urn generates the outcomes in the absence of susceptibility, $Y_k^0$. The urn is initialised with event weight $20p_{0,l}$ and non-event weight $20(1-p_{0,l})$, where $p_{0,l}=\exp\{\beta_0+0.10U+\beta_L(l)\}$. At each time point in the stratum, $Y_k^0$ is drawn with the current fraction of the event weight as success probability, and the weight of the drawn colour is then increased by one.
        \item At $V_k=0$, $Y_k=Y_k^0$. At $V_k=1$, if $Y_k^0=0$, an event is added independently with probability $a_l=(r-1)p_{0,l}/(1-p_{0,l})$, so that the marginal event probability under susceptibility is $rp_{0,l}$. Only the outcomes $Y_k^0$ reinforce the urn. Here $r=1$, hence no effect.
        \item Assumption \ref{ASS: ind out} (i) is violated, as the outcomes within a stratum are dependent. Under the null, however, the probability of any outcome path of an urn depends only on its number of events, and the joint probability is the product over the urns, so that the outcome vector is invariant under within-$L_k$ permutations, and Assumption \ref{ASS: null strong inter} holds.
    \end{itemize}
 
    \item \textbf{PolyaUrn\_MildEffect} and \textbf{PolyaUrn\_StrongEffect}:
    \begin{itemize}
        \item Same data-generating mechanism as for \textbf{PolyaUrn\_Null}, except $r=1.5$ and $r=2$, respectively.
    \end{itemize}
    
    \item \textbf{BurstEpisodes\_Null}:
    \begin{itemize}
        \item Treatment model remains unchanged
        \item An outcome onset variable was introduced:
        $$
            O_k\sim \mathrm{Bernoulli}(\mathrm{exp}(\zeta_0+\zeta_VV_k+\zeta_U U+\zeta_L(L_k))).
        $$
        If the onset variable was equal to 1, $O_k=1$, then in the next $c=6$ timepoints, the outcome was deterministically set to $1$, $Y_k= \dots = Y_{k+5}=1$. Every individual starts as inactive, and onsets are drawn only while the individual is not in an active episode. Episodes can cross the boundaries of $L_k$.

        The onset process is still multiplicative on the probability scale in the covariates, $\mathrm{exp}(\zeta_V)$ interpreted as the effect of the at risk indicator on the onset.

        \item The parameters of the onset model are set to 
        $$
        \zeta_0=-8.19, \quad \zeta_V=0, \quad \zeta_L=(0, 0.25, 0.5, 0.75), \quad \zeta_U=0.1.
        $$

        \item Assumption \ref{ASS: ind out} (i) is violated, as outcomes are sequentially dependent.
        \item Assumption \ref{ASS: stat all} is violated, as even conditional on the covariates, the expected potential outcome depends on the onset occurring in the last $c$ timepoints.
        \item Assumption \ref{ASS: stat all} was among the sufficient assumptions for all three methods, thus invalidating them all.
    \end{itemize}
    \item \textbf{BurstEpisodes\_ModerateEffect}:
    \begin{itemize}
        \item Same data-generating mechanism as for \textbf{BurstEpisodes\_Null}, except $\zeta_V=\mathrm{log}(1.8)$ and $\zeta_0=-8.71$. The onset risk ratio is $1.8$, and the $CRR_m$, computed from the expected counts over the first $30$ intervals, is approximately $1.8$.
        \item Assumptions \ref{ASS: ind out} (i) and \ref{ASS: stat all} remain violated.
    \end{itemize}
    \item \textbf{Heterogeneous\_SameDirection} and \textbf{Heterogeneous\_MixedDirection}:
    \begin{itemize}
        \item $U \in \{-1.5,1.5\}$ with equal probabilities, $\beta_U=0.10$, no $U\times L_k$ interaction, and the treatment model of \textbf{Independent\_Null}.
        \item Independent outcomes with probability $\exp\{\beta_0+0.10U+\beta_L(L_k)\}\,r(U)^{V_k}$. In \textbf{Heterogeneous\_SameDirection}, $r(-1.5)=1.25$, $r(1.5)=2$ and $\beta_0=-6.895164527675$; in \textbf{Heterogeneous\_MixedDirection}, $r(-1.5)=0.5$, $r(1.5)=1.5$ and $\beta_0=-6.516658426237$.
        \item Assumption \ref{ASS: Stat U} is violated. The intervention ratios are
        $$
        CRR_m=\frac{\sum_{u\in\{-1.5,1.5\}}e^{0.10u}r(u)}{\sum_{u\in\{-1.5,1.5\}}e^{0.10u}}=1.680832 \quad\text{and}\quad 1.074443,
        $$
        respectively, thus both scenarios are alternatives.
    \end{itemize}
\end{itemize}

\subsection{Simulation results}

The empirical rejection rates based on $5,000$ replications of each scenario are summarised in Table \ref{TAB: test}.
The average number of cases per simulated dataset ranged from about $400$ in the \textbf{BurstEpisodes} scenarios, in which the events occur in episodes, to about $1{,}970$ in the \textbf{Independent} scenarios. The fit of the standard $SCCS$ model failed in $157$ replications in total, $78$ of them in \textbf{ULInteraction\_AlignedStrongNull} and at most $25$ in any other scenario; its rejection rates are computed among the successful fits. Algorithms \ref{alg: hyper} and \ref{ALG: Dependent} had no failures.
In the \textbf{CorrelatedNoise} scenarios, the joint distribution is not invariant under arbitrary within-stratum permutations, and hence Assumption \ref{ASS: null strong inter} does not hold, invalidating Algorithm \ref{alg: hyper}. Moreover, as the stationarity assumption, Assumption \ref{ASS: stat all} does hold, the lack of invariance, implies that Assumption \ref{ASS: ind out}(i) also fails, invalidating the standard SCCS design.
The empirical rejection rates confirm this, and the violation grows with the strength of the correlation: for $\rho=0.3$, $0.7$ and $0.9$, the rejection rates under the null were $0.053$, $0.105$ and $0.272$ for the standard $SCCS$ procedure, and $0.051$, $0.103$ and $0.267$ for Algorithm \ref{alg: hyper}, whereas Algorithm \ref{ALG: Dependent}, which requires neither assumption, remained at the nominal level ($0.055$, $0.050$ and $0.051$).
In the \textbf{PolyaUrn} scenarios, the outcomes of an individual are dependent, but under the null their joint distribution is invariant under within-$L_k$ permutations, so that Assumption \ref{ASS: ind out}(i) fails, whereas Assumption \ref{ASS: null strong inter} holds. Accordingly, under the null, the standard $SCCS$ procedure rejected in $0.113$ of the replications, whereas Algorithms \ref{alg: hyper} and \ref{ALG: Dependent} maintained the nominal level ($0.051$ and $0.053$).
In the \textbf{Independent} scenarios, the sufficient assumptions of all three approaches hold, and we find the empirical rejection rates under the null at the nominal level ($0.049$, $0.047$ and $0.055$). Under the alternatives, the standard $SCCS$ procedure had the greatest power, followed by Algorithms \ref{alg: hyper} and \ref{ALG: Dependent}, with rejection rates of $0.924$, $0.823$ and $0.666$ for the small effect ($r=1.25$), and of $1.000$, $0.999$ and $0.954$ for the mild effect ($r=1.5$), consistent with its use of an approximately correctly specified model.
The standard $SCCS$ is susceptible to bias whenever the model is misspecified, for example, when there is interaction between the measured and unmeasured covariates: in the \textbf{ULInteraction} scenarios, its empirical rejection rate under the null increased with the strength of the interaction, from $0.275$ to $0.564$, and was $1.000$ when the same interaction also raised the treatment probabilities. In contrast, neither of the proposed testing algorithms rely on the no interaction assumption, Assumption \ref{ASS: no int} thus even in the \textbf{ULInteraction} scenarios the empirical rejection rates of Algorithm \ref{alg: hyper} ranged from $0.041$ to $0.052$, and those of Algorithm \ref{ALG: Dependent} from $0.051$ to $0.059$, except in \textbf{ULInteraction\_AlignedStrongNull}, where Algorithm \ref{ALG: Dependent} showed a finite-sample inflation, rejecting in $0.121$ of the replications. In this scenario, treatment is nearly deterministic in several $U\times L_k$ strata, with per-time-point treatment probabilities ranging from about $0.002$ to $0.85$, so that many informative strata contain only a few at-risk or not at-risk time points. This is a likely reason why the asymptotic approximation underlying Algorithm \ref{ALG: Dependent} performs poorly in this scenario, and why the fit of the standard $SCCS$ model failed more often.
In the \textbf{Heterogeneous} scenarios, Assumption \ref{ASS: Stat U} is violated and the null hypothesis is false, with $CRR_m=1.681$ and $1.074$. When all individual effects had the same direction, the rejection rates were $0.999$, $0.998$ and $0.969$ for the standard $SCCS$ procedure, Algorithm \ref{alg: hyper} and Algorithm \ref{ALG: Dependent}, respectively. When the effects had opposite directions in different individuals, the rejection rates differed markedly ($0.774$, $0.492$ and $0.074$). All three procedures summarise the individual contrasts, but weight the individuals differently: the standard $SCCS$ procedure and Algorithm \ref{alg: hyper} by the amount of information they contribute, Algorithm \ref{ALG: Dependent} equally across cases (Appendix \ref{ASS: algs}). Effects of opposite signs therefore cancel to a different extent, and none of the procedures has power against all departures from the null hypothesis.

\begin{sidewaystable}[!h]
\vspace{15cm}
\centering
\small
\renewcommand{\arraystretch}{1.2}

\setlength{\tabcolsep}{4pt}

\begin{tabular}{lcccc r}
\toprule
 & & \multicolumn{3}{c}{Rejection rate [95\% Monte Carlo CI]} & \\
\cmidrule(lr){3-5}
Scenario & $H_0$ true & SCCS & Algorithm \ref{alg: hyper} & Algorithm \ref{ALG: Dependent} & Failed SCCS fits\\
\midrule
Independent\_Null & yes & 0.049 [0.043, 0.055] & 0.047 [0.041, 0.053] & 0.055 [0.048, 0.061] & 10\\
Independent\_SmallEffect & no & 0.924 [0.917, 0.932] & 0.823 [0.812, 0.833] & 0.666 [0.653, 0.679] & 22\\
Independent\_MildEffect & no & 1.000 [0.999, 1.000] & 0.999 [0.998, 1.000] & 0.954 [0.948, 0.959] & 2\\
Independent\_StrongEffect & no & 1.000 [0.999, 1.000] & 1.000 [0.999, 1.000] & 0.999 [0.998, 1.000] & 0\\
\midrule
ULInteraction\_WeakNull & yes & 0.275 [0.263, 0.288] & 0.048 [0.042, 0.054] & 0.052 [0.046, 0.059] & 4\\
ULInteraction\_Null & yes & 0.564 [0.551, 0.578] & 0.052 [0.046, 0.058] & 0.051 [0.045, 0.058] & 0\\
ULInteraction\_AlignedNull & yes & 1.000 [0.999, 1.000] & 0.048 [0.042, 0.054] & 0.059 [0.053, 0.066] & 1\\
ULInteraction\_AlignedStrongNull & yes & 1.000 [0.999, 1.000] & 0.041 [0.035, 0.047] & 0.121 [0.112, 0.130] & 78\\
\midrule
CorrelatedNoise\_Rho30\_Null & yes & 0.053 [0.047, 0.060] & 0.051 [0.045, 0.058] & 0.055 [0.049, 0.062] & 4\\
CorrelatedNoise\_Rho30\_MildEffect & no & 1.000 [0.999, 1.000] & 0.999 [0.998, 1.000] & 0.949 [0.943, 0.955] & 3\\
CorrelatedNoise\_Null & yes & 0.105 [0.097, 0.114] & 0.103 [0.095, 0.112] & 0.050 [0.044, 0.056] & 1\\
CorrelatedNoise\_MildEffect & no & 1.000 [0.999, 1.000] & 0.995 [0.993, 0.997] & 0.912 [0.903, 0.919] & 0\\
CorrelatedNoise\_StrongEffect & no & 1.000 [0.999, 1.000] & 1.000 [0.999, 1.000] & 0.999 [0.997, 0.999] & 0\\
CorrelatedNoise\_Rho90\_Null & yes & 0.272 [0.260, 0.285] & 0.267 [0.254, 0.279] & 0.051 [0.045, 0.057] & 0\\
CorrelatedNoise\_Rho90\_MildEffect & no & 0.987 [0.984, 0.990] & 0.970 [0.965, 0.975] & 0.768 [0.756, 0.779] & 0\\
\midrule
PolyaUrn\_Null & yes & 0.113 [0.104, 0.122] & 0.051 [0.045, 0.057] & 0.053 [0.047, 0.060] & 0\\
PolyaUrn\_MildEffect & no & 0.999 [0.997, 1.000] & 1.000 [0.999, 1.000] & 0.952 [0.946, 0.958] & 1\\
PolyaUrn\_StrongEffect & no & 1.000 [0.999, 1.000] & 1.000 [0.999, 1.000] & 1.000 [0.999, 1.000] & 1\\
\midrule
BurstEpisodes\_Null & yes & 0.395 [0.381, 0.409] & 0.413 [0.399, 0.426] & 0.118 [0.110, 0.128] & 0\\
BurstEpisodes\_ModerateEffect & no & 0.995 [0.993, 0.997] & 0.977 [0.972, 0.981] & 0.802 [0.790, 0.813] & 0\\
\midrule
Heterogeneous\_SameDirection & no & 0.999 [0.998, 1.000] & 0.998 [0.997, 0.999] & 0.969 [0.964, 0.973] & 5\\
Heterogeneous\_MixedDirection & no & 0.774 [0.762, 0.785] & 0.492 [0.478, 0.506] & 0.074 [0.067, 0.082] & 25\\
\bottomrule
\end{tabular}
\caption{Empirical rejection rates of the two-sided tests ($\alpha = 0.05$) over $5,000$ replications by scenario, with $95\%$ Monte Carlo confidence intervals in brackets. The rejection rates and confidence intervals of the standard $SCCS$ method are computed among the successful fits, and the last column gives the number of replications in which the fit failed. Algorithms \ref{alg: hyper} and \ref{ALG: Dependent} had no failures.}
\label{TAB: test}
\end{sidewaystable}

\section{Extended time-varying confounding}\label{APP: L_k ext}
In most applications of the SCCS design, $L_k$ denotes age or seasonality, both of which may plausibly act as common causes of the treatment and the outcome, and vary as time progresses. However, because age and seasonality are deterministic functions of time, their trajectories are fully determined by their value at any single time point, up to the chosen time scale. Thus, conditional on an individual’s age, or on the calendar month, at time $k$, the corresponding value at any subsequent time $k'>k$ is fixed. Inversely, the current value of age or calendar month, $L_k$, contains all relevant information about its history $\overline{L}_k$, rendering the prior history $\overline{L}_{k-1}$ redundant. Consequently, Assumption \ref{ASS: trans conf} holds in this setting.

However, we can hypothesise other time-varying common causes of treatment and outcome, such as blood pressure, concurrent medications, physical activity, etc. For these variables, determinism over time does not hold, while $L_k$ can plausibly affect the outcome and the treatment at future times $k'>k$. For example, certain nonsteroidal anti-inflammatory drugs might be contraindicated if an individual has experienced continuously high blood pressure, while a history of high blood pressure can increase the risk for certain adverse outcomes, such as transient ischemic attacks or atrial fibrillation episodes. Therefore, not only $L_k$ acts as a confounder between $Y_k$ and $A_k$, but possibly the whole history of confounders, $\overline{L}_k$. More realistically, this history may be represented by a finite lag window, $L_{k-m_l+1}, \dots, L_k$, where $m_l\geq 1$ denotes the number of time points over which the confounder can exert forward effects on subsequent treatments and outcomes. Consequently, Assumption \ref{ASS: trans conf} does not hold, due to the direct effect of $L_{k-m_l+1}, \dots, L_{k-1}$ on $Y_k$, however, we can derive similar results under the modified version of it.
\begin{assumption}[General transient time-varying confounding]\label{ASS: gen trans conf}
    $$
    Y_k \independent \overline{L}_{k-m_l}|\overline{A}_k, L_{(k-m_l+1):k},  U
    $$
\end{assumption}

\begin{remark}
A covariate window $L_{(k-m_l+1):k}$ is defined only for $k \ge m_l$. Throughout this appendix we therefore restrict comparison times to $\mathcal T_{m_l} := \{m_l, \dots, t\}$.

\end{remark}

Similarly, the conditional rate ratio has to be adjusted for the extended effect of the time-varying confounders:
$$
c\text{-}RR(u,l_{(k-m_l+1):k})_k = \frac{E[Y_k^{\overline{a}^1_k }|U=u, L_{k-m_l+1}=l_{k-m_l+1}, \dots, L_k=l_k]}{E[Y_k^{\overline{a}^0_k }|U=u, L_{k-m_l+1}=l_{k-m_l+1}, \dots, L_k=l_k]},
$$

Stationarity is assumed conditional on the extended window of confounders as well, however, the justification of the assumption remains the same: As $L_k$ usually encompasses some measurement of time, conditional on the calendar time, the expected outcome does not depend on the time index started since the beginning of the follow-up.

\begin{assumption}[General stationarity of potential outcomes]\label{ASS: gen stat all}
    $$
    P(Y_{m_l}^{\overline{a}^1_{m_l}}=y|U=u, L_{1:m_l}=\overline{l})=\dots = P(Y_t^{\overline{a}^1_t}=y|U=u, L_{(t-m_l+1):t}=\overline{l})
    $$
    and
    
    $$
    P(Y_{m_l}^{\overline{a}^0_{m_l}}=y|U=u, L_{1:m_l}=\overline{l})=\dots = P(Y_t^{\overline{a}^0_t}=y|U=u, L_{(t-m_l+1):t}=\overline{l})
    $$

    for any $y\in \{0, 1, 2,\dots\}$ for all $u\in \mathcal{U}$, $\overline{l}\in \mathcal{L}^{m_l}$, where $\overline{a}^1_k \in  \{\overline{a}_k: \sum_{i=k-m+1}^{k} a_i >0 \}$ and $\overline{a}^0_k \in \{\overline{a}_k: \sum_{i=k-m+1}^{k} a_i =0 \}$, respectively.
\end{assumption}

Assumption \ref{ASS: gen stat all} is imposed only for $k \in \mathcal T_{m_l}$. For $k < m_l$, the window $L_{(k-m_l+1):k}$ would include covariate values at times $k \le 0$, before follow-up begins, which are not recorded. However, Assumption \ref{ASS: gen stat all} can be further generalised, by either relying on the convention $L_{k}=0$ for all $k<0$, or by specifying the functional form through which the measured covariates express their effect on the treatment and the outcome, e.g. through the sum $\sum_{k'=k-m_l+1}^k L_{k'}$, and stating Assumption \ref{ASS: gen stat all} in terms of the chosen summary variable.

To identify the $ c\mathrm{-}RR(u, l_{(k-m_l+1:k)})_k$, the positivity assumption has to be adjusted as well: at least two time points need to exist within the history of the same individual, where in the last $m_l$ time points, the history of the covariate $L_k$ took the same values, $l_{(k-m_l+1):k}$, however, the treatment histories at these two timepoints, $k^1$ and $k^0$, define the susceptibility indicators to take the values $1$ and $0$, respectively.

\begin{assumption}[General time point positivity]\label{ASS: gen pos}
\leavevmode
\begin{enumerate}[label=(\roman*)]
    \item For a given $u\in \mathcal{U}$ and $l_{(k-m_l+1):k}\in\mathcal L^{m_l}$ ,
    \begin{align*}
        &\sum_{k\in \mathcal{T}_{m_l}} \mathbb I\left( V_k=1,U=u, L_{(k-m_l+1):k}=l_{(k-m_l+1):k}\right)>0&
        &\text{and}&\\
        &\sum_{k \in \mathcal T_{m_l}} \mathbb I\left(V_k=0,U=u,L_{(k-m_l+1):k}=l_{(k-m_l+1):k} \right)>0.&
    \end{align*}

    \item Given $U=u$ and $\overline{L}_k=\overline{l}_k$,
        $P\left(\overline A_k=\overline a_k\mid\overline L_k=\overline l_k,U=u
        \right)>0$
    for all treatment regimes $\overline a_k$ of interest.
\end{enumerate}
\end{assumption}

\begin{proposition}\label{PROP: c-RR gen}
    Under Assumptions \ref{ASS: exch}, \ref{ASS: ind out} and \ref{ASS: gen trans conf}-\ref{ASS: gen pos} the $c\text{-}RR(u,l_{(k-m_l+1):k})_k$ is identified as
    $$
    \frac{E[Y_{k^1}|U=u, \overline{A}_t=\overline{a}_t, \overline{L}_t=\overline{l}_t, S_t>0]}{E[Y_{k^0}|U=u, \overline{A}_t=\overline{a}_t, \overline{L}_t=\overline{l}_t, S_t>0]},
    $$
    for some time points $k^1, k^0 \in \mathcal T_{m_l}$, where
    $\overline{a}_t, \overline{l}_t$ are defined such that 
    $\sum_{k=k^1-m+1}^{k^1} a_k >0$, $\sum_{k=k^0-m+1}^{k^0} a_k =0 $, and $l_{(k^1-m_l+1):k^1}=l_{(k^0-m_l+1):k^0}=l_{(k-m_l+1):k}$.
    Moreover, the $c\text{-}RR(u,l_{(k-m_l+1):k})$ is equivalently identified by
    $$
    \frac{\sum_{k \in \mathcal{K}_{1,l_{(k-m_l+1):k}}} E[Y_{k}|U=u, \overline{A}_t=\overline{a}_t, \overline{L}_t=\overline{l}_t, S_t>0]/|\mathcal{K}_{1,l_{(k-m_l+1):k}}|}{\sum_{k \in \mathcal{K}_{0,l_{(k-m_l+1):k}}} E[Y_{k}|U=u, \overline{A}_t=\overline{a}_t, \overline{L}_t=\overline{l}_t, S_t>0]/|\mathcal{K}_{0,l_{(k-m_l+1):k}}|},
    $$
    where $\mathcal{K}_{1,l_{(k-m_l+1):k}}:=\{k' \in \mathcal{T}_{m_l}: \sum_{i=k'-m+1}^{k'} a_i >0, l_{(k'-m_l+1):k'}=l_{(k-m_l+1):k}\}$, $\mathcal{K}_{0,l_{(k-m_l+1):k}}:=\{k' \in \mathcal{T}_{m_l}: \sum_{i=k'-m+1}^{k'} a_i =0, l_{(k'-m_l+1):k'}=l_{(k-m_l+1):k}\}$, for any $u$ and $l_{(k-m_l+1):k}$ for which the ratio is well-defined.
\end{proposition}

See the proof in Appendix \ref{APP_sub: gen c-RR proof}.

Finally, to ensure that the estimator of the marginal quantity, the $CRR_m$ is unbiased, when the confounder $L_k$ acts over the multiple time points, the semi-invariance of the $c-RR(u, l_{(k-m_l+1):k}$ should hold.

\begin{assumption}[General semi-invariance]\label{ASS: gen stat U}
Suppose that the support of the unmeasured covariates can be divided into two disjoint subsets, $\mathcal{U}' \dot\cup \mathcal{I}_U=\mathcal{U}$, and that the set of measured covariate window histories can be partitioned as $\mathcal L^{m_l} = \mathcal L'_{m_l} \,\dot\cup\, \mathcal I_{L,m_l}$, such that for all $k \in \mathcal T_{m_l}$:
    $$\frac{E[Y_k^{\overline{a}^1_k}|U=u, L_{(k-m_l+1):k}=l_{(k-m_l+1):k}]}{E[Y_k^{\overline{a}^0_k}|U=u,  L_{(k-m_l+1):k}=l_{(k-m_l+1):k}]}= \frac{E[Y_k^{\overline{a}^1_k}|U=u', L_{(k-m_l+1):k}=l'_{(k-m_l+1):k}]}{E[Y_k^{\overline{a}^0_k}|U=u',L_{(k-m_l+1):k}=l'_{(k-m_l+1):k}]}$$
    for all $u, u' \in \mathcal{U'}$ and $l_{(k-m_l+1):k},l'_{(k-m_l+1):k} \in (\mathcal{L}')^{m_l}$, where $\overline{a}^1_k \in  \{\overline{a}_k: \sum_{i=k-m+1}^{k} a_i >0 \}$ and $\overline{a}^0_k \in \{\overline{a}_k: \sum_{i=k-m+1}^{k} a_i =0 \}$, respectively and
    \begin{align*}
        E[Y_k^{a_k} \mid U = u, L_{(k-m_l+1):k} = l] = 0
    \end{align*}
    for all $(u, l) \in (\mathcal U \times \mathcal L^{m_l}) \setminus (\mathcal U' \times \mathcal L'_{m_l})$, for any $\overline{a}_k\in \{0,1\}^{k}$
\end{assumption}

\begin{proposition}\label{PROP: gen CRR}
    Under Assumptions \ref{ASS: rare event}, \ref{ASS: exch}, \ref{ASS: ind out}, and \ref{ASS: gen trans conf}-\ref{ASS: gen stat U}
    $$
    \frac{\sum_{k \in \mathcal{K}_{1,l_{(k-m_l+1):k}}} E[Y_{k}|U=u, \overline{A}_t=\overline{a}_t, \overline{L}_t=\overline{l}_t, S_t=s_t]/|\mathcal{K}_{1,l_{(k-m_l+1):k}}|}{\sum_{k \in \mathcal{K}_{0,l_{(k-m_l+1):k}}} E[Y_{k}|U=u, \overline{A}_t=\overline{a}_t, \overline{L}_t=\overline{l}_t, S_t=s_t]/|\mathcal{K}_{0,l_{(k-m_l+1):k}}|} \rightarrow CRR_m
    $$
        where $\mathcal{K}_{1,l_{(k-m_l+1):k}}:=\{k: \sum_{i=k-m+1}^{k} a_i >0, L_{(k-m_l+1):k}=l_{(k-m_l+1):k}\}$, $\mathcal{K}_{0,l_{(k-m_l+1):k}}:=\{k: \sum_{i=k-m+1}^{k} a_i =0, L_{(k-m_l+1):k}=l_{(k-m_l+1):k}\}$, for any $u \in \mathcal{U}'$, $l_{(k-m_l+1):k} \in\mathcal{L}'_{m_l}$.

    Moreover, the $CRR_m$ can be identified in the limit by
    \begin{align*}
            \hspace{-2em}\sum_{u, l_{(k-m_l+1):k} \in \mathcal{U}', \mathcal{L}'_{m_l}} \phi(u, l_{(k-m_l+1):k})
            \cdot \frac{\sum_{k \in \mathcal{K}_{1,l_{(k-m_l+1):k}}} E[Y_{k}|U=u, \overline{A}_t=\overline{a}_t, \overline{L}_t=\overline{l}_t, S_t=s_t]/|\mathcal{K}_{1,l_{(k-m_l+1):k}}|}{\sum_{k \in \mathcal{K}_{0,l_{(k-m_l+1):k}}} E[Y_{k}|U=u, \overline{A}_t=\overline{a}_t, \overline{L}_t=\overline{l}_t, S_t=s_t]/|\mathcal{K}_{0,l_{(k-m_l+1):k}}|},
    \end{align*}
    where $\phi: \mathcal{U}'\times \mathcal{L}'_{m_l} \rightarrow [0,1]$ is a generic probability mass function.
\end{proposition}

See the proof in Appendix \ref{APP_sub: proof gen CRRm}.

\section{Multivalued $V$}\label{APP: multi v}
For simplicity, we have so far defined $V$ as a binary variable: an indicator of at least one treatment during the previous $m$ time points. The corresponding target estimand is therefore the causal effect of treatment versus no treatment. In practice, however, investigators are often interested in comparing one active intervention with another, such as two vaccines or a vaccine with a placebo. Let $A_k \in \{0,1,2\}$ denote treatment at time $k$, where $0$ corresponds to no treatment, $1$ to type 1 treatment (e.g., placebo or comparator), and $2$ to type 2 treatment (the treatment of interest). Suppose that the reactogenicity windows for treatments 1 and 2 are $m_1$ and $m_2$, respectively, and define the minimum of these two intervals as $m_{\min} := \min(m_1, m_2)$.

The target estimand is the rate ratio of adverse events under the increased susceptibility induced by the  two treatment levels, defined as

$$
CRR_{m_{\min}} := \frac{E[\sum_{k=1}^{m_{\min}} Y_k^{\overline{a}_{m_{\min}}=(2,0, \dots, 0)}]/m_{\min}}{E[\sum_{k=1}^{m_{\min}} Y_k^{\overline{a}_{m_{\min}}=(1,0, \dots, 0)}]/m_{\min}}.
$$

The $CRR_{m_{\min}}$ quantifies the relative difference in increased susceptibility across treatment levels during a period in which both treatments induce biological processes that increase susceptibility. Suppose that $m_1 < m_2$. Then, for time points $k=m_1+1,\dots,m_2$,
$
Y_k^{\overline{a}_{m_2}=(1,0,\dots,0)}
=
Y_k^{\overline{a}_{m_2}=(0,\dots,0)},
$
by the definition of the reactogenicity window. Hence, over the period $(m_1+1,\dots,m_2)$, the corresponding contrast has already been estimated in a trial comparing treatment 2 with no treatment.

To accommodate multiple versions of treatment, the indicator for being within the reactogenicity window must also be generalised.

$$
V_{k}=\begin{cases}
    0 &\text{if } \sum_{j=k-m_1+1}^{k} \mathbb{I}(A_j=1)=0 \text{ and } \sum_{j=k-m_2+1}^{k} \mathbb{I}(A_j=2)=0,\\
    1 &\text{if } \sum_{j=k-m_1+1}^{k} \mathbb{I}(A_j=1)>0 \text{ and } \sum_{j=k-m_2+1}^{k} \mathbb{I}(A_j=2)=0,\\
    2 &\text{if } \sum_{j=k-m_1+1}^{k} \mathbb{I}(A_j=1)=0 \text{ and } \sum_{j=k-m_2+1}^{k} \mathbb{I}(A_j=2)>0,\\
    3 &\text{if } \sum_{j=k-m_1+1}^{k} \mathbb{I}(A_j=1)>0 \text{ and } \sum_{j=k-m_2+1}^{k} \mathbb{I}(A_j=2)>0.\\
\end{cases}
$$

The default option implemented in the SCCS \texttt{R} package assumes that, when two different active ingredients are administered, their combined effect on the outcome is multiplicative, rather than estimating a separate interaction term, $V_k=3$ \citep{farrington_self-controlled_2018}. This no-interaction assumption can be encoded by introducing two susceptibility indicators, $V_{1,k}$ and $V_{2,k}$:

\begin{align*}
    &V_{1,k}=\begin{cases}
    1 &\text{if } \sum_{j=k-m_1+1}^{k} \mathbb{I}(A_j=1)>0,\\
    0 &\text{if } \sum_{j=k-m_1+1}^{k} \mathbb{I}(A_j=1)=0,
\end{cases}&
&V_{2,k}=\begin{cases}
    1 &\text{if } \sum_{j=k-m_2+1}^{k} \mathbb{I}(A_j=2)>0,\\
    0 &\text{if } \sum_{j=k-m_2+1}^{k} \mathbb{I}(A_j=2)=0,
\end{cases}
\end{align*}

By explicitly assuming no interaction between the two susceptibility terms, that is, no interactions between the treatments, they can be included as separate terms in the log-Poisson model, yielding multiplicative effects.

Alternatively, if an individual receives both treatments within their respective reactogenicity windows, $V_k$ can be defined according to the most recent treatment received. Let $T_{1,k}$ and $T_{2,k}$ denote the most recent times before or at $k$ at which treatments 1 and 2 were received within their respective reactogenicity windows:
\begin{align*}
    &T_{1,k}=\max\{j\leq k : A_j=1, j\geq k-m_1+1\},&  &T_{2,k}=\max\{j\leq k : A_j=2, j\geq k-m_2+1\}.&
\end{align*}
We set $T_{a,k}$ to $-\infty$ whenever treatment $a$ was not received during its corresponding reactogenicity window. Then
$$
V_{k}=\begin{cases}
    0 &\text{if } T_{1,k}=T_{2,k}=-\infty,\\
    1 &\text{if } T_{1,k}>T_{2,k}\\
    2 &\text{if } T_{1,k}<T_{2,k}.
\end{cases}
$$

Since only one treatment can be received at any time $k$, the case $T_{1,k}=T_{2,k}>-\infty$ cannot occur, so $V_k$ is well-defined.

Whenever the reactogenicity windows are not equal, a treatment received more recently need not determine $V_k$ if it lies outside its own reactogenicity window with respect to the time point $k$. To make things concrete, suppose $m_2=m_1+10$, and an individual received treatments 1 and 2 at times $k-m_1-2$ and $k-m_1-5$, respectively.  Treatment 1 is outside its reactogenicity window at time $k$, thus $T_{1,k}=-\infty$, whereas treatment 2 remains inside the reactogenicity window $m_2$, so $T_{2,k}=k-m_2+5$. Therefore, $V_k=2$, even though treatment $2$ occurred prior to treatment $1$.

Finally, by recoding $V_k$, the model can incorporate different levels of susceptibility induced by repeated treatment within the reactogenicity window $m$. For simplicity, consider again a single treatment where $A_k=1$ denotes treatment at time $k$ and $A_k=0$ denotes no treatment at time $k$. Define 
$$
V_{k}=\sum_{j=k-m+1}^k A_j.
$$
That is, the susceptibility induced by single versus repeated treatment is being distinguished, but we do not specify a functional form relating these effects, for example, a multiplicative model. The multiplicative effect of repeated treatments can instead be encoded by introducing separate variables for the first, second, third, and subsequent treatments within the reactogenicity window:
$$V_k^{i}=\begin{cases}
    1 \text{ if } \sum_{j=k-m+1}^k A_j\geq i, \\
    0 \text{ if } \sum_{j=k-m+1}^k A_j< i.
\end{cases}
$$
Then, under a no-interaction assumption between the repeated treatments, the multiplicative effect follows.

Since none of these representations changes the structure of the DAG, except by including multiple $V_k$ variables at each time point, the graphical assumptions require no conceptual modification, they need only be restated with the appropriate notation. Similarly, the identification assumptions require only recoding. In particular, stationarity and positivity must hold for the two sets of interventions $\overline{a}_k^{v}$, and $\overline{a}_k^{v'}$, that induce $V_k=v$ and  $V_k=v'$ for the contrast of interest. Assumption \ref{ASS: Stat U}, namely semi-invariance, must hold for the corresponding ratio being identified.

\section{Proofs}\label{APP: PROOFS}
\subsection{Proof of Proposition \ref{PROP: c-RR(u,l)}}\label{APP_sub: proof c-rr(u,l)}
\begin{proof}
    \begin{align}
        &E[Y_k^{\bar a_k}\mid U=u,L_k=l_k]\notag\\
        &\overset{LOTP}{=}\sum_{\bar l_{k-1}} E[Y_k^{\bar a_k}\mid U=u,\bar L_k=(\bar l_{k-1},l_k)]\;P(\bar L_{k-1}=\bar l_{k-1}\mid U=u,L_k=l_k)\notag\\
        &\overset{\mathrm{Assumption}\; \ref{ASS: exch}}{=}\sum_{\bar l_{k-1}} E[Y_k^{\bar a_k}\mid \bar A_k=\bar a_k,U=u,\bar L_k=(\bar l_{k-1},l_k)]\;P(\bar L_{k-1}=\bar l_{k-1}\mid U=u,L_k=l_k)\notag\\
        &\overset{\mathrm{Consistency}}{=}\sum_{\bar l_{k-1}} E[Y_k\mid \bar A_k=\bar a_k,U=u,\bar L_k=(\bar l_{k-1},l_k)]\;P(\bar L_{k-1}=\bar l_{k-1}\mid U=u,L_k=l_k)\notag\\
        &\overset{\mathrm{Assumption}\; \ref{ASS: trans conf}}{=} E[Y_k\mid \bar A_k=\bar a_k,U=u,L_k=l_k]\notag\\
        &\overset{\mathrm{Assumption}\; \ref{ASS: trans conf}}{=} E[Y_k\mid \bar A_k=\bar a_k,U=u,\bar L_k=\bar l_k]\notag\\
        &\overset{\mathrm{Assumption}\; \ref{ASS: ind out}(ii)}{=} E[Y_k\mid \bar A_t=\bar a_t,U=u,\bar L_k=\bar l_k]\notag\\
        &\overset{\mathrm{Assumption}\; \ref{ASS: ind out}(iii)}{=} E[Y_k\mid \bar A_t=\bar a_t,U=u,\bar L_t=\bar l_t]\notag\\
        &\overset{\mathrm{LOTP}}{=}E[Y_k|\overline{A}_t=\overline{a}_t, U=u, \overline{L}_t=\overline{l}_t, S_t>0]P(S_t>0|\overline{A}_t=\overline{a}_t, U=u, \overline{L}_t=\overline{l}_t)\notag\\
        &+E[Y_k|\overline{A}_t=\overline{a}_t, U=u, \overline{L}_t=\overline{l}_t, S_t=0]P(S_t=0|\overline{A}_t=\overline{a}_t, U=u, \overline{L}_t=\overline{l}_t)\notag\\
        &=E[Y_k|\overline{A}_t=\overline{a}_t, U=u, \overline{L}_t=\overline{l}_t, S_t>0]P(S_t>0|\overline{A}_t=\overline{a}_t, U=u, \overline{L}_t=\overline{l}_t)\label{EQ: Case prob}.
    \end{align}
    
    Thus
    $$
    E[Y_{k^1}^{\overline{a}_{k^1}}|U=u, L_{k^1}=l] = E[Y_{k^1}|\overline{A}_t=\overline{a}_t, U=u, \overline{L}_t=\overline{l}_t, S_t>0]P(S_t>0|\overline{A}_t=\overline{a}_t, U=u, \overline{L}_t=\overline{l}_t),
    $$
    where $\overline{a}_t=(\overline{a}_{k^1},  A_{k^1+1}=a_{k^1+1}, \dots, A_t=a_t)$, such that $\overline{a}_{k^1} \in \{\overline{a}_{k^1}': \sum_{k=k^1-m+1}^{k^1} a'_k >0 \}$ and $\overline{l}_t=(l_1, \dots, l_{k^1}=l, \dots, l_t)$ and similarly
    $$
    E[Y_{k^0}^{\overline{a}_{k^0}}|U=u, L_{k^0}=l] = E[Y_{k^0}|\overline{A}_t=\overline{\dot{a}}_t, U=u, \overline{L}_t=\overline{\dot{l}}_t, S_t>0]P(S_t>0|\overline{A}_t=\overline{\dot{a}}_t, U=u, \overline{L}_t=\overline{\dot{l}}_t),
    $$
    where $\overline{\dot{a}}_t=(\overline{\dot{a}}_{k^0}, A_{k^0+1}=\dot{a}_{k^0+1}, \dots, A_t=\dot{a}_t)$, $\overline{\dot{l}}_t=(\dot{l}_1, \dots, \dot{l}_{k^0}=l, \dot{l}_{k^0+1}, \dots, \dot{l}_t)$.

    By Assumption \ref{ASS: stat all} 
    $$
    \frac{E[Y_k^{\overline{a}_k^1}|U=u, L_k=l]}{E[Y_k^{\overline{a}_k^0}|U=u, L_k=l]}=
    \frac{E[Y_{k^1}^{\overline{a}_{k^1}}|U=u, L_{k^1}=l]}{E[Y_{k^0}^{\overline{a}_{k^0}}|U=u, L_{k^0}=l]},
    $$
    therefore, by the results above,
    \begin{align*}
       &\frac{E[Y_k^{\overline{a}_k^1}|U=u, L_k=l]}{E[Y_k^{\overline{a}_k^0}|U=u, L_k=l]} \\
       &=\frac{E[Y_{k^1}|\overline{A}_t=\overline{a}_t, U=u, \overline{L}_t=\overline{l}_t, S_t>0]P(S_t>0|\overline{A}_t=\overline{a}_t, U=u, \overline{L}_t=\overline{l}_t)}{E[Y_{k^0}|\overline{A}_t=\overline{\dot{a}}_t, U=u, \overline{L}_t=\overline{\dot{l}}_t, S_t>0]P(S_t>0|\overline{A}_t=\overline{\dot{a}}_t, U=u, \overline{L}_t=\overline{\dot{l}}_t)}.
    \end{align*}
    If $\overline{\dot{l}}_t=\overline{l}_t$ and $\overline{\dot{a}}_t=\overline{a}_t$, then the conditional probabilities of being a case cancel out in the expression. Due to the specific interventions and the conditioning set, the restrictions  $\overline{a}_{k^1} \in  \{\overline{a}_{k^1}': \sum_{k=k^1-m+1}^{k^1} a'_k >0 \}$, $\overline{a}_{k^0}' \in  \{\overline{a}_{k^0}: \sum_{k=k^0-m+1}^{k^0} a'_k =0 \}$, and $l_{k^1}=l_{k^0}=l$ must hold, but this is possible when 

    \begin{align*}
        \overline{\dot{l}}_t=\overline{l}_t&=(l_1, \dots, l_{k^0}=l , \dots,  l_{k^1}=l, \dots, l_t),\\
        \overline{\dot{a}}_t=\overline{a}_t&=(a_1, \dots, a_{k_0-m}, a_{k_0-m+1}=0 , \dots a_{k_0}=0, a_{k_0+1}, \dots,  a_{k_1-m+1}, \dots , a_{k_1}, a_{k_1+1}, \dots, a_t),
    \end{align*}
    such that $\sum_{k=k^1-m+1}^{k^1} a_k >0$.

    Note that in the example above $k_0+m<k_1$, but $k_1 \in \{k_0+1, \dots k_0+m-1\}$ or $k_1<k_0$ could hold as well.

    Since $P(S_t>0|\overline{A}_t=\overline{a}_t, U=u, \overline{L}_t=\overline{l}_t)$ is invariant in the time index $k$
    and by Assumption \ref{ASS: stat all} $E[Y_k^{\overline{a}_k}|U=u, L_k=l]$ is also invariant in $k$, it follows from Equation \eqref{EQ: Case prob}, that for a fixed $U$, for any $k \in \mathcal{K}_{a,l}$ the expectation $E[Y_k=y| \overline{A}_t=\overline{a}_t, U=u, \overline{L}_t=\overline{l}_t]$, is invariant in the time index $k$. Therefore,
    $$
    E[Y_{k^1}|\overline{A}_t=\overline{a}_t, U=u, \overline{L}_t=\overline{l}_t]=E[Y_{k}|\overline{A}_t=\overline{a}_t, U=u, \overline{L}_t=\overline{l}_t],
    $$
    for any $k \in \mathcal{K}_{1,l}$, and 
    $$
    E[Y_{k^0}|\overline{A}_t=\overline{a}_t, U=u, \overline{L}_t=\overline{l}_t]=E[Y_{k}|\overline{A}_t=\overline{a}_t, U=u, \overline{L}_t=\overline{l}_t],
    $$
    for any $k \in \mathcal{K}_{0,l}$, which concludes the second half of our proof.
\end{proof}

\subsection{Proof of Proposition \ref{PROP: c-RR(u,l) St=st}}\label{APP_sub: St=st proof}
\begin{proof}
Identical to the proof of Proposition \ref{PROP: c-RR(u,l)} we can show
$$
 E[Y_k^{\overline{a}_k}|U=u, L_k=l]=E[Y_k|\overline{A}_t=\overline{a}_t, U=u, \overline{L}_t=\overline{l}_t].
$$

    In the following, we will show that $E[Y_k|S_t=s, \overline{H}_t=\overline{h}_t]\rightarrow E[Y_k|\overline{H}_t=\overline{h}]\frac{s}{\lambda}$ where $\sum_{k \in  \{1, \dots , t\}} E[Y_k|\overline{H}_t=\overline{h}_t] \rightarrow \lambda(\overline{h_t})$ as $t \rightarrow \infty$.

    Consider the probability generating function (PGF) of any $Y_k$ conditional on the history $\overline{h}_t$. For any $|z|\leq 1$ it is equal to 
    \begin{align*}
        E[z^{Y_k}|\overline{H}_t&=\overline{h_t}]=\sum_y P(Y_k=y|\overline{H}_t=\overline{h}_t)z^y\\
        &=P(Y_k=0|\overline{H}_t=\overline{h}_t)+zP(Y_k=1|\overline{H}_t=\overline{h}_t)+\sum_{y>1}P(Y_k=y|\overline{H}_t=\overline{h}_t)z^y\\
        &=1-P(Y_k>0|\overline{H}_t=\overline{h}_t)+zP(Y_k=1|\overline{H}_t=\overline{h}_t)+\sum_{y>1}P(Y_k=y|\overline{H}_t=\overline{h}_t)z^y\\
        &=1+(z-1)P(Y_k=1|\overline{H}_t=\overline{h}_t)+\sum_{y>1}P(Y_k=y|\overline{H}_t=\overline{h}_t)(z^y-1)\\
        &=1+(z-1)E(Y_k|\overline{H}_t=\overline{h}_t)+\sum_{y>1}P(Y_k=y|\overline{H}_t=\overline{h}_t)(z^y-1-y(z-1)).
    \end{align*}

    By the triangle inequality and by the fact that $|z|\leq 1$ we have 
    $$
    |z^y-1-y(z-1)|\leq |z^y-1|+y|z-1|\leq 2+2y,
    $$
    which is less than $3y$ for any $y>1$.

    Therefore, the PGF of $Y_k$ conditional on $\overline{H}_t=\overline{h}_t$ is equal to 
    $$
    1+(z-1)E(Y_k|\overline{H}_t=\overline{h}_t)+r(z),
    $$
    where $r(z)$ is a remainder term bounded by $|r(z)|\leq 3  E[Y_k \mathbb{I}(Y_k>1)| \overline{H}_t=\overline{h}_t]$ for any $|z|\leq 1$

    The PGF of $S_t=\sum_{k=1}^t Y_k$, which are independent conditional on the full history $\overline{H}_t=\overline{h}_t$, by Assumption \ref{ASS: ind out} (i), can be written as 
    $$
    E[z^{S_t}]=\prod_{k=1}^t E[z^{Y_k}]= \prod_{k=1}^t\big(  1+(z-1)E(Y_k|\overline{H}_t=\overline{h}_t)+r(z)\big).
    $$
    By taking the logarithm, we have
    $$
    \sum_{k=1}^t\mathrm{log}\big(  1+(z-1)E(Y_k|\overline{H}_t=\overline{h}_t)+r(z)\big)
    $$

    Note that $|r(z)|\leq E[Y_k \mathbb{I}(Y_k>1)| \overline{H}_t=\overline{h}_t]\leq E[Y_k| \overline{H}_t=\overline{h}_t]$,
    thus by Assumption \ref{ASS: rare event} (i), the perturbation goes to $0$, which allows us to rewrite it as 
    $$
    \mathrm{log}\Big( E\big[z^{S_t}\big] \Big)= \sum_{k=1}^t(z-1)E(Y_k|\overline{H}_t=\overline{h}_t)+ \sum_{k=1}^tr(z)+O\Big(\sum_{k=1}^t \big( (z-1)E(Y_k|\overline{H}_t=\overline{h}_t)+r(z)\big)^2 \Big).
    $$
    Since $|z|\leq 1$ and $|r(z)|\leq 3 E(Y_k|\overline{H}_t=\overline{h}_t)$, it follows that 
    $$\sum_{k=1}^t \big( (z-1)E(Y_k|\overline{H}_t=\overline{h}_t)+r(z)\big)^2 \leq  \sum_{k=1}^t 25 E(Y_k|\overline{H}_t=\overline{h}_t)^2.$$
     As Assumptions \ref{ASS: rare event} (i) and (ii) imply $\sum_{k \in  \{1, \dots , t\}} E[Y_k|\overline{H}_t=\overline{h}_t]^2 \rightarrow 0$, the error term $O\Big(\sum_{k=1}^t \big( (z-1)E(Y_k|\overline{H}_t=\overline{h}_t)+r(z)\big)^2 \Big)$ goes to $0$.

    Now, we need to show that not only the expectation $E[Y_k \mathbb{I}(Y_k>1)| \overline{H}_t=\overline{h}_t]$ is negligible, but also the sum of these expectations across all $k= 1, \dots , t$.

    \begin{align}
        E[Y_k \mathbb{I}(Y_k>1)| \overline{H}_t=\overline{h}_t] &\leq 2  E[(Y_k-1) \mathbb{I}(Y_k>1)| \overline{H}_t=\overline{h}_t]\notag \\
        &\leq 2 ( E[Y_k| \overline{H}_t=\overline{h}_t] -P(Y_k>0|\overline{H}_t=\overline{h}_t))\notag\\
        &=  2 ( E[Y_k|Y_k>0,  \overline{H}_t=\overline{h}_t] -1)P(Y_k>0|\overline{H}_t=\overline{h}_t)\notag \\
        &\leq 2 P(Y_k>0|\overline{H}_t=\overline{h}_t) \sup_k(  E[Y_k|Y_k>0,  \overline{H}_t=\overline{h}_t] -1)\notag\\
        &\leq 2 E(Y_k|\overline{H}_t=\overline{h}_t) \sup_k(  E[Y_k|Y_k>0,  \overline{H}_t=\overline{h}_t] -1)\label{EQ: greater 2 order}
    \end{align}
Summing over $k$ and taking the limit $t \rightarrow \infty$ gives 
$$
\sum_k E[Y_k \mathbb{I}(Y_k>1)| \overline{H}_t=\overline{h}_t] \leq 2 \lambda(\overline{h}_t) \sup_k(  E[Y_k|Y_k>0,  \overline{H}_t=\overline{h}_t] -1),
$$
by Assumption \ref{ASS: rare event} (ii) and by Assumption \ref{ASS: rare event} (iii), the right hand side goes to $0$.

Therefore the sum
$$
\sum_{k=1}^tr(z) \leq 3 \sum_{k=1}^t E[Y_k \mathbb{I}(Y_k>1)| \overline{H}_t=\overline{h}_t]
$$

converges to $0$.

Finally, by Assumption \ref{ASS: rare event} (ii) 
$$
 \sum_{k=1}^t(z-1)E(Y_k|\overline{H}_t=\overline{h}_t) \rightarrow (z-1) \lambda(\overline{h}_t).
$$

Therefore, the PGF of $S_t$ equals to
$$
\mathrm{exp}((z-1) \lambda(\overline{h}_t))+o(1),
$$
which is the Poisson PGF with vanishing error term.

By the definition of conditional probability, and from the fact that conditional on $\overline{H}_t=\overline{h}_t$, the $Y_k$-s are independent, we can rewrite 
$$
E[Y_k|S_t=s, \overline{H}_t=\overline{h}_t]=\frac{\sum_{y>0}y P(Y_k=y|\overline{H}_t=\overline{h}_t)P(S_{t,-k}=s-y|\overline{H}_t=\overline{h}_t)}{P(S_t=s|\overline{H}_t=\overline{h}_t)},
$$
where $S_{t, -k}=\sum_{k'\neq k} Y_{k'}$.

Separate the numerator to 
$$
P(Y_k=1|\overline{H}_t=\overline{h}_t)P(S_{t,-k}=s-1|\overline{H}_t=\overline{h}_t)+\sum_{y>1}y P(Y_k=y|\overline{H}_t=\overline{h}_t)P(S_{t,-k}=s-
y|\overline{H}_t=\overline{h}_t).
$$
The second part of the sum is upper bounded by 
\begin{align*}
    \sum_{y>1}y P(Y_k=y|\overline{H}_t=\overline{h}_t)P(S_{t,-k}=s-y|\overline{H}_t=\overline{h}_t)&\leq \sum_{y>1}y P(Y_k=y|\overline{H}_t=\overline{h}_t)\\
    &= E[Y_k \mathbb{I}(Y_k>1)|\overline{H}_t=\overline{h}_t],
\end{align*}
which is, by Assumption \ref{ASS: rare event} (iii) and Equation \eqref{EQ: greater 2 order} of order $o(E[Y_k |\overline{H}_t=\overline{h}_t])$.

For $y=1$, as
$$
E[Y_k |\overline{H}_t=\overline{h}_t] = P(Y_k=1|\overline{H}_t=\overline{h}_t) + \sum_{y>1}y P(Y_k=y|\overline{H}_t=\overline{h}_t),
$$
and we just showed that the latter term is of order $o(E[Y_k |\overline{H}_t=\overline{h}_t])$, it follows that 
$$
P(Y_k=1|\overline{H}_t=\overline{h}_t)=E[Y_k |\overline{H}_t=\overline{h}_t](1+o(1)).
$$
By identical reasoning as for $S_t$, the distribution of the $S_{t, -k}$ can be approximated by a Poisson with a vanishing error term, thus
\begin{align*}
    P(S_{t, -k}=s-1)&= \mathrm{exp}(-(\lambda(\overline{h}_t)-E[Y_k|\overline{H}_t=\overline{h}_t]))\frac{(\lambda(\overline{h}_t)-E[Y_k|\overline{H}_t=\overline{h}_t])^{s-1}}{(s-1)!}\\
    &=\mathrm{exp}(-\lambda(\overline{h}_t))\frac{(\lambda(\overline{h}_t))^{s-1}}{(s-1)!}(1+o(1)),
\end{align*}
therefore, the numerator is equal to 
$$
E[Y_k |\overline{H}_t=\overline{h}_t]\mathrm{exp}(-\lambda(\overline{h}_t))\frac{(\lambda(\overline{h}_t))^{s-1}}{(s-1)!}(1+o(1))
$$

By similar reasoning as for $S_{t, -k}$, 
$$
P(S_{t}=s)=\mathrm{exp}(-\lambda(\overline{h}_t))\frac{(\lambda(\overline{h}_t))^{s}}{s!}(1+o(1)),
$$
thus 
\begin{align*}
    E[Y_k|S_t=s, \overline{H}_t=\overline{h}_t]&= \frac{E[Y_k |\overline{H}_t=\overline{h}_t]\mathrm{exp}(-\lambda(\overline{h}_t))\frac{(\lambda(\overline{h}_t))^{s-1}}{(s-1)!}(1+o(1))}{\mathrm{exp}(-\lambda(\overline{h}_t))\frac{(\lambda(\overline{h}_t))^{s}}{s!}(1+o(1))}\\
    &= E[Y_k |\overline{H}_t=\overline{h}_t]\frac{s}{\lambda(\overline{h}_t)}(1+o(1)).
\end{align*}

Therefore, as $t \rightarrow \infty$, 
$$
 E[Y_k|S_t=s, \overline{H}_t=\overline{h}_t] \rightarrow E[Y_k |\overline{H}_t=\overline{h}_t]\frac{s}{\lambda(\overline{h}_t)}
$$

For the same fixed history $\overline{h}_t$ and a fixed $s$ both in the numerator and the denominator
\begin{align*}
    \frac{E[Y_{k^1}|\overline{H}_t=\overline{h}_t, S_t=s]}{E[Y_{k^0}|\overline{H}_t=\overline{h}_t, S_t=s]} &\rightarrow \frac{E[Y_{k^1} |\overline{H}_t=\overline{h}_t]\frac{s}{\lambda(\overline{h}_t)}}{E[Y_{k^0} |\overline{H}_t=\overline{h}_t]\frac{s}{\lambda(\overline{h}_t)}}\\
    &= \frac{E[Y_{k^1} |\overline{H}_t=\overline{h}_t]}{E[Y_{k^0} |\overline{H}_t=\overline{h}_t]},
\end{align*}
which, combined with Equation \eqref{EQ: Case prob}, concludes the first half of the proof.

    By Assumption \ref{ASS: stat all} and Equation \eqref{EQ: Case prob}, it follows that for a fixed $U$, for any $k \in \mathcal{K}_{a,l}$ the expectation $P[Y_k=y|\overline{H}_t=\overline{h}_t]$ is invariant in the time index $k$. 
    Thus
    
    \begin{align*}
        \frac{\sum_{k \in \mathcal{K}_{1,l}} E[Y_{k}|\overline{H}_t=\overline{h}_t, S_t=s]/|\mathcal{K}_{1,l}|}{\sum_{k \in \mathcal{K}_{0,l}} E[Y_{k}|\overline{H}_t=\overline{h}_t, S_t=s]/|\mathcal{K}_{0,l}|}&\rightarrow \frac{\sum_{k \in \mathcal{K}_{1,l}} E[Y_{k}|\overline{H}_t=\overline{h}_t]\frac{\lambda(\overline{h}_t)}{s}/|\mathcal{K}_{1,l}|}{\sum_{k \in \mathcal{K}_{0,l}} E[Y_{k}|\overline{H}_t=\overline{h}_t]\frac{\lambda(\overline{h}_t)}{s}/|\mathcal{K}_{0,l}|}\\
        &=\frac{\sum_{k \in \mathcal{K}_{1,l}} E[Y_{k}|\overline{H}_t=\overline{h}_t]/|\mathcal{K}_{1,l}|}{\sum_{k \in \mathcal{K}_{0,l}} E[Y_{k}|\overline{H}_t=\overline{h}_t]/|\mathcal{K}_{0,l}|} \\
        &=\frac{ E[Y_{k^1}|\overline{H}_t=\overline{h}_t]}{E[Y_{k^0}|\overline{H}_t=\overline{h}_t]},
    \end{align*}
    which concludes the proof.
\end{proof}

\subsection{Proof of Proposition \ref{PROP: CRR}}\label{APP_sub: proof CRR}
\begin{proof}
By Assumption \ref{ASS: Stat U} and the law of total probability
\begin{align*}
    E[Y_k^{\overline{a}_k}]&=\sum_{u, l \in \mathcal{U}', \mathcal{L}'} E[Y_k^{\overline{a}_k}|U=u, L_k=l]P(U=u, L_k=l)\\
    &+\sum_{u, l \in \mathcal{I}_U, \mathcal{L}} E[Y_k^{\overline{a}_k}|U=u, L_k=l]P(U=u, L_k=l)\\
    &+\sum_{u, l \in \mathcal{U}, \mathcal{I}_l} E[Y_k^{\overline{a}_k}|U=u, L_k=l]P(U=u, L_k=l)\\
    &+\sum_{u, l \in \mathcal{I}_U, \mathcal{I}_l} E[Y_k^{\overline{a}_k}|U=u, L_k=l]P(U=u, L_k=l)\\
    &= \sum_{u, l \in \mathcal{U}', \mathcal{L}'} E[Y_k^{\overline{a}_k}|U=u, L_k=l]P(U=u, L_k=l).
\end{align*}
Therefore
    \begin{align*}
        \frac{\sum_{k=1}^{m}E[Y_k^{\overline{a}_m=(1, 0, \dots, 0)}]}{\sum_{k=1}^{m}E[Y_k^{\overline{a}_m=( 0, \dots, 0)}]}&\overset{LOTP}{=}\frac{\sum_{k=1}^{m}\sum_{u, l \in \mathcal{U}', \mathcal{L}'} E[Y_k^{\overline{a}_m=(1, 0, \dots, 0)}|U=u, L_k=l]P(U=u, L_k=l)}{\sum_{k=1}^{m}\sum_{u, l \in \mathcal{U}', \mathcal{L}'} E[Y_k^{\overline{a}_m=(0, \dots, 0)}|U=u, L_k=l]P(U=u, L_k=l)}.
    \end{align*}

    By Assumption \ref{ASS: stat all} and from the fact that the reactogenicity window is equal to $m$, $E[Y_k^{\overline{a}_m=(0, \dots, 0)}|U=u, L_k=l]=E[Y_1^{\overline{a}_m=(0, \dots, 0)}|U=u, L_1=l]$ and $E[Y_k^{\overline{a}_m=(1, 0, \dots, 0)}|U=u, L_k=l]=E[Y_1^{\overline{a}_m=(1, 0, \dots, 0)}|U=u, L_1=l]$, thus
    \begin{align*}
        &\frac{\sum_{k=1}^{m}\sum_{u, l \in \mathcal{U}', \mathcal{L}'} E[Y_k^{\overline{a}_m=(1, 0, \dots, 0)}|U=u, L_k=l]P(U=u, L_k=l)}{\sum_{k=1}^{m}\sum_{u, l \in \mathcal{U}', \mathcal{L}'} E[Y_k^{\overline{a}_m=(0, \dots, 0)}|U=u, L_k=l]P(U=u, L_k=l)}\\
        &=\frac{\sum_{u, l \in \mathcal{U}', \mathcal{L}'} E[Y_1^{\overline{a}_m=(1, 0, \dots, 0)}|U=u, L_1=l]\sum_{k=1}^{m}P(U=u, L_k=l)}{\sum_{u, l \in \mathcal{U}', \mathcal{L}'} E[Y_1^{\overline{a}_m=(0, \dots, 0)}|U=u, L_1=l]\sum_{k=1}^{m}P(U=u, L_k=l)}.
    \end{align*}

    Using Assumption \ref{ASS: Stat U} $E[Y_1^{\overline{a}_m=(1, 0, \dots, 0)}|U=u, L_1=l] = \alpha E[Y_1^{\overline{a}_m=(0, \dots, 0)}|U=u, L_1=l]$, for some $\alpha \in [0, \infty)$, that is invariant in $u$ and $l$, thus
    \begin{align*}
        &\frac{\sum_{u, l \in \mathcal{U}', \mathcal{L}'} E[Y_1^{\overline{a}_m=(1, 0, \dots, 0)}|U=u, L_1=l]\sum_{k=1}^{m}P(U=u, L_k=l)}{\sum_{u, l \in \mathcal{U}', \mathcal{L}'} E[Y_1^{\overline{a}_m=(0, \dots, 0)}|U=u, L_1=l]\sum_{k=1}^{m}P(U=u, L_k=l)}\\
        &=\frac{\sum_{u, l \in \mathcal{U}', \mathcal{L}'}  \alpha E[Y_1^{\overline{a}_m=(0, \dots, 0)}|U=u, L_1=l]\sum_{k=1}^{m}P(U=u, L_k=l)}{\sum_{u, l \in \mathcal{U}', \mathcal{L}'} E[Y_1^{\overline{a}_m=(0, \dots, 0)}|U=u, L_1=l]\sum_{k=1}^{m}P(U=u, L_k=l)}\\
        &= \alpha \frac{\sum_{u, l \in \mathcal{U}', \mathcal{L}'} E[Y_1^{\overline{a}_m=(0, \dots, 0)}|U=u, L_1=l]\sum_{k=1}^{m}P(U=u, L_k=l)}{\sum_{u, l \in \mathcal{U}', \mathcal{L}'} E[Y_1^{\overline{a}_m=(0, \dots, 0)}|U=u, L_1=l]\sum_{k=1}^{m}P(U=u, L_k=l)}\\
        &=\alpha = \frac{E[Y_1^{\overline{a}_m=(1, 0, \dots, 0)}|U=u, L_1=l]}{E[Y_1^{\overline{a}_m=(0, \dots, 0)}|U=u, L_1=l]},
    \end{align*}
    for any $u \in \mathcal{U}'$ and $l \in \mathcal{L}'$. 

    By Assumption \ref{ASS: stat all}
    $$
    \alpha :=  \frac{E[Y_1^{\overline{a}_m=(1, 0, \dots, 0)}|U=u, L_1=l]}{E[Y_1^{\overline{a}_m=(0, \dots, 0)}|U=u, L_1=l]}= \frac{E[Y_k^{\overline{a}^1_k }|U=u, L_k=l]}{E[Y_k^{\overline{a}^0_k }|U=u, L_k=l]}, 
    $$
    where $\overline{a}^1_k \in  \{\overline{a}_k: \sum_{i=k-m+1}^{k} a_i >0 \}$ and $\overline{a}^0_k \in \{\overline{a}_k: \sum_{i=k-m+1}^{k} a_i =0 \}$, respectively. By Proposition \ref{PROP: c-RR(u,l) St=st}, the proof of the first half of the result is concluded.

    In the first part of the proof, we have shown that the $CRR_m$ is equal to $\alpha$, therefore for any $\phi: \mathcal{U}'\times \mathcal{L}' \rightarrow [0,1]$
    \begin{align*}
        CRR_m&=\alpha\\
        &=\frac{E[Y_{k}^{\overline{a}_k^1}|U=u, L_k=l]}{E[Y_{k}^{\overline{a}_k^0}|U=u, L_k=l]}\\
        &\overset{LOTP}{=}\alpha \sum_{u,l \in \mathcal{U}', \mathcal{L'}} \phi(u,l)\\
        &=\sum_{u,l \in \mathcal{U}', \mathcal{L'}} \phi(u,l) \alpha\\
        &=\sum_{u,l \in \mathcal{U}', \mathcal{L'}} \phi(u,l) \cdot \frac{\sum_{k \in \mathcal{K}_{1,l}} E[Y_k|\overline{A}_t=\overline{a}_t, U=u, \overline{L}_t=\overline{l}_t, S_t=s_t ]/|\mathcal{K}_{1,l}|}{\sum_{k \in \mathcal{K}_{0,l}}E[Y_k|\overline{A}_t=\overline{a}_t, U=u, \overline{L}_t=\overline{l}_t, S_t=s_t ])/|\mathcal{K}_{0,l}|},
    \end{align*}
    where in the last line we used the result of the first half of the proof.
\end{proof}

\subsection{Proof of Proposition \ref{PROP: test hyper}}\label{APP: proof hyper}
\begin{proof}
    As in Algorithm \ref{alg: hyper}, we index the observed cases by $i=1, \dots, n$. For individual $i$ and stratum $l$ define
    \begin{align*}
        &\mathcal{K}_{1,il}:=\{k: L_{ik}=l, V_{ik}=1\}, \qquad \mathcal{K}_{0,il}:=\{k: L_{ik}=l, V_{ik}=0\},\\
        &\mathcal{K}_{il}:=\mathcal{K}_{1,il}\cup \mathcal{K}_{0,il},
    \end{align*}
    so that $|\mathcal{K}_{il}|=|\mathcal{K}_{1,il}|+|\mathcal{K}_{0,il}|$, and let
    $$
    S_{\mathcal{K}_{il}}:=\sum_{k\in \mathcal{K}_{il}} Y_{ik}, \qquad S_{\mathcal{K}_{1,il}}:=\sum_{k\in \mathcal{K}_{1,il}} Y_{ik}
    $$
    denote the total number of events, and the number of events in at-risk intervals, of individual $i$ in stratum $l$. The informative strata of individual $i$ are $\mathcal{L}_i^+:=\{l: |\mathcal{K}_{1,il}|>0,|\mathcal{K}_{0,il}|>0\}$.

    Under Assumption \ref{ASS: bin}, $Y_{ik} \in \{0,1\}$ for all $i$ and $k$. Conditional on $(U_i, \overline{L}_{it}, \overline{A}_{it})$, the intervals $\mathcal{K}_{1, il}$ and $\mathcal{K}_{0, il}$ are fixed for all $l$.

    Under Hypothesis \ref{HYP: null}
    \begin{align*}
        \overline{Y}_{it}^{\overline{a}_t}|\overline{L}_{it}, U_i&\overset{d}{=} \overline{Y}_{it}|\overline{L}_{it}, U_i
    \end{align*}
    and under Assumption \ref{ASS: exch} and consistency
    \begin{align*}
        \overline{Y}_{it}^{\overline{a}_t}|\overline{L}_{it}, U_i&\overset{d}{=} \overline{Y}_{it}|\overline{L}_{it}, U_i, \overline{A}_{it},
    \end{align*}
    thus Assumption \ref{ASS: null strong inter} holds for individual $i$ conditional on $\overline{A}_{it}$ as well.

    Since $S_{\mathcal{K}_{il}}$ is a deterministic function of $\overline{Y}_{it}$, it follows that
    $$
    P(\overline{Y}_{it}=\overline{y}_{it}|\overline{L}_{it}=\overline{l}_{it}, U_i=u_i, \overline{A}_{it}=\overline{a}_{it}, S_{\mathcal{K}_{il}}=s_{\mathcal{K}_{il}})=\frac{P(\overline{Y}_{it}=\overline{y}_{it}|\overline{L}_{it}=\overline{l}_{it}, U_i=u_i, \overline{A}_{it}=\overline{a}_{it})}{P(S_{\mathcal{K}_{il}}=s_{\mathcal{K}_{il}}|\overline{L}_{it}=\overline{l}_{it}, U_i=u_i, \overline{A}_{it}=\overline{a}_{it})},
    $$
    for any $\overline{y}_{it}$ compatible with $s_{\mathcal{K}_{il}}$ and $0$ otherwise. Since interchangeability of the outcome variables holds in the numerator, it follows that the outcomes are interchangeable conditional on $\overline{L}_{it}=\overline{l}_{it}, U_i=u_i, \overline{A}_{it}=\overline{a}_{it}, S_{\mathcal{K}_{il}}=s_{\mathcal{K}_{il}}$ as well.

    Denote the vector of the outcomes of individual $i$ in stratum $l$ with $\mathbf{Y}_{\mathcal{K}_{il}}:=(Y_{ik})_{k\in\mathcal{K}_{il}}$. Interchangeability implies that any vector $\mathbf{Y}_{\mathcal{K}_{il}}$ which is compatible with $S_{\mathcal{K}_{il}}=s_{\mathcal{K}_{il}}$ has the same probability. The number of ways we can construct a vector of length $|\mathcal{K}_{il}|$ with $s_{\mathcal{K}_{il}}$ many $1$-s and $|\mathcal{K}_{il}|-s_{\mathcal{K}_{il}}$ many $0$-s is ${|\mathcal{K}_{il}|\choose s_{\mathcal{K}_{il}}}$, thus each of these vectors has the probability
    $$
    P(\mathbf{Y}_{\mathcal{K}_{il}}=\mathbf{y}|\overline{L}_{it}=\overline{l}_{it}, U_i=u_i, \overline{A}_{it}=\overline{a}_{it}, S_{\mathcal{K}_{il}}=s_{\mathcal{K}_{il}})=\frac{1}{{|\mathcal{K}_{il}|\choose s_{\mathcal{K}_{il}}}}.
    $$

    We are interested in the probability of observing $s_{\mathcal{K}_{1,il}}$ events in the $|\mathcal{K}_{1,il}|$ at-risk positions of individual $i$ in stratum $l$, which is equal to
    $$
    P(S_{\mathcal{K}_{1,il}}=s_{\mathcal{K}_{1,il}}|\overline{L}_{it}=\overline{l}_{it}, U_i=u_i, \overline{A}_{it}=\overline{a}_{it}, S_{\mathcal{K}_{il}}=s_{\mathcal{K}_{il}})=\frac{{|\mathcal{K}_{1,il}|\choose s_{\mathcal{K}_{1,il}}}{|\mathcal{K}_{0,il}|\choose s_{\mathcal{K}_{il}}- s_{\mathcal{K}_{1,il}}}}{{|\mathcal{K}_{il}|\choose s_{\mathcal{K}_{il}}}},
    $$
    which is equal to the probability mass function of a Hypergeometric random variable, with parameters $|\mathcal{K}_{il}|$, $|\mathcal{K}_{1,il}|$, and $s_{\mathcal{K}_{il}}$, therefore we have shown that
    $$
    S_{\mathcal{K}_{1,il}}|\overline{L}_{it}=\overline{l}_{it}, U_i=u_i, \overline{A}_{it}=\overline{a}_{it}, S_{\mathcal{K}_{il}}=s_{\mathcal{K}_{il}} \sim \mathrm{Hypergeometric}( |\mathcal{K}_{il}|, |\mathcal{K}_{1,il}|, s_{\mathcal{K}_{il}}).
    $$

    For $l\notin\mathcal{L}_i^+$ the at-risk count is deterministic given $\mathcal{K}_{1,il}$, $\mathcal{K}_{0,il}$ and $s_{\mathcal{K}_{il}}$: $S_{\mathcal{K}_{1,il}}=0$ if $\mathcal{K}_{1,il}=\emptyset$, and $S_{\mathcal{K}_{1,il}}=s_{\mathcal{K}_{il}}$ if $\mathcal{K}_{0,il}=\emptyset$. Moreover, because the compatible vectors of the outcomes act separately within the different strata of individual $i$, conditional on the stratum totals of individual $i$, the joint distribution satisfies
    $$
    P(\mathbf{S}_{\mathcal{K}_{1,i}}=\mathbf{s}_{\mathcal{K}_{1,i}}|\overline{L}_{it}=\overline{l}_{it}, U_i=u_i, \overline{A}_{it}=\overline{a}_{it}, \mathbf{S}_{\mathcal{K}_i}=\mathbf{s}_{\mathcal{K}_i})
    =
    \prod_{l\in\mathcal{L}_i^+} \frac{{|\mathcal{K}_{1,il}|\choose s_{\mathcal{K}_{1,il}}}{|\mathcal{K}_{0,il}|\choose s_{\mathcal{K}_{il}}- s_{\mathcal{K}_{1,il}}}}{{|\mathcal{K}_{il}|\choose s_{\mathcal{K}_{il}}}},
    $$
    where $\mathbf{S}_{\mathcal{K}_{1,i}}:=(S_{\mathcal{K}_{1,il}})_{l\in\mathcal{L}_i^+}$ is the vector of the numbers of events of individual $i$ in the at-risk intervals of the informative strata, and $\mathbf{S}_{\mathcal{K}_i}:=(S_{\mathcal{K}_{il}})_{l:\,\mathcal{K}_{il}\neq\emptyset}$ is the vector of the total numbers of events of individual $i$ in all strata, without considering treatment.

    The observed centred contribution of individual $i$ in stratum $l\in\mathcal{L}_i^+$ is
    $$
    D_{\mathcal{K}_{il}}
    =
    S_{\mathcal{K}_{1,il}} - s_{\mathcal{K}_{il}}\frac{|\mathcal{K}_{1,il}|}{|\mathcal{K}_{1,il}|+|\mathcal{K}_{0,il}|}.
    $$
    For $l\notin\mathcal{L}_i^+$ the same expression is identically $0$ by the previous paragraph, thus these strata can be omitted from the test statistic.

    Since
    $$
    E[S_{\mathcal{K}_{1,il}}|\overline{L}_{it}=\overline{l}_{it}, U_i=u_i, \overline{A}_{it}=\overline{a}_{it}, \mathbf{S}_{\mathcal{K}_i}=\mathbf{s}_{\mathcal{K}_i}]
    =
    s_{\mathcal{K}_{il}}\frac{|\mathcal{K}_{1,il}|}{|\mathcal{K}_{1,il}|+|\mathcal{K}_{0,il}|},
    $$
    it follows that
    $$
    E[D_{\mathcal{K}_{il}}|\overline{L}_{it}=\overline{l}_{it}, U_i=u_i, \overline{A}_{it}=\overline{a}_{it}, \mathbf{S}_{\mathcal{K}_i}=\mathbf{s}_{\mathcal{K}_i}]=0.
    $$

    The observed test statistic is
    $$
    D
    =
    \sum_{i=1}^n \sum_{l \in \mathcal{L}_i^+}  D_{\mathcal{K}_{il}}.
    $$

    For Monte Carlo replicate $b$, Algorithm \ref{alg: hyper} generates
    $$
    s_{\mathcal{K}_{1,il}}^{(b)}
    \sim
    \mathrm{Hypergeometric}(|\mathcal{K}_{il}|,|\mathcal{K}_{1,il}|,s_{\mathcal{K}_{il}}),
    $$
    independently across individuals $i=1,\dots,n$ and strata $l\in\mathcal{L}_i^+$, and sets
    $$
    D_{\mathcal{K}_{il}}^{(b)}
    =
    s_{\mathcal{K}_{1,il}}^{(b)}
    -
    s_{\mathcal{K}_{il}}\frac{|\mathcal{K}_{1,il}|}{|\mathcal{K}_{1,il}|+|\mathcal{K}_{0,il}|},
    \qquad
    D^{(b)}
    =
    \sum_{i=1}^n \sum_{l \in \mathcal{L}_i^+}  D_{\mathcal{K}_{il}}^{(b)}.
    $$

    As $D$ depends on the data of all individuals, denote the conditioning event of individual $i$, and of the whole sample, by
    $$
    \mathcal{C}_i:=\{\overline{L}_{it}=\overline{l}_{it}, U_i=u_i, \overline{A}_{it}=\overline{a}_{it}, \mathbf{S}_{\mathcal{K}_i}=\mathbf{s}_{\mathcal{K}_i}\},
    \qquad
    \mathcal{C}:=\bigcap_{i=1}^n \mathcal{C}_i.
    $$
    Since individuals are independent from each other, conditional on $\mathcal{C}$ the vectors $\mathbf{S}_{\mathcal{K}_{1,1}},\dots,\mathbf{S}_{\mathcal{K}_{1,n}}$ are independent, and the conditional distribution of $\mathbf{S}_{\mathcal{K}_{1,i}}$ given $\mathcal{C}$ is equal to its conditional distribution given $\mathcal{C}_i$. Therefore, by the preceding conditional distributional result,
    $$
    P(\mathbf{S}_{\mathcal{K}_{1,1}}=\mathbf{s}_{\mathcal{K}_{1,1}},\dots,\mathbf{S}_{\mathcal{K}_{1,n}}=\mathbf{s}_{\mathcal{K}_{1,n}}|\mathcal{C})
    =
    \prod_{i=1}^n\prod_{l\in\mathcal{L}_i^+} \frac{{|\mathcal{K}_{1,il}|\choose s_{\mathcal{K}_{1,il}}}{|\mathcal{K}_{0,il}|\choose s_{\mathcal{K}_{il}}- s_{\mathcal{K}_{1,il}}}}{{|\mathcal{K}_{il}|\choose s_{\mathcal{K}_{il}}}},
    $$
    which is exactly the joint distribution from which Algorithm \ref{alg: hyper} draws $\{s_{\mathcal{K}_{1,il}}^{(b)}: i=1,\dots,n,l\in\mathcal{L}_i^+\}$. As $D$ and $D^{(b)}$ are the same function of the observed and of the simulated at-risk counts, respectively,
    $$
    (D^{(b)}|\mathcal{C})
    \overset{d}{=}
    (D|\mathcal{C}).
    $$
    Since the Monte Carlo replicates are generated independently conditional on $\mathcal{C}$,
    $$
    D,D^{(1)},\ldots,D^{(B)}
    $$
    are conditionally exchangeable under Hypothesis \ref{HYP: null}.

    The Monte Carlo $p$-value can therefore be written as
    $$
    p_{\text{two-sided}}^{B}
    =
    \frac{\sum_{b=1}^B \mathbb{I}(|D^{(b)}|\geq |D|)+1}{B+1}.
    $$

    By conditional exchangeability, the rank of $|D|$ among $|D|,|D^{(1)}|,\dots,|D^{(B)}|$ is uniform when ties are broken at random.
    Using the weak inequality $|D^{(b)}|\geq |D|$ can only increase $\sum_{b=1}^B \mathbb{I}(|D^{(b)}|\geq |D|)+1$ in the presence of ties. Hence, for every integer $c$,
    $$
    P\left(\sum_{b=1}^B \mathbb{I}(|D^{(b)}|\geq |D|)+1\leq c\,\middle|\,\mathcal{C}\right)
    \leq
    \frac{c}{B+1}.
    $$

    It follows that
    \begin{align*}
    P(p_{\text{two-sided}}^{B}\leq\alpha|\mathcal{C})
    &=P\left(\sum_{b=1}^B \mathbb{I}(|D^{(b)}|\geq |D|)+1\leq \alpha(B+1)\,\middle|\,\mathcal{C}\right)\\
    &=P\left(\sum_{b=1}^B \mathbb{I}(|D^{(b)}|\geq |D|)+1\leq \left\lfloor \alpha(B+1) \right\rfloor\,\middle|\,\mathcal{C}\right)\\
    &\leq \frac{ \left\lfloor \alpha(B+1) \right\rfloor}{ B+1}\\
    &\leq \alpha.
    \end{align*}

    Therefore, the testing procedure in Algorithm \ref{alg: hyper} is a valid
    level-$\alpha$ test of Hypothesis \ref{HYP: null}.
\end{proof}\subsection{Proof of Corollary \ref{COR: int group}}\label{APP_sub: proof of cor}

\begin{proof}
    By the law of total probability
    $$
   E[Y_k^{\overline{a}_k^{1'}}]= \sum_{u,\overline{l}_k} E[Y_k^{\overline{a}_k^{1'}}|U=u, \overline{L}_k=\overline{l}_k]P(U=u, \overline{L}_k=\overline{l}_k).
    $$
    By Assumption \ref{ASS: exch}, positivity and consistency
    $$
    E[Y_k^{\overline{a}_k^{1'}}|U=u, \overline{L}_k=\overline{l}_k]=E[Y_k|U=u, \overline{L}_k=\overline{l}_k, \overline{A}_k=\overline{a}_k^{1'}].
    $$
    It follows the low of total probability, and the definition of $V_k$ and $\overline{a}_k^{1'}$
    \begin{align*}
        &E[Y_k|U=u, \overline{L}_k=\overline{l}_k, \overline{A}_k=\overline{a}_k^{1'}]\\
        &=E[Y_k|U=u, \overline{L}_k=\overline{l}_k, \overline{A}_k=\overline{a}_k^{1'}, V_k=1]P(V_k=1|U=u, \overline{L}_k=\overline{l}_k, \overline{A}_k=\overline{a}_k^{1'})\\
        &+E[Y_k|U=u, \overline{L}_k=\overline{l}_k, \overline{A}_k=\overline{a}_k^{1'}, V_k=0]P(V_k=0|U=u, \overline{L}_k=\overline{l}_k, \overline{A}_k=\overline{a}_k^{1'})\\
        &=E[Y_k|U=u, \overline{L}_k=\overline{l}_k, \overline{A}_k=\overline{a}_k^{1'}, V_k=1] \cdot 1\\
        &+E[Y_k|U=u, \overline{L}_k=\overline{l}_k, \overline{A}_k=\overline{a}_k^{1'}, V_k=0] \cdot 0\\
        &=E[Y_k|U=u, \overline{L}_k=\overline{l}_k, \overline{A}_k=\overline{a}_k^{1'}, V_k=1].
    \end{align*}
    By Assumption \ref{ASS: trans risk}
    $$
    E[Y_k|U=u, \overline{L}_k=\overline{l}_k, \overline{A}_k=\overline{a}_k^{1'}, V_k=1]=E[Y_k|U=u, \overline{L}_k=\overline{l}_k, \overline{A}_k=\overline{a}_k^{1''}, V_k=1],
    $$
    therefore, taking the steps above in reverse, and using the definition of $\overline{a}_k^{1''}$
    $$
    E[Y_k|U=u, \overline{L}_k=\overline{l}_k, \overline{A}_k=\overline{a}_k^{1''}, V_k=1] = E[Y_k^{\overline{a}_k^{1''}}|U=u, \overline{L}_k=\overline{l}_k],
    $$
    which shows the first part of the claim.
    
    Finally 
    \begin{align*}
        &E[Y_k^{\overline{a}_k^{1'}}]= \sum_{u,\overline{l}_k} E[Y_k^{\overline{a}_k^{1'}}|U=u, \overline{L}_k=\overline{l}_k]P(U=u, \overline{L}_k=\overline{l}_k)\\
        &= \sum_{u,\overline{l}_k} E[Y_k^{\overline{a}_k^{1''}}|U=u, \overline{L}_k=\overline{l}_k]P(U=u, \overline{L}_k=\overline{l}_k) = E[Y_k^{\overline{a}_k^{1''}}],
    \end{align*}
    which shows the first part of our claim.

    The proof of the second half of the corollary follows analogously. 
\end{proof}

\subsection{Proof of Proposition \ref{PROP: c-RR gen}}\label{APP_sub: gen c-RR proof}
\begin{proof}
    \begin{align}
        &E[Y_k^{\overline{a}_k}|U=u, L_{(k-m_l+1):k}=l_{(k-m_l+1):k}]\\
        &\overset{\mathrm{Assumption}\; \ref{ASS: exch}, \; \ref{ASS: gen pos}}{=} E[Y_k^{\overline{a}_k}|\overline{A}_k=\overline{a}_k, U=u, L_{(k-m_l+1):k}=l_{(k-m_l+1):k}]\notag\\
        &\overset{\mathrm{Consistency}}{=} E[Y_k|\overline{A}_k=\overline{a}_k, U=u, L_{(k-m_l+1):k}=l_{(k-m_l+1):k}]\notag\\
        &\overset{\mathrm{Assumption}\; \ref{ASS: gen trans conf}}{=} E[Y_k|\overline{A}_k=\overline{a}_k, U=u, \overline{L}_k=\overline{l}_k]\notag\\
        &\overset{\mathrm{Assumption}\; \ref{ASS: ind out} \; (ii)}{=} E[Y_k|\overline{A}_t=\overline{a}_t, U=u, \overline{L}_k=\overline{l}_k]\notag\\
        &\overset{\mathrm{Assumption}\; \ref{ASS: ind out} \; (iii)}{=} E[Y_k|\overline{A}_t=\overline{a}_t, U=u, \overline{L}_t=\overline{l}_t]\notag\\
        &\overset{\mathrm{LOTP}}{=}E[Y_k|\overline{A}_t=\overline{a}_t, U=u, \overline{L}_t=\overline{l}_t, S_t>0]P(S_t>0|\overline{A}_t=\overline{a}_t, U=u, \overline{L}_t=\overline{l}_t)\notag\\
        &+E[Y_k|\overline{A}_t=\overline{a}_t, U=u, \overline{L}_t=\overline{l}_t, S_t=0]P(S_t=0|\overline{A}_t=\overline{a}_t, U=u, \overline{L}_t=\overline{l}_t)\notag\\
        &=E[Y_k|\overline{A}_t=\overline{a}_t, U=u, \overline{L}_t=\overline{l}_t, S_t>0]P(S_t>0|\overline{A}_t=\overline{a}_t, U=u, \overline{L}_t=\overline{l}_t).
    \end{align}

    The remaining of the proof is identical to the proof of Proposition \ref{PROP: c-RR(u,l)}, with the indices $k^1$, $k^0$ being adjusted to the extended time-varying confounding setting, as well as the subsets $\mathcal{K}_{1, l_{(k-m_l+1):k}}$ and $\mathcal{K}_{0, l_{(k-m_l+1):k}}$
\end{proof}

\subsection{Proof of Proposition \ref{PROP: gen CRR}}\label{APP_sub: proof gen CRRm}
\begin{proof}
   Fix $k \in \mathcal T_{m_l}$. Since

    $$\mathcal U \times \mathcal L^{m_l} = (\mathcal U' \times \mathcal L'_{m_l}) \,\dot\cup\, (\mathcal I_U \times \mathcal L^{m_l}) \,\dot\cup\, (\mathcal U' \times \mathcal I_{L,m_l}),$$
    
    the law of total probability gives

    $$E[Y_k^{a_k}] = \sum_{(u,l) \in \mathcal U' \times \mathcal L'_{m_l}} E[Y_k^{a_k} \mid U = u, L_{(k-m_l+1):k} = l] P(U = u, L_{(k-m_l+1):k} = l),$$
    
    because the sums over the other two sets vanish by Assumption \ref{ASS: gen stat U}.

    The remaining part of the proof is an adaptation of the proof of Proposition \ref{PROP: CRR}, with replacing Assumption \ref{ASS: stat all} with Assumption \ref{ASS: gen stat all}.
\end{proof}

\subsection{Proof of Lemma \ref{APP PROP: small bias}}\label{APP_sub: proof lemma}
\begin{proof}
    Using $r_s$, the deviation of the bias from $1$ can be rewritten as 
    $$
    \left|
\mathrm{bias}(p^1,p^0,s, \overline{h}_t)-1
\right|= \left|\frac{(p^1-p^0)(r_s-1)}{1+p^0(r_s-1)}\right|,
    $$
which again immediately shows the two cases when the bias vanishes.

When the ratio $r_s$ is well-defined, it is non-negative, thus the denominator is greater than or equal to $1-p^0$, which is always non-negative (positive for all $p^0\in [0,1)$), thus 
$$
    \left|
\mathrm{bias}(p^1,p^0,s, \overline{h}_t)-1
\right|= \left|\frac{(p^1-p^0)(r_s-1)}{1+p^0(r_s-1)}\right|\leq \left|\frac{(p^1-p^0)(r_s-1)}{1-p^0}\right|=\frac{|p^1-p^0|\cdot|r_s-1|}{1-p^0},
$$
therefore, whenever $\left|r_s-1\right| \leq \epsilon$,
$$\left| \mathrm{bias}(p^1,p^0,s, \overline{h}_t)-1 \right|
\leq
\frac{|p^1-p^0|}{1-p^0}\epsilon.$$
\end{proof}

\subsection{Proof of Proposition \ref{APP PROP: S mode}}\label{APP_sub: proof S mode}
\begin{proof}
    As the outcomes are independent, and binary, their sum is Poisson-Binomial distributed. Since $S_{t, -k^1, -k^0}\mid H_t=h_t$ is Poisson-Binomial, its PMF is ultra log-concave \citep{tang_poisson_2023}, therefore, for $1\leq s \leq t-3$
    \begin{equation}\label{APP: EQ: ultra log}
    \begin{aligned}
        &P(S_{t, -k^1, -k^0}=s\mid H_t=h_t)^2\\
    &\geq P(S_{t, -k^1, -k^0}=s-1\mid H_t=h_t)P(S_{t, -k^1, -k^0}=s+1\mid H_t=h_t)\frac{s+1}{s}\frac{t-1-s}{t-2-s}
    \end{aligned}
    \end{equation}

    Equation \ref{APP: EQ: ultra log} implies, whenever both ratios are well-defined,
    $$
    r_s=\frac{ P(S_{t, -k^1, -k^0}=s-2\mid H_t=h_t)}{ P(S_{t, -k^1, -k^0}=s-1\mid H_t=h_t)}\leq \frac{ P(S_{t, -k^1, -k^0}=s-1\mid H_t=h_t)}{ P(S_{t, -k^1, -k^0}=s\mid H_t=h_t)}=r_{s+1}
    $$
    therefore $r_s$ is non-decreasing, and hence $\{s: |r_s-1|<\epsilon\}$ is an interval. Let $m=\max(\mathcal{M}_{S_t^-})$ denote the largest mode of $S_{t, -k^1, -k^0}\mid H_t=h_t$. As $m$ is a mode, $r_{m+1}\leq 1$, with equality if and only if $m-1$ is also a mode, and as $m$ is the largest mode, $r_{m+2}>1$ whenever $r_{m+2}$ is well-defined. Since $r_s$ is non-decreasing, $|r_s-1|<\epsilon$ for some $s\leq m+1$ implies $1-\epsilon<r_s\leq r_{m+1}\leq 1$, while $|r_s-1|<\epsilon$ for some $s\geq m+2$ implies $1<r_{m+2}\leq r_s<1+\epsilon$. Therefore, whenever $\{s: |r_s-1|<\epsilon\}$ is non-empty, it must include $m+1$ or $m+2$.
 
    Suppose the distribution is not unimodal. Then, as the support of a Poisson-Binomial distribution consists of consecutive integers, there must exist some value $\tilde{s}$ with $P(S_{t, -k^1, -k^0}=\tilde{s}\mid H_t=h_t)>0$ such that
    $$
    P(S_{t, -k^1, -k^0}=\tilde{s}-1\mid H_t=h_t)\geq P(S_{t, -k^1, -k^0}=\tilde{s}\mid H_t=h_t) \leq P(S_{t, -k^1, -k^0}=\tilde{s}+1\mid H_t=h_t),
    $$
    that would contradict Equation \ref{APP: EQ: ultra log}.
    
    Similarly, suppose that there exists a value $\dot{s}$, for which
    $$
    P(S_{t, -k^1, -k^0}=\dot{s}-1\mid H_t=h_t)= P(S_{t, -k^1, -k^0}=\dot{s}\mid H_t=h_t) = P(S_{t, -k^1, -k^0}=\dot{s}+1\mid H_t=h_t),
    $$
    such that all three of them are in the mode set. Substituting them into Equation \ref{APP: EQ: ultra log} leads to a contradiction, thus, the distribution of $S_{t, -k^1, -k^0}\mid H_t=h_t$ is unimodal, and it has at most two adjacent modes. 

     By unimodality it follows that for values less than $\mathcal{M}_{S_t^-}$, the distribution of $S_{t, -k^1, -k^0}\mid H_t=h_t$ is increasing, while for values greater than the modes, it is decreasing. Moreover, as the factor $\frac{s+1}{s}\frac{t-1-s}{t-2-s}$ in Equation \ref{APP: EQ: ultra log} is strictly greater than $1$, two adjacent masses that are equal and positive must both be modes: otherwise, by unimodality, both would lie on the same side of the mode set, their neighbour on the side of the mode set would have at least the same mass, and substituting these three masses into Equation \ref{APP: EQ: ultra log} would lead to a contradiction. Hence, on its support, the PMF of $S_{t, -k^1, -k^0}\mid H_t=h_t$ is strictly increasing up to $\min(\mathcal{M}_{S_t^-})$ and strictly decreasing from $\max(\mathcal{M}_{S_t^-})$ on. The same holds for $S_{t, -k^0}\mid H_t=h_t$ and $S_t\mid H_t=h_t$, which are also Poisson-Binomial.
 
    Let $s = \min( \mathcal{M}_{S_t^-})$, and consider $S_{t, -k^0}$, the total number of events except at time $k^0$. Values with zero probability cannot be modes, thus take any $s'<s$ with $P(S_{t, -k^0}=s'\mid H_t=h_t)>0$, which implies $P(S_{t, -k^1, -k^0}=s'\mid H_t=h_t)>0$. Then under Assumption \ref{ASS: ind out}, by the strict monotonicity of the PMF of $S_{t, -k^1, -k^0}\mid H_t=h_t$ below $s$,
    \begin{align*}
        P(S_{t, -k^0}=s'\mid H_t=h_t)&=P(S_{t, -k^1, -k^0}=s'-1\mid H_t=h_t)p^1\\
        &+P(S_{t, -k^1, -k^0}=s'\mid H_t=h_t)(1-p^1)\\
        &\leq P(S_{t, -k^1, -k^0}=s'\mid H_t=h_t)p^1\\
        &+P(S_{t, -k^1, -k^0}=s'+1\mid H_t=h_t)(1-p^1)\\
        &=P(S_{t, -k^0}=s'+1 \mid H_t=h_t),
    \end{align*}
    implying that the mode of $S_{t, -k^0}\mid H_t=h_t$ cannot be any $s'$ that is less than $s$. Similarly, for any $s'$ that is less than the minimum of the mode set of $S_{t, -k^0}\mid H_t=h_t$, which is known to be greater than or equal to $s=\min( \mathcal{M}_{S_t^-})$
    \begin{align*}
        P(S_{t}=s'\mid H_t=h_t)&=P(S_{t, -k^0}=s'-1\mid H_t=h_t)p^0\\
        &+P(S_{t, -k^0}=s'\mid H_t=h_t)(1-p^0)\\
        &\leq P(S_{t, -k^0}=s'\mid H_t=h_t)p^0\\
        &+P(S_{t, -k^0}=s'+1\mid H_t=h_t)(1-p^0)\\
        &=P(S_{t}=s'+1 \mid H_t=h_t),
    \end{align*}
    therefore, the $\min(\mathcal{M}_{S_t})\geq \min(\mathcal{M_{S_t^-}})$.

    Now take $s'-1>s=\max(\mathcal{M}_{S_{t}^-})$, then
    \begin{align*}
        P(S_{t, -k^0}=s'\mid H_t=h_t)&=P(S_{t, -k^1, -k^0}=s'-1\mid H_t=h_t)p^1\\
        &+P(S_{t, -k^1, -k^0}=s'\mid H_t=h_t)(1-p^1)\\
        &\leq P(S_{t, -k^1, -k^0}=s'-2\mid H_t=h_t)p^1\\
        &+P(S_{t, -k^1, -k^0}=s'-1\mid H_t=h_t)(1-p^1)\\
        &=P(S_{t, -k^0}=s'-1 \mid H_t=h_t),
    \end{align*}
    therefore $S_{t, -k^0}\mid H_t=h_t$ cannot have a mode, that is greater than $\max(\mathcal{M}_{S_t^-})+1$. Once again, by taking any $s'-1 > \max(\mathcal{M}_{S_t^-})+1$, we know that the PMF of $S_{t, -k^0}\mid H_t=h_t$ is strictly decreasing , thus
    \begin{align*}
        P(S_{t}=s'\mid H_t=h_t)&=P(S_{t, -k^0}=s'-1\mid H_t=h_t)p^0\\
        &+P(S_{t, -k^0}=s'\mid H_t=h_t)(1-p^0)\\
        &\leq P(S_{t, -k^0}=s'-2\mid H_t=h_t)p^0\\
        &+P(S_{t, -k^0}=s'-1\mid H_t=h_t)(1-p^0)\\
        &=P(S_{t}=s'-1 \mid H_t=h_t),
    \end{align*}
    therefore $\max(\mathcal{M}_{S_t})\leq \max(\mathcal{M}_{S_t^-})+2$. As $S_t\mid H_t=h_t$ is also Poisson-Binomial, $\mathcal{M}_{S_t}$ consists of at most two adjacent values, hence $\mathcal{M}_{S_t}\subset \{s', s'+1, s'+2\}$ if $\mathcal{M}_{S_t^-}=\{s'\}$, and $\mathcal{M}_{S_t}\subset \{s', s'+1, s'+2, s'+3\}$ if $\mathcal{M}_{S_t^-}=\{s', s'+1\}$.
 
    Finally, by the proof of Lemma \ref{APP PROP: small bias},
    $$
    \left|\mathrm{bias}(p^1,p^0,s, \overline{h}_t)-1\right|= |p^1-p^0|\frac{|r_s-1|}{1+p^0(r_s-1)},
    $$
    where the denominator is positive whenever the bias is well-defined, and the right-hand side is non-increasing in $r_s$ for $r_s\leq 1$ and non-decreasing in $r_s$ for $r_s\geq 1$. As $r_s$ is non-decreasing in $s$, with $r_{m+1}\leq 1<r_{m+2}$ (the latter whenever well-defined), the bias is minimised at $s=m+1$ or $s=m+2$, where $m=\max(\mathcal{M}_{S_t^-})$, if $\mathcal{M}_{S_t^-}=\{s', s'+1\}$, then $r_{s'+2}=1$, and the bias is equal to $1$ at $s=s'+2$. In both cases, every mode of $S_t$ is at most two event counts away from the value for which the bias is minimised, which concludes our proof.
\end{proof}

\subsection{Proof of Proposition \ref{PROP: test boot}}\label{APP: proof boot}
\begin{proof}
    Individuals $i=1, \dots,  n$ are i.i.d. copies of $(U, \overline{L}_t, \overline{A}_t, \overline{Y}_t)$, generated under the FFRCISTG model of Section \ref{SEC: dat gen}. To avoid ambiguity we index each individual with $i=1, \dots, n$.
    Similarly, we modify the notation of each covariate stratum $l$, to
    \begin{align*}
        &\mathcal{K}_{1,il}=\{k: L_{ik}=l, V_{ik}=1\},& &\mathcal{K}_{0,il}=\{k: L_{ik}=l, V_{ik}=0\}&,
    \end{align*}
    so $\mathcal{K}_{1,il} \cup \mathcal{K}_{0,il}=\{k: L_{ik}=l\}$, and denote the informative strata of individual $i$ by $\mathcal{L}_i^+:=\{l:|\mathcal{K}_{1,il}|>0,|\mathcal{K}_{0,il}|>0\}$, as in Algorithm \ref{ALG: Dependent}.
    \begin{lemma}\label{LEM: equal obs exp}
        Under Assumptions \ref{ASS: stat all}, \ref{ASS: bin}, \ref{ASS: exch}, \ref{ASS: ind out} (ii), (iii), \ref{ASS: trans conf}, and Hypothesis \ref{HYP: null}
        \begin{align*}
            E[Y_{ik}|\overline{A}_{it}=\overline{a}_{it},U_i=u_i, \overline{L}_{it}=\overline{l}_{it}]
        \end{align*}
        is constant for every $k \in \mathcal{K}_{1,il} \cup \mathcal{K}_{0,il}$ for fixed $u_i$ and $l$.
    \end{lemma}
    \begin{proof}
        By Hypothesis \ref{HYP: null} and Assumption \ref{ASS: exch}
        \begin{align*}
            Y_{ik} \independent \overline{A}_{ik}| U_i, L_{ik}, \overline{L}_{i(k-1)},
        \end{align*}
        which combined with Assumption \ref{ASS: trans conf}, by the intersection graphoid axiom results in
        \begin{align*}
             Y_{ik} \independent (\overline{A}_{ik}, \overline{L}_{i(k-1)})| U_i, L_{ik},
        \end{align*}
        Thus, applying Hypothesis \ref{HYP: null} and decomposition
        \begin{align*}
            Y_{ik}^{\overline{a}_{ik}} \independent \overline{L}_{i(k-1)}| U_i, L_{ik}.
        \end{align*}
        Therefore,
        \begin{align*}
            E[Y_{ik}^{\overline{a}_{ik}}|U_i=u_i, L_{ik}=l]&=E[Y_{ik}^{\overline{a}_{ik}}|U_i=u_i, \overline{L}_{ik}=\overline{l}_{ik}]\\
            &\overset{\mathrm{Assumption}\; \ref{ASS: exch}, \; \mathrm{consistency}}{=}E[Y_{ik}|U_i=u_i, \overline{A}_{ik}=\overline{a}_{ik}, \overline{L}_{ik}=\overline{l}_{ik}]\\
            &\overset{\mathrm{Assumption}\; \ref{ASS: ind out}(ii)}{=}E[Y_{ik}|U_i=u_i, \overline{A}_{it}=\overline{a}_{it}, \overline{L}_{ik}=\overline{l}_{ik}]\\
            &\overset{\mathrm{Assumption}\; \ref{ASS: ind out}(iii)}{=}E[Y_{ik}|U_i=u_i, \overline{A}_{it}=\overline{a}_{it}, \overline{L}_{it}=\overline{l}_{it}].
        \end{align*}
         As by Assumptions \ref{ASS: stat all}, \ref{ASS: bin} and Hypothesis \ref{HYP: null} for all $k \in \mathcal{K}_{1,il} \cup \mathcal{K}_{0,il}$ $E[Y_{ik}^{\overline{a}_{ik}}|U_i=u_i, L_{ik}=l]$ is invariant in $k$ for any intervention $\overline{a}_k$, the proof is concluded.
    \end{proof}

    Conditional on $\overline{A}_{it}=\overline{a}_{it},U_i=u_i, \overline{L}_{it}=\overline{l}_{it}$, the sets $\mathcal{K}_{0, il}$ and $\mathcal{K}_{1, il}$, and hence their cardinalities are fixed. Therefore
    \begin{equation}
    \label{EQ: D_i exp 0}
        \begin{aligned}
            &E[D_{i}| \overline{A}_{it}=\overline{a}_{it},U_i=u_i, \overline{L}_{it}=\overline{l}_{it}]
            =\frac{1}{\sum_{l \in \mathcal{L}_i^+} w_{il}}\sum_{l \in \mathcal{L}_i^+} w_{il} \cdot E[D_{\mathcal{K}_{il}}|\overline{A}_{it}=\overline{a}_{it},U_i=u_i, \overline{L}_{it}=\overline{l}_{it}]\\
            &=\frac{1}{\sum_{l \in \mathcal{L}_i^+} w_{il}}\cdot \\
            &\sum_{l \in \mathcal{L}_i^+} w_{il}
            \left(\frac{1}{|\mathcal{K}_{1,il}|}\sum_{k \in \mathcal{K}_{1,il}}E[Y_{ik}|\overline{A}_{it}=\overline{a}_{it},U_i=u_i, \overline{L}_{it}=\overline{l}_{it}]
            -\frac{1}{|\mathcal{K}_{0,il}|}\sum_{k \in \mathcal{K}_{0,il}}E[Y_{ik}|\overline{A}_{it}=\overline{a}_{it},U_i=u_i, \overline{L}_{it}=\overline{l}_{it}] \right)\\
            &=\frac{1}{\sum_{l \in \mathcal{L}_i^+} w_{il}}\sum_{l \in \mathcal{L}_i^+} w_{il} \left(\frac{|\mathcal{K}_{1,il}|}{|\mathcal{K}_{1,il}|}E[Y_{ik}|\overline{A}_{it}=\overline{a}_{it},U_i=u_i, \overline{L}_{it}=\overline{l}_{it}]-\frac{|\mathcal{K}_{0,il}|}{|\mathcal{K}_{0,il}|}E[Y_{ik}|\overline{A}_{it}=\overline{a}_{it},U_i=u_i, \overline{L}_{it}=\overline{l}_{it}] \right)\\
            &=0,
        \end{aligned}
    \end{equation}
    where $w_{il}$ denotes the individual specific weights for the stratum $l$, and the last line follows from Lemma \ref{LEM: equal obs exp}.

    Since the outcomes are non-negative, in fact binary by Assumption \ref{ASS: bin}, $S_{it}=0$ implies $Y_{ik}=0$ for all $k$, hence
    \begin{align*}
        D_{i}\mathbb{I}(S_{it}>0)=D_i.
    \end{align*}

    Then, for any histories for which $P(S_{it}>0|\overline{A}_{it}=\overline{a}_{it},U_i=u_i, \overline{L}_{it}=\overline{l}_{it})>0$ holds
    \begin{align*}
        E[D_{i}|\overline{A}_{it}=\overline{a}_{it},U_i=u_i, \overline{L}_{it}=\overline{l}_{it}, S_{it}>0] &=\frac{E[D_i\mathbb{I}(S_{it}>0)|\overline{A}_{it}=\overline{a}_{it},U_i=u_i, \overline{L}_{it}=\overline{l}_{it}]}{P(S_{it}>0|\overline{A}_{it}=\overline{a}_{it},U_i=u_i, \overline{L}_{it}=\overline{l}_{it})}\\
        &=\frac{E[D_i|\overline{A}_{it}=\overline{a}_{it},U_i=u_i, \overline{L}_{it}=\overline{l}_{it}]}{P(S_{it}>0|\overline{A}_{it}=\overline{a}_{it},U_i=u_i, \overline{L}_{it}=\overline{l}_{it})}=0,
    \end{align*}
    where the last line follows from Equation \ref{EQ: D_i exp 0}. Taking expectations over the covariate histories gives
    \begin{align*}
        E[D_i| S_{it}>0]=0.
    \end{align*}

    Under Assumption \ref{ASS: bin} each average in $D_{\mathcal{K}_{il}}$ lies in $[0,1]$, therefore $|D_{\mathcal{K}_{il}}| \leq 1$. As $D_{i}$ is a convex combination of the stratum specific differences $D_{\mathcal{K}_{il}}$, $|D_i|\leq 1$ as well. From $D_i \in [-1,1]$ it follows that $\mathrm{Var}(D_i|S_{it}>0)\leq 1$
    As individuals are assumed to be i.i.d., the individual-level differences are, $D_i$, are i.i.d. as well, and bounded, with $E[D_i|S_{it}>0]=0$, by Theorem 2.1 of \citet{bickel_asymptotic_1981} the bootstrap distribution
    \begin{align*}
        \{\sqrt{n}D^{0(b)}: b=1, \dots, B\}
    \end{align*}
    consistently estimates the sampling distribution of $\sqrt{n}D$ under Hypothesis \ref{HYP: null} as $n\rightarrow \infty, B\rightarrow \infty$. Consequently, $p_{\text{two-sided}}^{B}$ is asymptotically uniform under Hypothesis \ref{HYP: null}, and the test has asymptotic level $\alpha$.
\end{proof}
\subsection{Proof of Proposition \ref{PROP: pois}}\label{APP_Sub: pois proof}

\begin{proof}
Fix corresponding histories of treatment and covariates $\overline H_t=\overline h_t$. We leave the dependence of $Y_k$ on the resolution $t$ implicit

Let $\Delta_t$ denote the time unit,  $\mathcal I_t$ denote the time intervals belonging to \(I\), therefore
\begin{equation}\label{EQ: int partition}
  |\mathcal I_t|\Delta_t= |I|,  
\end{equation}

that is, the length of the fixed interval $I$ is equal to the number of sub-intervals $|\mathcal I_t|$ multiplied by their respective length.

For a fixed number of timepoints $t$, that is for a fixed resolution, define the expected outcome conditional on the history, at time $k$ and the total expected number of outcome in the interval $I$
$$
\mu_{tk}
:=
E[Y_k\mid\overline H_t=\overline h_t],
\qquad
\Lambda_{I,t}
:=
\sum_{k\in\mathcal I_t}\mu_{tk}.
$$
By the assumed resolution invariance, 
\begin{equation}\label{EQ: res inv}
   \Lambda_{I,t}=\Lambda_I(\overline h). 
\end{equation}

By the identification result established in the proof of Proposition \ref{PROP: c-RR(u,l)},
$$
\mu_{tk}
=
E\left[
Y_k^{\overline a_k}| U=u,L_k=l_k
\right].
$$
By Assumption \ref{ASS: stat all} and Corollary \ref{COR: int group}, this expectation depends on $k$ and $\overline a_k$ only through the corresponding values $v_k,u,l_k$. Make this dependence explicit by the notation $\mu_{tk}=e_t(v_k,u,l_k)$, where the subscript \(t\) distinguishes expectations at different resolutions.

For an interval $I$ in which the values of of $v_k=v,u=u,l_k=l$ are constant across time,
$$
\Lambda_{I,t}
=
|\mathcal I_t|e_t(v,u,l).
$$
It follows from Equations \eqref{EQ: int partition} and \eqref{EQ: res inv} that
\begin{equation}\label{EQ: v,u,l ratio}
    \frac{e_t(v,u,l)}{\Delta_t}
=
\frac{\Lambda_{I,t}}{|\mathcal I_t|\Delta_t}
=
\frac{\Lambda_I(\overline h)}{|I|}
=:r(v,u,l).
\end{equation}

Choose reference levels $u^*\in\mathcal{U'}$ and $l^*\in\mathcal{L'}$. By Assumptions \ref{ASS: Stat U} and \ref{ASS: no int} 
\begin{equation}
\begin{aligned}
e_t(v,u,l)
=&
\frac{e_t(v,u^*,l^*)}{e_t(0,u^*,l^*)}
\frac{e_t(0,u,l^*)}{e_t(0,u^*,l^*)}
\frac{e_t(0,u^*,l)}{e_t(0,u^*,l^*)}
e_t(0,u^*,l^*).
\end{aligned}
\label{EQ: Pois multi rep}
\end{equation}
This is the multiplicative representation developed in Appendix \ref{APP: Pois reg}.

Dividing \eqref{EQ: Pois multi rep} by $\Delta_t$,
\begin{equation}
    \begin{aligned}
        r(v,u,l)
        =
        \frac{r(v,u^*,l^*)}{r(0,u^*,l^*)}
        \frac{r(0,u,l^*)}{r(0,u^*,l^*)}
        \frac{r(0,u^*,l)}{r(0,u^*,l^*)}
        r(0,u^*,l^*).
    \end{aligned}
\end{equation}

Define
$$
\begin{aligned}
\lambda_0
&:=\log r(0,u^*,l^*),\\[1mm]
\alpha_v
&:=\log\left\{
\frac{r(v,u^*,l^*)}{r(0,u^*,l^*)}
\right\},\\[1mm]
\beta_u
&:=\log\left\{
\frac{r(0,u,l^*)}{r(0,u^*,l^*)}
\right\},\\[1mm]
\beta_l
&:=\log\left\{
\frac{r(0,u^*,l)}{r(0,u^*,l^*)}
\right\}.
\end{aligned}
$$
Then
$$
r(v,u,l)
=
\exp(\lambda_0+\alpha_v+\beta_l+\beta_u),
$$
and, by \eqref{EQ: v,u,l ratio},
\begin{equation}\label{EQ: Pois model sigma_I}
    \Lambda_I(\overline h)
    =
    \exp(\lambda_0+\alpha_v+\beta_l+\beta_u+\sigma_I),
    \qquad
    \sigma_I:=\log|I|.
\end{equation}

Then the correct mean model is established and it is left to prove the convergence to the Poisson distribution. 

For notational convenience, write $E_t$, $P_t$, $\mathcal{L}_t$ for expectation, probability, and distribution conditional on the fixed history $\overline H_t=\overline h_t$.

Define the event indicators and their probabilities by
$$
B_k:=I(Y_k>0),
\qquad
q_{tk}:=P_t(Y_k>0),
$$
and let
$$
B_{I,t}:=\sum_{k\in\mathcal I_t}B_k,
\qquad
Q_{I,t}:=\sum_{k\in\mathcal I_t}q_{tk}.
$$
By Assumption \ref{ASS: ind out}(i), the $Y_k$-s, and therefore the $B_k$-s, are mutually independent conditional on the full history. Consequently, Le Cam’s inequality \citep{lecam_distribution_1965, steele_cams_1994}, gives

\begin{equation}\label{EQ: le cam}
    \sum_{s=0}^{\infty}
\left|
P_t(B_{I,t}=s)
-
\frac{e^{-Q_{I,t}}Q_{I,t}^{\,s}}{s!}
\right|
\leq
2\sum_{k\in\mathcal{I}_t}q_{tk}^{2}.
\end{equation}

Equivalently,
\begin{equation}\label{EQ: TV le cam}
    d_{\mathrm{TV}}\left(
\mathcal L_t(B_{I,t}),
\operatorname{Poisson}(Q_{I,t})
\right)
\leq
\sum_{k\in\mathcal I_t}q_{tk}^{2}
\leq
\sum_{k\in\mathcal I_t}\mu_{tk}^{2}.
\end{equation}

The factor $2$ in \eqref{EQ: le cam} disappears in \eqref{EQ: TV le cam} because of the factor $1/2$ in the definition of total variation distance.

It remains to control the difference between the original event count $S_{I,t}$ and the indicator sum $B_{I,t}$, and between their respective means, as $Y_k$ can take values greater than $1$. Define
$$
\varepsilon_t
:=
\sup_{k:q_{tk}>0}
\left\{E_t[Y_k\mid Y_k>0]-1\right\},
$$
with $\varepsilon_t=0$ if all $q_{tk}=0$. By Assumption \ref{ASS: rare event} (iii), $\varepsilon_t\rightarrow0$. 

By the law of total expectation
$$
\begin{aligned}
0\leq\mu_{tk}-q_{tk}
&=
q_{tk}\left\{E_t[Y_k\mid Y_k>0]-1\right\}\\
&\leq\varepsilon_t q_{tk}\\
&\leq\varepsilon_t\mu_{tk}.
\end{aligned}
$$

Summing over the intervals \(k\in\mathcal I_t\) gives
\begin{equation}\label{EQ: interval sum}
    0\leq\Lambda_{I,t}-Q_{I,t}
\leq\varepsilon_t\Lambda_{I,t}.
\end{equation}

Since $Y_k$ is a nonnegative integer, $Y_k-B_k$ is also a nonnegative integer. Thus,

\begin{align*}
P_t(S_{I,t}\neq B_{I,t})
&=
P_t(S_{I,t}-B_{I,t}\geq1)\\
&\leq E_t[S_{I,t}-B_{I,t}]\\
&=\Lambda_{I,t}-Q_{I,t}.
\end{align*}

By the coupling inequality \citep{steele_cams_1994},
\begin{equation}\label{EQ: coupling steele}
        \begin{aligned}
        d_{\mathrm{TV}}\!\left(
        \mathcal L_t(S_{I,t}),
        \mathcal L_t(B_{I,t})
        \right)
        &\leq P_t(S_{I,t}\neq B_{I,t})\\
        &\leq\Lambda_{I,t}-Q_{I,t}\\
        &\leq\varepsilon_t\Lambda_{I,t}.
    \end{aligned}
\end{equation}

By the resolution invariance $\Lambda_{I,t}=\Lambda_I(\overline{h}_t)$.

Since $\Lambda_I(\overline{h}_t)<\infty$ and $\varepsilon_t\rightarrow0$, the bound in Equation \eqref{EQ: interval sum} gives
$$
0\leq\Lambda_I-Q_{I,t}
\leq\varepsilon_t\Lambda_I
\rightarrow0.
$$
Therefore,
\begin{equation}\label{EQ: Q to Lambda}
    Q_{I,t}\rightarrow\Lambda_I(\overline{h}_t).
\end{equation}

By the triangle inequality and the bounds established in Equations \eqref{EQ: TV le cam}and \eqref{EQ: coupling steele},
\begin{equation}\label{EQ: triangle TV}
    \begin{aligned}
    &d_{\mathrm{TV}}\left(
    \mathcal L_t(S_{I,t}),
    \mathrm{Poisson}(Q_{I,t})
    \right)\\
    &\leq
    d_{\mathrm{TV}}\left(
    \mathcal L_t(S_{I,t}),
    \mathcal L_t(B_{I,t})
    \right)
    +
    d_{\mathrm{TV}}\left(
    \mathcal L_t(B_{I,t}),
    \mathrm{Poisson}(Q_{I,t})
    \right)\\
    &\leq
    \varepsilon_t\Lambda_I
    +
    \sum_{k\in\mathcal I_t}\mu_{tk}^{2}\\
    &\leq
    \varepsilon_t\Lambda_I
    +
    \left(\max_{1\leq k\leq t}\mu_{tk}\right)
    \sum_{k\in\mathcal I_t}\mu_{tk}\\
    &=
    \Lambda_I
    \left(
    \varepsilon_t+\max_{1\leq k\leq t}\mu_{tk}
    \right)
    \rightarrow0.
    \end{aligned}
\end{equation}
Here, the convergence follows from $\varepsilon_t\rightarrow0$, Assumption \ref{ASS: rare event} (i), and the finiteness of $\Lambda_I(\overline{h}_t)$.

For every fixed $s\in\{0,1,2,\dots\}$, the difference between the probabilities assigned to the singleton $\{s\}$ is bounded by the total variation distance. Consequently, Equation \eqref{EQ: triangle TV} implies
\[
P_t(S_{I,t}=s)
=
\frac{e^{-Q_{I,t}}Q_{I,t}^{\,s}}{s!}+o(1).
\]
Using Equation \eqref{EQ: Q to Lambda} and continuity of the Poisson probabilities in their parameter,
\begin{equation}\label{EQ: S pois conv}
    P_t(S_{I,t}=s)
\longrightarrow
\frac{e^{-\Lambda_I}\Lambda_I^{\,s}}{s!}
\qquad
\text{for every }s\in\{0,1,2,\ldots\}.
\end{equation}

Thus $S_{I,t}$ converges to a $\mathrm{Poisson}(\Lambda_I(\overline{h}_t))$ random variable.

By Equation \eqref{EQ: Pois model sigma_I},
$$
\mathcal L\left(
S_{I,t}\mid\overline H_t=\overline h_t
\right)
\rightarrow
\mathrm{Poisson}\!\left(
\exp\{\lambda_0+\alpha_v+\beta_l+\beta_u+\sigma_I\}
\right).
$$

We still need to identify the treatment parameter and establish independence of the counts in non-overlapping intervals. By the definitions of \(\alpha_v\) and \(r(v,u,l)\), and the common rate ratio implied by Assumption \ref{ASS: Stat U},
$$
\exp(\alpha_1)
=
\frac{r(1,u,l)}{r(0,u,l)}
=
\frac{e_t(1,u,l)}{e_t(0,u,l)}
$$
for every supported non-immune stratum with positive baseline rate. The identification argument of Proposition \ref{PROP: c-RR(u,l)} shows that this common conditional rate ratio equals $\mathrm{CRR}_m$.

Finally, let $I$ and $I$ be two non-overlapping intervals, with disjoint index sets $\mathcal I_t$ and $\mathcal I'_t$ fixed conditional on the treatment and covariate histories. By Assumption \ref{ASS: ind out}(i), the counts
$$
S_{I,t}=\sum_{k\in\mathcal I_t}Y_k,
\qquad
S_{I',t}=\sum_{k\in\mathcal I'_t}Y_k
$$
are functions of disjoint subsets of conditionally independent outcomes. Therefore,
$$
S_{I,t} \independent S_{I',t}|\overline{H}_t
$$
at every resolution $t$.
\end{proof}

\section{Justification of the multiplicative model}\label{APP: pies}
Here, we justify the use of a multiplicative model through the sufficient-component cause model \citep{hernan_causal_2025, rothman1976causes}. For simplicity, we refer to a single time point and omit the time index $k$, which is justified by Assumption \ref{ASS: stat all}. Furthermore, we consider the simple setting where all variables are binary, the extension to multivariate discrete covariates is straightforward. 

Let $\omega_0, \omega_1, \dots, \omega_8, U, L, V \in \{0,1\}$ be mutually independent binary variables. As in the main text, $V$ indicates treatment-induced susceptibility, while $L$ and $U$ represent measured and unmeasured covariates. The variables $\omega_0, \dots, \omega_8$ are exogenous variables. We consider a binary outcome $Y$. The key assumption is that the outcome $Y$ occurs if and only if three events happen at the same time: $Y_U$, $Y_L$, and $Y_V$. We refer to these three events as constituent outcomes. Each constituent outcome can only be caused by the variable in the subscript and a set of exogenous variables. The variables affecting a constituent outcome cannot affect another constituent outcome. For example, the occurrence of the event $Y_U$ only depends on the value of the variable $U$ and on the value of $\omega_0,\omega_1,\omega_2$, and these variables do not cause $Y_V$ or $Y_L$.

\begin{figure}
    \centering
    \includegraphics[width=0.5\linewidth]{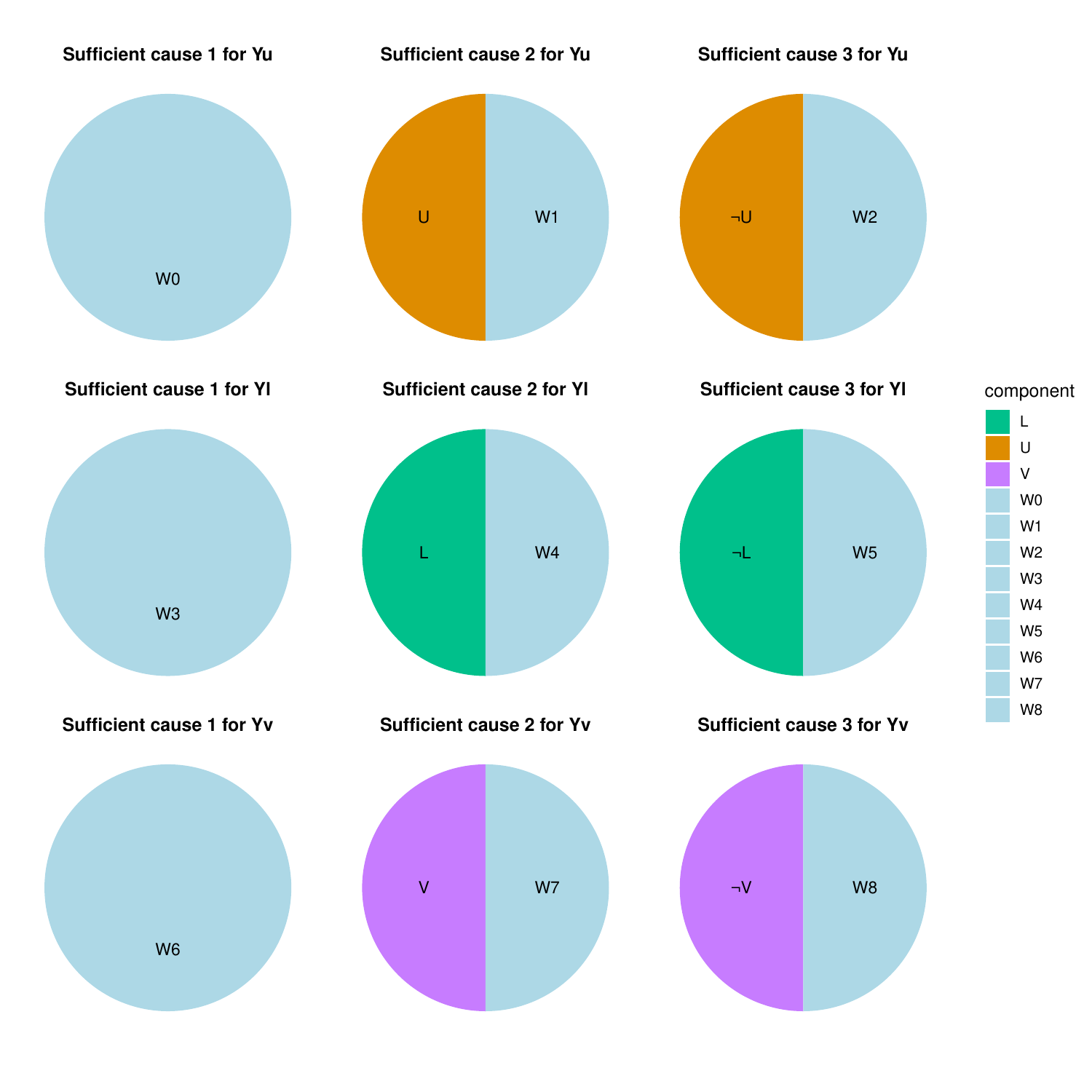}
    \caption{Sufficient causes for the constituent outcomes induced by $U$, $L$, and $V$. For example, the constituent outcome induced by the measured covariate occurs if $\omega_3=1$, or if $L=1$ and $\omega_4=1$, or if $L=0$ and $\omega_5=1$.}
    \label{FIG: pieset9}
\end{figure}
\begin{figure}
    \centering
    \includegraphics[width=0.5\linewidth]{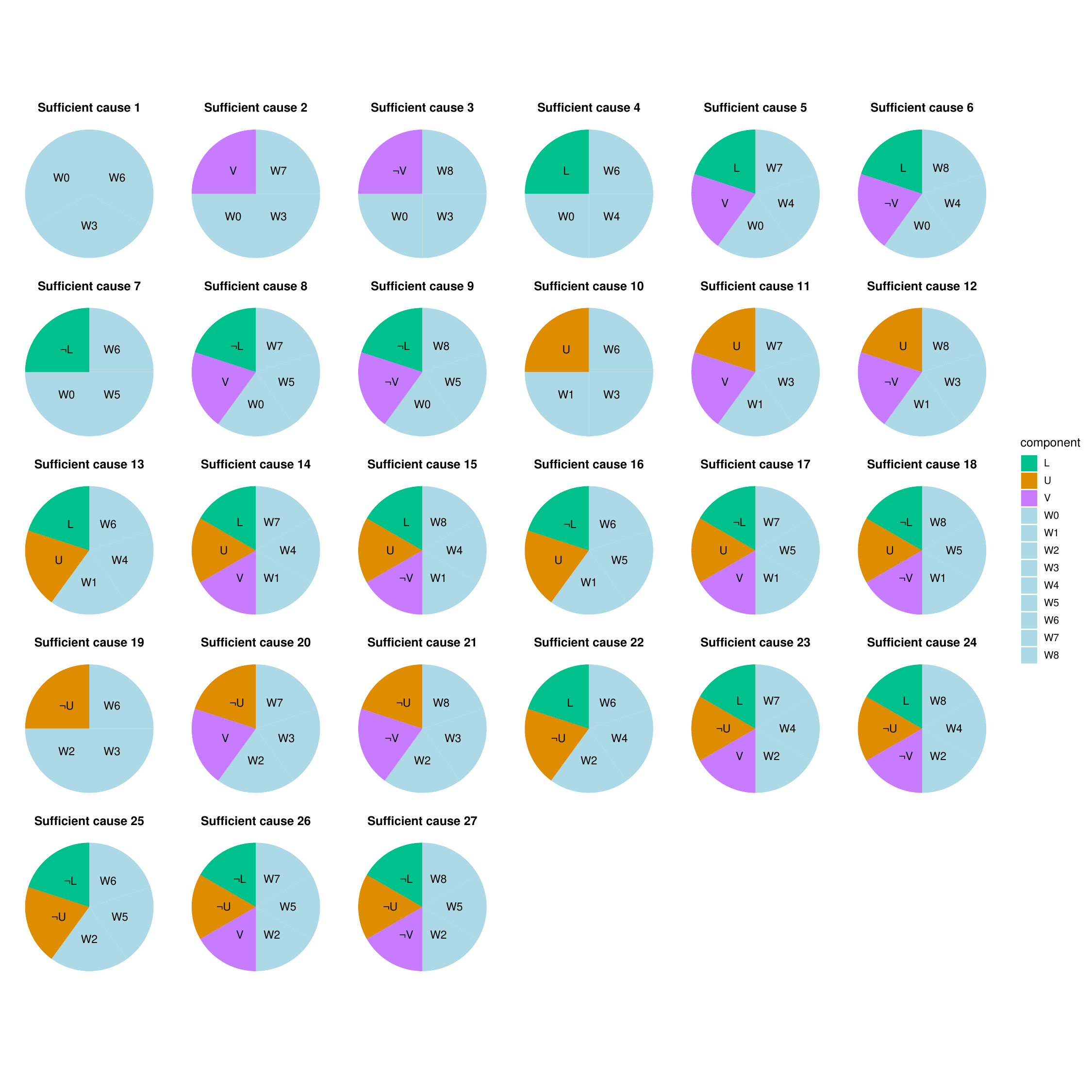}
    \caption{Sufficient cause models for the outcome} 
    \label{FIG: pies 27}
\end{figure}

Figure \ref{FIG: pieset9} represents the sufficient causes of the constituent outcomes. If any causal pie in a given row is completed, then the corresponding constituent outcome occurs. Since all three constituent outcomes are necessary for the outcome $Y$ to occur, at least one causal pie must hold true in each row. Thus, the sufficient causes for the outcome $Y$ are obtained by combining one completed causal pie from each row. These combinations are represented in Figure \ref{FIG: pies 27}.

For example, ``Sufficient cause 8’’ in Figure \ref{FIG: pies 27} corresponds to the following mechanism: The condition $\omega_0=1$ ensures that $Y_U$ occurs regardless of the value of $U$, the condition $\omega_5=1 \wedge \neg L=1$ ensures that $Y_L$ occurs, and the condition $\omega_7=1 \wedge V=1$ ensures that $Y_V$ occurs. Since all three constituent outcomes are necessary, the outcome $Y$ occurs when these three conditions hold simultaneously.

The sufficient-component cause representation implies
\begin{align*}
    P(Y=1|U=u, L=l, V=v)&=P(Y_U\wedge Y_L\wedge Y_V|U=u, L=l, V=v)\\
    &=P((\omega_u=1)\wedge(\omega_l=1)\wedge(\omega_v=1)|U=u, L=l, V=v),
\end{align*}
where we denote with $\omega_u, \omega_l$, and  $\omega_v$ denote the exogenous factors that complete the sufficient causes for $Y_U, Y_L$, and $Y_V$, respectively, under the covariate values being conditioned on. For example, when $L=1$, the relevant exogenous factors for $Y_L$ are $\omega_3$ and $\omega_4$, so $\omega_{l=1}=\neg (\omega_3=0 \wedge \omega_4=0)$.

By mutual independence of the exogenous variables and their independence from $U$, $L$, and $V$,
\begin{align*}
    &P((\omega_u=1)\wedge(\omega_l=1)\wedge(\omega_v=1)|U=u, L=l, V=v)\\
    &=P(\omega_u=1|U=u)P(\omega_l=1|L=l)P(\omega_v=1|V=v).
\end{align*}
Thus, the probability of the outcome $Y$, equivalently, the probability that at least one of the causal pies in Figure \ref{FIG: pies 27} is completed, factorizes into the product of a function of $U$, a function of $L$, and a function of $V$. This shows that our data generating mechanism induces a multiplicative model for the outcome.

Intuitively, the multiplicative model is induced because we posited that the outcome occurs when three independent mechanisms, each depending only on one of the three variables $U,L,V$, result in a ``event" at the same time. The outcome is the result of the alignment of three independent processes. This might be a reasonable assumption for certain outcomes like vaccine-induced immune thrombocytopenia and thrombosis (VITT). It has been suggested for example that VITT arises when an individual carries a specific genetic mutation, is exposed to adenoviruses like the one used as vector in some COVID-19 vaccines, and at the same time a somatic hypermutation has occurred \citep{wang_adenoviral_2026}.

\section{Justification of the Poisson regression}\label{APP: Pois reg}

As the frequency of the data collection increases, for example, instead of daily data collection, the data is collected every minute, it is reasonable to aim to approximate the data-generating mechanism of the outcome by a continuous-time process. However, even when the data is collected with sparse periodicity, the underlying data-generating mechanism may follow a continuous time process, with the data collection introducing bias to its estimation. In the following, we will describe how Assumptions \ref{ASS: stat all}-\ref{ASS: no int} and \ref{ASS: trans risk}-\ref{ASS: trans conf} ensure that this underlying continuous-time process is a Poisson process, which is log-linear in the covariates.

Suppose that the data are generated according to the DAG in Figure \ref{FIG: DAG}. As in Assumption \ref{ASS: rare event}, suppose that as the number of time points increases, the instantaneous expectations decrease, and the outcomes $Y_k$ converge to ``essentially" Bernoulli distributed random variables, see proof of Proposition \ref{PROP: c-RR(u,l) St=st} for details.
For example, suppose that the follow-up period lasts for 180 days, and each time point $k$ corresponds to 1 hour in that period, resulting in $t=4320$. Due to the small length of each of the time points, it is reasonable to treat the probability of thrombosis $P(Y_k>1|Y_k>0) \approx 0$ for all $k$.
 By Assumption \ref{ASS: ind out} (i), each outcome is independent from the joint distribution of all other outcomes, conditional on the time-fixed covariate ($U$) and the full histories of the treatment and the covariates $(\overline{A}_t, \overline{L}_t)$.

Assumption \ref{ASS: stat all} implies that for a given level of $\overline{a}_k \in \{0,1\}^k$, the expectation of $Y_k^{\overline{a}_k}$ is fixed across time. That is, $E[Y_k^{\overline{a}_k}|U=u, L_k=l]$ does not depend on $k$, instead depending on the values of $\overline{a}_k \in \{0,1\}^k, u \in \mathcal{U}, l \in \mathcal{L}$ only. Since $V^{\overline{a}_k}$ is deterministic, and the effect of treatment on $Y_k$ is fully mediated through $V_k$, see Corollary \ref{COR: int group}, the counterfactual conditional expectation can be equivalently expressed as a function of $v$, $u$, and $l$. Denote this function as $e(v,u,l)$.

Assumption \ref{ASS: Stat U} states that for any values of $u \in \mathcal{U}', l\in \mathcal{L}'$, the rate ratio is equal to some constant $\alpha \in [0, \infty)$. Thus, for all $u \in \mathcal{U}', l\in \mathcal{L}'$,
$$
\frac{e(1,u,l)}{e(0,u,l)}=\alpha,
$$
which implies that 
$$
e(1,u,l)=\alpha\cdot e(0,u,l),
$$
that is, the effect of being susceptible is proportional within a fixed level of $u$ and $l$.

For any $l, u \in \mathcal{L}' \times \mathcal{U}'$, we can define the ratio of the expected potential outcome in a population with $U=u$ and $L_k=l$ under not being at-risk, versus the expected potential outcome under not being at-risk, versus some reference levels $L_k=l^*$, $U=u^*$. That is 

$$
\beta(u, l)=\frac{e(0,u, l)}{e(0,u^*, l^*)}.
$$
From Assumption \ref{ASS: Stat U} it follows that
$$
\beta(u, l)=\frac{e(0,u, l)}{e(0,u^*, l^*)}=\frac{e(1,u, l)}{e(1,u^*, l^*)}
$$
thus $\beta(u, l)$ describes the ratio of the expected potential outcomes between a population in which the unmeasured covariate is equal to $u$ and the time-varying covariate is equal to $l$ versus a population in which the unmeasured covariate is equal to the reference level $u^*$, and the time-varying covariate is equal to the reference level $l^*$ for any fixed level of the at-risk indicator $V_k$.

To make the model multiplicative in $U$ and $L_k$ separately, we had to impose Assumption \ref{ASS: no int} as well.

Assumption \ref{ASS: no int} is plausible whenever we assume that $U$ and $L_k$ act on the potential outcomes via independent pathways. For example, if $U$ is the individual-level confounder, and $L_k$ denotes age, if the effect of age on thrombosis is the same for all individuals, without interacting with the individual-level factors, Assumption \ref{ASS: no int} holds.

By Assumption \ref{ASS: Stat U}, Assumption \ref{ASS: no int} implies that no interaction holds under being at-risk as well.

By Assumption \ref{ASS: no int}
$$
\frac{e(0, u, l)}{e(0, u^*, l^*)}= \frac{e(0, u^*, l)}{e(0, u^*, l^*)}\frac{e(0, u, l^*)}{e(0, u^*, l^*)},
$$
which ratios on the right-hand side we can denote as $\beta(l)$ and $\beta(u)$, respectively.

Under these results on proportionality, any $e(v, u, l)$ can be expressed with respect to the reference level, as 
\begin{equation}\label{EQ: prop}
    e(v, u, l)=
    \alpha(v) \cdot  \beta(u) \cdot \beta(l) \cdot  e(0, u^*, l^*),
\end{equation}
where $\alpha(1)=\alpha$ and $\alpha(0)=1$.

Denote the baseline reference history as $\overline{h}_t^0:=(\overline{a}_t=0, u^*, \overline{l}_k=l^*)$. 
Similarly to the proof of Proposition \ref{PROP: c-RR(u,l) St=st}, it follows that as $t \rightarrow \infty$ the sum of the outcomes under the baseline history converge to a Poisson distributed random variable, with parameter $\lambda(\overline{h}_t^0)$, which is equal to the limit $t \cdot e(0, u^*, l^*)$, due to the homogeneity across time. Now, define an arbitrary subsequence of the time points $I(t)$, such that the number of time points is non-zero measure of all time points as $t \rightarrow \infty$, that is $\frac{|I(t)|}{t}\rightarrow \sigma(I(t))$, where $\sigma(I(t))>0$. Since the sum of the expectation across the subsequence is equal to $|I(t)|\cdot e(0, u^*, l^*)$, by identical reasoning as for the full history, it follows that the sum of the outcomes in the subsequence converge to a Poisson random variable, with parameter $ \frac{|I(t)|}{t}\cdot t \cdot e(0, u^*, l^*) \rightarrow \sigma(I(t)) \cdot \lambda(\overline{h}_t^{0})$.

Now take any interval $I(t)$, for which the non-zero measure limit holds, and the values of the at-risk indicator and the covariates are the same for each $k \in I(t)$, however, not necessarily corresponding to baseline. Based on Equation \eqref{EQ: prop}, for each $k \in I(t)$ we can rewrite the expectation, based on the corresponding history of the covariates and $V$ as $E[Y^{\overline{a}_k}|U=u, \overline{L}_k=\overline{l}_k]=e(v, u,l)= \alpha(v) \cdot  \beta(u) \cdot \beta(l) \cdot  e(0, u^*, l^*)$. By the previous paragraph it follows that the sum of these outcomes in the interval is converging to a Poisson-distributed random variable, with parameter 
$$
    \alpha(v) \cdot  \beta(u) \cdot \beta(l) \cdot  e(0, u^*, l^*) \cdot t \cdot  \frac{|I(t)|}{t} \rightarrow \alpha(v) \cdot  \beta(u) \cdot \beta(l) \cdot \sigma(I(t)) \cdot \lambda(\overline{h}_t^{0}).
$$

Then, by rescaling $\sigma(I(t))$ to some reference unit of time, e.g. $\sigma'(I(t))$ denoting the length of the interval $I(t)$ in days, and accordingly rescaling $\lambda'(\overline{h}_t^{0})$ to denote the baseline risk in one day, we showed the number of events in a given interval can be modelled according to the proposed log-linear Poisson model.

Assumptions \ref{ASS: pos} and \ref{ASS: trans risk}-\ref{ASS: trans conf} were needed, to interpret $\alpha(v)$ as target estimand $CRR_m$, as shown in the proof of Proposition \ref{PROP: CRR}.

A formal proof of Proposition \ref{PROP: pois} is provided in Appendix \ref{APP_Sub: pois proof}.

\section{Bias introduced by conditioning on exact event numbers}\label{APP: bias}
In the proof of Proposition \ref{PROP: c-RR(u,l) St=st}, we relied on the asymptotic Assumption \ref{ASS: rare event} to show that the bias introduced by conditioning on the exact number of events, that is, selection bias, vanishes as the data-generating process approaches its continuous-time limit. In this section, we provide a non-asymptotic result on this bias that holds even when the outcome is not rare, and hence when the data-generating process does not approximate a Poisson process. We show that values of $S_t=s$ for which the bias is large are unlikely to be observed. This provides intuition for why the SCCS design can have negligible bias even when the data are collected on a discrete time scale when the analysis conditions on the observed event count $S_t=s$. Since we are not relying on the asymptotic rarity, in the following we assume that $Y_k$ is binary for all $k$, that is, that Assumption \ref{ASS: bin} holds.

Using notation identical to the main text, the ratio
$$
\frac{P(Y_{k^1}=1|  \overline{H}_t=\overline{h}_t)}{P(Y_{k^0}=1| \overline{H}_t=\overline{h}_t)},
$$
which by Assumption \ref{ASS: exch}, \ref{ASS: pos} and consistency identifies the $c\text{-}RR(u,l)_k$, can be rewritten as 
    $$
    \frac{E[Y_{k^1}|  \overline{H}_t=\overline{h}_t, S_t=s]}{E[Y_{k^0}|  \overline{H}_t=\overline{h}_t, S_t=s]}\cdot 
    \mathrm{bias}(p^1, p^0, s, \overline{h}_t),
    $$
with the bias term defined as 

    \begin{align*}
        &\mathrm{bias}(p^1, p^0, s, \overline{h}_t)=\frac{P(S_{t, -k^1, -k^0}=s-2| \overline{H}_t=\overline{h}_t)p^1+P(S_{t, -k^1, -k^0}=s-1|  \overline{H}_t=\overline{h}_t)(1-p^1)}{P(S_{t, -k^1, -k^0}=s-2| \overline{H}_t=\overline{h}_t)p^0+P(S_{t, -k^1, -k^0}=s-1| \overline{H}_t=\overline{h}_t)(1-p^0)}
    \end{align*}
    where 
    $$
    S_{t, -k^1, -k^0}=\sum_{\substack{k=1 \\ k \neq k^1,\; k \neq k^0}}^{t} Y_k, 
    $$
    and $p^1=P(Y_{k^1}=1|  \overline{H}_t=\overline{h}_t)$ and $p^0=P(Y_{k^0}=1| \overline{H}_t=\overline{h}_t)$.

We wish to minimize the deviation of the bias term from 1, which corresponds to the absence of bias. First, note that if $p_0=p_1$, as under the null hypothesis of no treatment effect, Hypothesis \ref{HYP: null}, then the bias term is equal to 1. This agrees with the causal representation: if the arrows from $V_k$ to $Y_k$ are absent from the DAG in Figure \ref{FIG: DAG}, corresponding to no causal effect of exposure on the outcome, then conditioning on $S_t=s$ no longer induces selection bias.

There is, however, a second mechanism by which the bias term can vanish: if
$$
P(S_{t,-k^1,-k^0}=s-2\mid \overline H_t=\overline h_t)
=
P(S_{t,-k^1,-k^0}=s-1\mid \overline H_t=\overline h_t),
$$
then the numerator and denominator in $\mathrm{bias}(p^1, p^0, s, \overline{h}_t)$ are equal, regardless of the values of $p_1$ and $p_0$, and the bias term is equal to $1$. Thus, even under a non-null treatment effect, the conditioning bias disappears at the values of $s$ where the probability mass function of the reduced event count $S_{t,-k^1,-k^0}$ is locally flat.

This observation suggests a way to characterize regions of small bias.

\begin{lemma}\label{APP PROP: small bias}
Suppose that Assumption \ref{ASS: bin} holds and that $p^0<1$.

Fix a history $H_t=h_t$.
Define the adjacent mass ratio
$$
r_s=\frac{P(S_{t, -k^1, -k^0}=s-2 \mid H_t=h_t)}{P(S_{t, -k^1, -k^0}=s-1 \mid H_t=h_t)},
$$
whenever $P(S_{t, -k^1, -k^0}=s-1 \mid H_t=h_t)>0$.
Then for any $\epsilon>0$,
$$\left|r_s -1 \right|\leq \epsilon
\implies
\left| \mathrm{bias}(p^1,p^0,s, \overline{h}_t)-1 \right|
\leq
\frac{|p^1-p^0|}{1-p^0}\epsilon.$$
\end{lemma}

See the proof in Appendix \ref{APP_sub: proof lemma}

Lemma \ref{APP PROP: small bias} formalises our claim from before, that in the locally flat regions of $S_{t, -k^1, -k^0}$ the bias is small. What is left to show, that the locally flat regions of $S_{t, -k^1, -k^0}$ is just a single region, close to the mode of $S_t$.

\begin{proposition}\label{APP PROP: S mode}
    Suppose that Assumptions \ref{ASS: ind out} and \ref{ASS: bin} hold.

    Then $S_{t, -k^1, -k^0}\mid H_t=h_t$ is Poisson-Binomial, thus its probability mass function is log-concave. Therefore the adjacent PMF ratio $r_s$ is nondecreasing in $s$. Hence the set of values of $s$ for which the adjacent masses are close, ($\{s: |r_s-1|<\epsilon\}$) is an interval, which includes the $m+1$ or $m+2$, where $m$ is the largest mode of $S_{t, -k^1, -k^0}\mid H_t=h_t$, whenever this set is not empty.

    Denote the set of modes of $S_{t, -k^1, -k^0}\mid H_t=h_t$ and $S_t|\overline{H}_t=\overline{h}_t$ with $\mathcal{M}_{S_t^-}$ and $\mathcal{M}_{S_t}$, respectively. If
    $$
    \mathcal{M}_{S_t^-}=\{s'\},
    $$
    then 
    $$
    \mathcal{M}_{S_t}\subset \{s', s'+1, s'+2\},
    $$
    and if
    $$
    \mathcal{M}_{S_t^-}=\{s', s'+1\},
    $$
    then 
    $$
    \mathcal{M}_{S_t}\subset \{s', s'+1, s'+2, s'+3\},
    $$
    where the cardinality of both $\mathcal{M}_{S_t^-}$ and $\mathcal{M}_{S_t}$ at most is $2$. Therefore, the mode(s) of $S_t$ are at most two event counts away from the value, for which the bias is minimised.
\end{proposition}

See the proof in Appendix \ref{APP_sub: proof S mode}.

\begin{figure}
    \centering
    \includegraphics[width=0.9\linewidth]{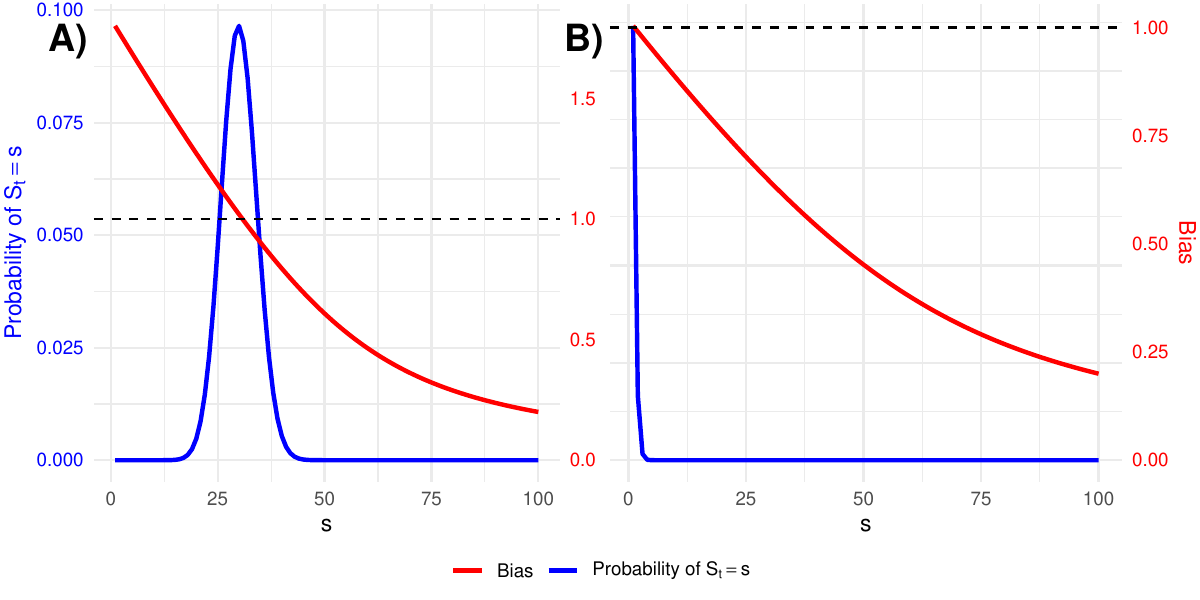}
    \caption{Illustrative scenario for the finite timepoint case. In both scenarios we considered $50$-$50$ treated and untreated timepoint, without any $U$ or $L$ covariates. Since the timepoints are exchangeable, the order of treatment does not matter. We plotted the probability $P(S_{100}=s|\overline{H}_{100}=\overline{h}_t)$ and the bias $\mathrm{bias}(p^1, p^0, s, \overline{h}_{100})$. In Subfigure $A)$ we set the probability of the outcome under treatment to $p^1=0.5$ at all treated time points, and to $p^0=0.1$ for all untreated timepoints. In Subfigure $B)$ $p^1=0.005$ and $p^0=0.001$. As the probabilities decrease, the distribution of $S_t$ approximates a Poisson distribution, in which the ratio of the number of events with non-negligible probabilities and the total number of timepoints $t$, goes to $0$.}
    \label{FIG: finite bias}
\end{figure}

By Lemma \ref{APP PROP: small bias} and Proposition \ref{APP PROP: S mode} it follows that the region in which the bias is minimised includes $\max(\mathcal{M}_{S_t^-})+1$ or $\max(\mathcal{M}_{S_t^-})+2$, and it is at most two event counts away from the most probable value(s) of $S_t$. Thus the values of $s$ for which the bias is small are those that occur with high probability. Equivalently, the values of $s$ for which the bias is large occur with small probability, as illustrated in Figure \ref{FIG: finite bias}.

\section{Bootstrap test}\label{APP: boot test}
We introduced an alternative to the standard $SCCS$ method in Section \ref{SEC: test}, an exact test, which, although it can only test the existence of a treatment effect, but not estimate it, relies on considerably less restrictive assumptions than the parametric methods of the $SCCS$ framework. Nonetheless, we imposed exchangeability between the outcomes, therefore restricting the dependence structure over time. In the following we introduce a different test, Algorithm \ref{ALG: Dependent} which is asymptotically valid under unspecified dependence structure.

{\spacingset{1}
\begin{algorithm}[!ht]
\caption{Person-level bootstrap test}
\label{ALG: Dependent}
\begin{algorithmic}[1]
\Require Observed cases $i=1, \dots, n$; informative strata $l\in \mathcal{L}_i^+=\{l:|\mathcal{K}_{1,il}|>0,|\mathcal{K}_{0,il}|>0\}$ with:
\Statex \hspace{\algorithmicindent} $\mathcal{K}_{1, il}$ = number of at-risk intervals for given $l$ and individual $i$
\Statex \hspace{\algorithmicindent} $\mathcal{K}_{0, il}$ = number of non-risk intervals for given $l$ and individual $i$
\Statex \hspace{\algorithmicindent} $s_{\mathcal{K}_{1,il}}$ = observed number of events in at-risk intervals for given $l$ and individual $i$
\Statex \hspace{\algorithmicindent} $s_{\mathcal{K}_{0,il}}$ = observed number of events in not at-risk intervals for given $l$ and individual $i$
\Require Number of Monte Carlo replicates $B$
\Ensure Observed statistic $D$, Monte Carlo $p$-value, null centred bootstrap distribution $\{D^{(b)}\}_{b=1}^B$

\For{$i\in \{1, \dots, n\}$ and $l\in\mathcal L_i^+$}
    \State Compute the observed difference score:
    \[
    D_{\mathcal{K}_{il}} := \frac{s_{\mathcal{K}_{1,il}}}{|\mathcal{K}_{1,il}|}-\frac{s_{\mathcal{K}_{0,il}}}{|\mathcal{K}_{0,il}|}
    \]
    \State Compute the inverse variance weight
    \[
    w_{il} := \left(
    \frac{|\mathcal{K}_{1,il}|+  |\mathcal{K}_{0,il}|}{|\mathcal{K}_{1,il}| \cdot |\mathcal{K}_{0,il}|} \right)^{-1}
    \]
\EndFor

\For{$i\in \{1, \dots, n\}$}
    \State Compute the person-level contrast
    \[
    D_i := \frac{\sum_{l\in \mathcal{L}_i^+} w_{il} \cdot D_{\mathcal{K}_{il}}}{\sum_{l \in \mathcal{L}_i^+} w_{il}}
    \]
\EndFor

\State Compute the observed test statistic:
\[
D := \frac{1}{n}\sum_{i=1}^n  D_{i}
\]
\State Center the person-level contrasts:
\[
D_i^0 := D_i -  D
\]

\For{$b=1,\dots,B$}
    \State Sample $n$ individuals with replacement from $\{D_1^0,\dots,D_n^0\}$
    \State Compute the null bootstrap statistics $ D^{0(b)}$
\EndFor

\State Compute the two-sided $p$-value
    \[
    p_{\text{two-sided}}^{B}:=\frac{\sum_{b=1}^B \mathbb{I}(|D^{0(b)}|\geq |D|)+1}{B+1}
    \]

\State \Return $ D$, $p_{\text{two-sided}}^{B}$, $\{D^{0(b)}\}_{b=1}^B$
\end{algorithmic}
\end{algorithm}
}

\begin{proposition}\label{PROP: test boot}
    Suppose that Assumptions \ref{ASS: stat all}, \ref{ASS: bin}, \ref{ASS: exch}, \ref{ASS: ind out} (ii), (iii), and \ref{ASS: trans conf} hold, the testing procedure proposed in Algorithm \ref{ALG: Dependent} is an asymptotically valid level-$\alpha$ test of Hypothesis \ref{HYP: null}.
\end{proposition}

Informally, since by Assumption \ref{ASS: stat all}, the event rates are invariant over time, and within each individual there is no unmeasured confounding between the outcomes at treated and untreated times, within each stratum the event rate under treatment and under no treatment should have the same expectation, under the null hypothesis. See the proof in Appendix \ref{APP: proof boot}.

\section{Censored event times}\label{APP: death}
A key assumption of the SCCS design is that the length of the observation period is independent of the outcome process. This condition is satisfied if we assume all individuals are observed for the total duration of the follow-up period, that is, if we assume that no censoring occurs. However, in most practical settings this is implausible. In the following, let $C_k$ indicate whether a censoring event has occurred at time $k$.

Alternatively, whenever censoring is independent from the outcome \citep{aalen_stochastic_2008}, that is, if early termination occurs from reasons that are not attributable to the outcome process, conditioning on the time of the censoring, or alternatively, on the length of the follow-up periods, we do not induce spurious dependence between the treatment and the outcome. 

\begin{assumption}[Independent censoring]\label{ASS_app: censi ind discrete}
    $$\overline{C}_k \independent \overline{Y}_k|\overline{A}_k, \overline{L}_k, U,$$
for all $k=1, \dots, t$.    
\end{assumption}

Consider an individual whose follow-up ends at time $\tau\leq t$, and let $S_\tau=\sum_{j=1}^{\tau} Y_j$ denote the number of events during the observed follow-up. Since $S_\tau$ and $Y_k$, for any $k\leq \tau$, are deterministic functions of $\overline{Y}_\tau$, it follows from Assumption \ref{ASS_app: censi ind discrete} at time $\tau$, using decomposition and weak union from the graphoid axioms, that
$$
Y_k\independent \overline{C}_\tau | S_\tau, \overline{A}_\tau, \overline{L}_\tau, U.
$$
 
Under  this independence condition, we can rewrite
$$
E[Y_k|\overline{A}_\tau=\overline{a}_\tau, \overline{L}_\tau=\overline{l}_\tau, U=u, S_\tau=s]=E[Y_k|\overline{A}_\tau=\overline{a}_\tau, \overline{L}_\tau=\overline{l}_\tau, U=u, S_\tau=s, \overline{C}_\tau=\overline{c}_\tau],
$$
for $k\leq\tau$, that is, conditioning on the end of the follow-up does not alter the conditional expectations on which the within-individual contrasts of Section \ref{SEC: ident} are based, with $t$ replaced by $\tau$.

However, rare adverse outcomes are of interest because of their severity, which eventually can lead to a level of health deterioration that prevents the individuals from being followed up on, or to death, both of which results in censoring. In these scenarios, the end of the follow-up, that is censoring, is informative about the timing of the event, thus, by conditioning on the follow-up period, we introduce bias to the treatment effect estimate. 

\citet{farrington_self-controlled_2018} introduce a general solution for dependent censoring by incorporating a parametric weight function in the likelihood to account for the information provided by the time elapsed between the outcome(s) and censoring. The estimation of the weights are done in a separate procedure from the SCCS fitting procedure, and it is discussed in detail in Section 7.2.7 of \citet{farrington_self-controlled_2018}. While the method, in principle solves the issue of event-dependent censoring, it is only applicable for non-recurrent events, which is often not the case in the data of interest. In addition the following must hold

\begin{assumption}[Conditional independence of censoring and treatment]\label{ASS_app: no trt on C}
    $$
    \overline{A}_k \independent \overline{C}_k | \overline{Y}_k, \overline{L}_k, U
    $$
\end{assumption}

Under Assumption \ref{ASS_app: no trt on C} it is ruled out that the treatment has any effect on the censoring that is not mediated through the outcome, or that there is any source of confounding other than $U$ or $L_k$ between the censoring and the treatment processes. It is reasonable to expect that the treatment and the censoring is independent, whenever the outcome is defined such that it captures all serious adverse events that can be plausibly caused by the treatment.

In the following, we will discuss a special case in which the treatment effect can be estimated without bias, even if the outcome induces censoring.

\subsection{Linear censoring hazard}\label{APP_sub: lin censoring}
We have shown in Appendix \ref{APP: Pois reg}, that as we approach continuous time, the event rate at any interval in which the covariates $V_k, L_k, U$ take constant values is described by a Poisson random variable with parameter 
$$
\lambda = \alpha(v)\beta(u)\beta(l)\sigma(I(t))\lambda(\overline{h}_t^0),
$$
where $\sigma(I(t))$ is the measure of length of the interval, and $\lambda(\overline{h}_t^0)$ is the reference (baseline) event rate per unit time. In the following, we assume that, conditional on the history, the events follow an inhomogeneous Poisson process with the corresponding intensity.
Thus, the intensity at any time $k$ can be written as 
$$
\lambda(k)=\alpha(v(k))\beta(u(k))\beta(l(k))\lambda^0,
$$
where $\lambda^0$ is the baseline intensity, while $v(k), u(k), l(k)$, are the values that the covariates are taking at time $k$.

Suppose that besides the time-varying confounder $L_k$, there is a nuisance parameter $\eta(t)$, that captures the effect of time on the event rate, that is, given an interval, $(a, b)$, at which the covariates  are taking a fixed value, the expected number of events of the inhomogeneous Poisson process in that interval is 
$$
\alpha(v)\beta(u)\beta(l)\lambda^0 \int_{a}^{b} \eta(k)\; \mathrm{d}k,
$$
or equivalently, the intensity at any time $k$ is equal to
$$
\lambda(k)=\alpha(v(k))\beta(u(k))\beta(l(k))\lambda^0 \eta(k).
$$

In this scenario, the intensity of the outcome process does not only change with the observed and unobserved covariates, but it is also a function of time $\eta(k)$, which has an assumed multiplicative effect on the outcome intensity.

Following \citet{farrington_self-controlled_2018} suppose that the continuous time censoring hazard at time $k$ is defined as 
$$
\mu(k|\overline{y}_k, \overline{h}_k)=\mu(k, u)+\theta \cdot s_k,
$$
where $s_k$ denotes the total number of events until time $k$. Note that by Assumption \ref{ASS_app: no trt on C} the censoring hazard does not depend on the treatment history $\overline{a}_k$, and while in principle it could be a function of the time-varying confounder $\overline{l}_k$, for simplicity we suppressed that dependence.

Suppose that the time-scale is chosen such that the follow-up starts at time $0$. Then the probability of remaining uncensored until time $c$ is 
$$
\mathrm{exp}\left(-\int_0^c \mu(k, u)+\theta \cdot s_k \;\mathrm{d}k  \right).
$$

Clearly, the first factor does not depend on the outcomes occurring before time $c$, while the second factor of the probability of remaining uncensored can be written as,
$$
\mathrm{exp}\left(-\theta \int_0^c s_k \; \mathrm{d}k \right)=\mathrm{exp}\left(- \sum_{j=1}^{s_c}(c-k_j)\right),
$$
where $s_c$ is the number of events until time $c$, and the $k_j$-s are the times of the events. The equality holds, because $s_k$ is a step function. Therefore, the probability of being uncensored until time $c$, given the history of the covariates $\overline{h}_c$ and the timing of the events $k_1, \dots , k_{s_c}$ is 
$$
\mathrm{exp}\left(-\int_0^c \mu(k, u) \; \mathrm{d}k\right)\mathrm{exp}(-\theta(c\cdot s_c))\prod_{j=1}^{s_c}\mathrm{exp}(\theta \cdot k_j).
$$

Without event dependent censoring, the conditional likelihood of the event times are 
\begin{align*}
    P(k_1, \dots , k_{s_c}&|\overline{H}_t=\overline{h}_t, S_c=s_c)=\frac{\prod_{j=1}^{s_c} \lambda(k_j) }{\underset{0\leq k'_1 \leq \dots \leq k'_{s_c}\leq c}{\int \dots \int} \prod_{j=1}^{s_c} \lambda(k'_j) \; \mathrm{d} k'_1 \dots \mathrm{d}k'_{s_c}}\\
    &=\frac{\prod_{j=1}^{s_c}\alpha(v(k_j))\beta(u(k_j))\beta(l(k_j))\lambda^0 \eta(k_j) }{\underset{0\leq k'_1 \leq \dots \leq k'_{s_c}\leq c}{\int \dots \int} \prod_{j=1}^{s_c} \alpha(v(k'_j))\beta(u(k'_j))\beta(l(k'_j))\lambda^0 \eta(k'_j)  \; \mathrm{d} k'_1 \dots \mathrm{d}k'_{s_c}}
\end{align*}

However, when the censoring hazard depends on prior outcome events, the conditional likelihood of the event times, given also that the follow-up ends at time $c$, becomes
$$
P(k_1, \dots , k_{s_c}|\overline{H}_t=\overline{h}_t, S_c=s_c)=\frac{\prod_{j=1}^{s_c} \lambda(k_j)\mathrm{exp}(\theta \cdot k_j) }{\underset{0\leq k'_1 \leq \dots \leq k'_{s_c}\leq c}{\int \dots \int} \prod_{j=1}^{s_c} \lambda(k'_j)\mathrm{exp}(\theta \cdot k'_j) \; \mathrm{d} k'_1 \dots \mathrm{d}k'_{s_c}},
$$
where we simplified fraction with the non-time-varying terms of the probability of remaining uncensored.

By defining a modified nuisance function of the time effect as
$$
\eta'(k)=\eta(k)\mathrm{exp}(\theta k),
$$
the conditional likelihood becomes
$$
\frac{\prod_{j=1}^{s_c}\alpha(v(k_j))\beta(u(k_j))\beta(l(k_j))\lambda^0 \eta'(k_j) }{\underset{0\leq k'_1 \leq \dots \leq k'_{s_c}\leq c}{\int \dots \int} \prod_{j=1}^{s_c} \alpha(v(k'_j))\beta(u(k'_j))\beta(l(k'_j))\lambda^0 \eta'(k'_j)  \; \mathrm{d} k'_1 \dots \mathrm{d}k'_{s_c}},
$$
which is identical to the standard SCCS likelihood, except for the modified time effect term. Therefore, when the censoring hazard is linear in the number of outcome events, and the model of the continuous time effect is flexible enough to capture $\eta'(k)$, the standard SCCS can still be used to estimate the at-risk effect $\alpha(v)$ (and the covariate effect $\beta(l)$), only the time effect will be biased by the event dependent censoring.

\section{Rationale for summaries in Algorithms \ref{alg: hyper} and \ref{ALG: Dependent}}\label{ASS: algs}

We introduced two testing approaches in Section \ref{SEC: test} and Appendix \ref{APP: boot test}, which provide valid inference on the null hypothesis of no effect, even when crucial modelling assumptions of the SCCS design fail. In the following, we justify the choice of the summary methods across individuals, and the levels of the time-varying confounder $L_k=l$.

\subsection{Power of the test statistics}\label{APP_subsub: hyper power}
Suppose that the family of alternatives is defined such that for each individual with $U=u$ and for each stratum $L_k=l$, the odds ratio is invariant, that is
$$
\frac{P(Y_k=1| V_k=1, L_k=l, U=u)/P(Y_k=0| V_k=1, L_k=l, U=u)}{P(Y_k=1| V_k=0, L_k=l, U=u)/P(Y_k=0| V_k=0, L_k=l, U=u)}=\mathrm{exp}({\beta}),
$$
for all $u,l \in \mathcal{U'}\times \mathcal{L}'$ and for all timepoints $k$.
The null hypothesis of no at-risk effect then corresponds to $\beta =0$.

For simplicity, under Assumption \ref{ASS: stat all}, denote with $p_1$ and $p_0$ the probabilities $P(Y_k=1| V_k=1, L_k=l, U=u)$ and $P(Y_k=1| V_k=0, L_k=l, U=u)$, where we omitted $u,l$ to reduce notational burden. Under the assumption of independent outcomes, in this family, the conditional distribution of the events happening at risk times $\mathcal{K}_{1,l}$ is
\begin{align*}
    &P_\beta(S_{\mathcal{K}_{1,l}}=s_{\mathcal{K}_{1,l}}| |\mathcal{K}_{1,l}|, |\mathcal{K}_{0,l}|, s_{\mathcal{K}_l})\\
    &\propto {|\mathcal{K}_{1,l}| \choose s_{\mathcal{K}_{1,l}}} p_1^{s_{\mathcal{K}_{1,l}}} (1-p_1)^{|\mathcal{K}_{1,l}| -s_{\mathcal{K}_{1,l}}} {|\mathcal{K}_{0,l}| \choose s_{\mathcal{K}_{l}}- s_{\mathcal{K}_{1,l}}} p_0^{s_{\mathcal{K}_{0,l}}} (1-p_0)^{|\mathcal{K}_{0,l}| -(s_{\mathcal{K}_{l}}-s_{\mathcal{K}_{1,l}})} \\
    &\propto  {|\mathcal{K}_{1,l}| \choose s_{\mathcal{K}_{1,l}}} {|\mathcal{K}_{0,l}| \choose s_{\mathcal{K}_{l}}- s_{\mathcal{K}_{1,l}}} \left(\frac{p_1}{1-p_1}\right)^{s_{\mathcal{K}_{1,l}}} \left(\frac{p_0}{1-p_0}\right)^{-s_{\mathcal{K}_{1,l}}}\\
    &={|\mathcal{K}_{1,l}| \choose s_{\mathcal{K}_{1,l}}} {|\mathcal{K}_{0,l}| \choose s_{\mathcal{K}_{l}}- s_{\mathcal{K}_{1,l}}} \mathrm{exp}(\beta \cdot  s_{\mathcal{K}_{1,l}}).
\end{align*}
Therefore, the conditional likelihood of the observed data, given the stratum totals, can be written according to the one-parameter exponential family
$$
L(\underline{s_{\mathcal{K}_{1,l}}}, \beta) = h(s_{\mathcal{K}_{1,l}})\cdot\mathrm{exp}\left(\beta \cdot \sum_{i=1}^n\sum_{l\in\mathcal{L}_i^+} s_{\mathcal{K}_{1,il}} -A(\beta)\right),
$$
where, as in Algorithm \ref{alg: hyper}, $s_{\mathcal{K}_{1,il}}$ denotes $s_{\mathcal{K}_{1,l}}$ of individual $i$.

Thus, in the family of alternatives, the likelihood ratio is monotone in the sufficient statistics $\sum_{i=1}^n\sum_{l\in\mathcal{L}_i^+} s_{\mathcal{K}_{1,il}}$, and thus the test based on the $\sum_{i=1}^n\sum_{l\in\mathcal{L}_i^+} s_{\mathcal{K}_{1,il}}$ is uniformly most powerful among tests that condition on the stratum totals, and hence uniformly most powerful unbiased, against one sided composite alternatives in the characterized family. Since by conditioning on the treatment regimes and the number of total events, the expected number of events at-risk is fixed, thus centring by
$$
s_{\mathcal{K}_{l}}\frac{|\mathcal{K}_{1,l}|}{|\mathcal{K}_{1,l}|+|\mathcal{K}_{0,l}|}
$$
does not modify the rejection rates of the test statistics in the proposed family.

The two-sided test can be shown to have local asymptotic power, relying on the local asymptotic normality of the test statistics.

\subsection{Algorithm \ref{ALG: Dependent}}\label{APP_sub: dependent}
Due to the unknown dependence structure between outcomes, we cannot simulate from an exact distribution under the null. However, by Hypothesis \ref{HYP: null}, Assumption \ref{ASS: exch}, and consistency, ensures that
$$E[Y_k|U=u, \overline{L}_k=\overline{l}_k, \overline{A}_{k}=\overline{a}_{k}]=E[Y_k|U=u, \overline{L}_k=\overline{l}_k, \overline{A}_{k}=\overline{a}'_{k}].
$$
for any $k$, which enables us to show that under time invariance of the expectation, Assumption \ref{ASS: stat all}, the individual-level difference of the expected rate of the outcome should be $0$. While this allows us to derive an asymptotically valid test, we cannot ensure uniform power against all alternatives, however, we argue for the chosen weights by describing the scenarios under which the weights are optimal.

\subsubsection{Power of the test statistics}
Similarly to the subsection \ref{APP_subsub: hyper power}, we aim to show that the proposed test statistic has optimal power for a certain well-defined set of alternatives. When the outcomes are dependent, no optimal weights can be found for the difference scores $D_{\mathcal{K}_{l}}$. Therefore, the optimality of the test is derived under the assumption of independent outcomes.
Suppose that for all $u, l \in \mathcal{U}' \times \mathcal{L}'$
$$
E[D_{\mathcal{K}_l}]=E\left[\frac{S_{\mathcal{K}_{1,l}}}{|\mathcal{K}_{1,l}|}-\frac{S_{\mathcal{K}_{0,l}}}{|\mathcal{K}_{0,l}|}\right]=\delta,
$$
for some fixed $\delta$, where $\delta=0$ corresponds to the null hypothesis.

By the central limit theorem, as the number of at-risk and not at-risk time points goes to infinity, each difference can be approximated by
$$
D_{\mathcal{K}_l}\sim \mathcal{N}(\delta, v_{l}),
$$
where $v_l$ is the $l$-strata specific variance of the estimator, where the individual specific subscript $i$ is suppressed, as we suppose that the variance of the estimator $\widehat{D_{\mathcal{K}_l}}$ is approximately equal for each individual.

Consider all linear estimators of $\delta$ using data from a single individual
$$
\widehat{D}_i= \sum_l w^*_l \widehat{D}_{\mathcal{K}_l}.
$$
To ensure that the estimator is unbiased $\sum_l w^*_l =1$ must hold. Thus, by the Cauchy-Schwarz inequality
\begin{align*}
    1=\left(\sum_l w^*_l \right)^2 = \left(\sum_l w^*_l \sqrt{v_l}\frac{1}{\sqrt{v_l}}\right)^2 \leq \left(\sum_l (w^*_l)^2 v_l\right) \left(\sum_l \frac{1}{v_l}\right),
\end{align*}
therefore, the variance of the linear estimator $\sum_l w^*_l \widehat{D}_{\mathcal{K}_l}$ is bounded below by
$$
\mathrm{Var}(\widehat{D}_i)= \sum_l (w^*_l)^2 v_l\geq \frac{1}{\sum_l \frac{1}{v_l}},
$$
where we used the independence between the estimators $\widehat{D}_{\mathcal{K}_l}$.

The lower bound is reached when for all $l$ $w^*_l \propto (v_l)^{-1}$, which implies inverse variance weighting. Then, since the estimators $\widehat{D}_{\mathcal{K}_l}$ are sample means of essentially Bernoulli distributed random variables, under the equal event rates implied by the null hypothesis, the variance $v_l$ is $p_l(1-p_l)\left(\frac{1}{|\mathcal{K}_{1,l}|}+\frac{1}{|\mathcal{K}_{0,l}|}\right)$, where $p_l$ denotes the common event probability in stratum $l$, which can be estimated by
$$
\frac{s_{\mathcal{K}_{1,l}}+s_{\mathcal{K}_{0,l}}}{|\mathcal{K}_{1,l}|+|\mathcal{K}_{0,l}|} \cdot \left(1-\frac{s_{\mathcal{K}_{1,l}}+s_{\mathcal{K}_{0,l}}}{|\mathcal{K}_{1,l}|+|\mathcal{K}_{0,l}|}\right)\cdot  \frac{1}{|\mathcal{K}_{1,l}|} + \frac{s_{\mathcal{K}_{1,l}}+s_{\mathcal{K}_{0,l}}}{|\mathcal{K}_{1,l}|+|\mathcal{K}_{0,l}|} \cdot \left(1-\frac{s_{\mathcal{K}_{1,l}}+s_{\mathcal{K}_{0,l}}}{|\mathcal{K}_{1,l}|+|\mathcal{K}_{0,l}|}\right)\cdot  \frac{1}{|\mathcal{K}_{0,l}|}.
$$

However, in practice, for finite data, the estimate of the true event rate under the null $$
\frac{s_{\mathcal{K}_{1,l}}+s_{\mathcal{K}_{0,l}}}{|\mathcal{K}_{1,l}|+|\mathcal{K}_{0,l}|},
$$
might have significant variance, thus it is not included in the weights.
 Thus, the weights proposed in Algorithm \ref{ALG: Dependent} are derived, they are optimal when $p_l$ is the same across the strata of an individual, and otherwise approximate the inverse variance weights.

\end{document}